\documentclass[%
  twocolumn, 
  floatfix,
  aps, 
  prx, 
  final,
  amsmath, amsfonts, amssymb
]{revtex4-2}

\usepackage{graphicx}
\usepackage{hyperref}
\hypersetup{
  colorlinks=true,
  linkcolor=blue, 
  citecolor=blue,
}
\usepackage[utf8]{inputenc}
\usepackage{physics}
\begin{document}

\title{Strength and weakness of disease-induced herd immunity in networks}
\author{Takayuki Hiraoka}
\email{takayuki.hiraoka@aalto.fi}
\author{Zahra Ghadiri}
\author{Abbas K.~Rizi}
\author{Mikko Kivel\"{a}}
\author{Jari Saram\"{a}ki}
\affiliation{Department of Computer Science, Aalto University, 00076 Espoo, Finland} 

\begin{abstract}
When a fraction of a population becomes immune to an infectious disease, the population-wide infection risk decreases nonlinearly due to collective protection, known as herd immunity.  Some studies based on mean-field models suggest that natural infection in a heterogeneous population may induce herd immunity more efficiently than homogeneous immunization. However, we theoretically show that this is not necessarily the case when the population is modeled as a network instead of using the mean-field approach. We identify two competing mechanisms driving disease-induced herd immunity in networks: the biased distribution of immunity toward socially active individuals enhances herd immunity, while the topological localization of immune individuals weakens it. The effect of localization is stronger in networks embedded in a low-dimensional space, which can make disease-induced immunity less effective than random immunization. Our results highlight the role of networks in shaping herd immunity and call for a careful examination of model predictions that inform public health policies.
\end{abstract}

\maketitle

\section*{Introduction}
A key challenge in infectious disease control is to protect a population by conferring immunity~\cite{anderson1991infectious, orenstein2000measles, reichert2001japanese, drolet2015populationlevel, toor2021lives, shattock2024contribution}. When some individuals gain immunity and become no longer susceptible to a disease, those without immunity also enjoy a reduced risk of infection because they are less likely to come into contact with others who can transmit the disease. Due to this indirect protection, the presence of immune individuals has a nonlinear impact on the overall level of protection in the population. This concept is known in epidemiology as \emph{herd immunity}~\cite{anderson1985vaccination, anderson1990immunisation, anderson1992concept, fine2011herd}. In particular, it is often presumed that there is a critical proportion of immune individuals---the herd immunity threshold---above which the chain of transmission cannot be sustained, and hence, the disease cannot invade the population.

Although the term ``herd immunity'' was coined and a rudimentary understanding of the phenomenon emerged around 1920, it was not until the 1970s that a quantitative theory of herd immunity was developed through mathematical modeling~\cite{fine1993herd, fine2011herd, jones2020history, morens2022concept}. The main focus was on estimating the vaccine coverage necessary for disease elimination. A simple expression for estimating the herd immunity threshold was obtained using a basic model that assumes immunization of a homogeneous, fully mixed population~\cite{anderson1982directly}. At the same time, much theoretical effort has been devoted to bridging the gap between such simplifying assumptions and the heterogeneity of real-world populations~\cite{fox1971herd, anderson1982directly, schenzle1984agestructured, may1984spatial, anderson1984spatial, anderson1985vaccination, hethcote1987epidemiological}. Such population heterogeneity can be leveraged to design efficient, targeted vaccination strategies. 

In general, immunity to an infectious disease can be acquired not only by vaccination but also by previous infection. In their seminal paper~\cite{kermack1927contribution}, Kermack and McKendrick derived what is now known as the final size equation, which implies the notion of \emph{disease-induced herd immunity} where an epidemic of a disease that confers immunity after recovery can end before the susceptible population is exhausted. More recently, disease-induced herd immunity has gained renewed attention, particularly in the context of the COVID-19 pandemic~\cite{britton2020mathematical, neipel2020powerlaw, lu2021datadriven, gomes2022individual, aguas2022herd}, for which no vaccine was available at the early stage. 
If one assumes a homogeneous and fully mixed population, disease-induced and vaccination-induced herd immunity are mathematically equivalent. However, this equivalence breaks down when variation in contact patterns is introduced.
Britton et al.~\cite{britton2020mathematical} showed that the threshold for disease-induced herd immunity in a heterogeneous population is considerably lower than expected in the homogeneous case; similar conclusions were drawn by several studies that adopted data-driven modeling approaches for COVID-19~\cite{neipel2020powerlaw, lu2021datadriven, gomes2022individual, aguas2022herd} and more recently for mpox~\cite{murayama2024accumulation, xiridou2024fading}. The essential reason for the lower threshold is that individuals with more contacts are more likely to get infected and become immune in an outbreak; epidemic spread thus effectively acts as targeted immunization. 

However, these results are derived using stratified mass-action models defined by differential equations or network models where nodes are randomly linked at every time step. In such modeling approaches, even if population heterogeneities are considered in terms of metapopulations defined by age, household, or spatial separation~\cite{diekmann1990definition, sattenspiel1995structured, lloyd1996spatial, ball1997epidemics, riley2007largescale, colizza2008epidemic, house2008deterministic, ajelli2010comparing, belik2011natural, zachreson2022effects}, the microscopic structure of persistent interactions between individuals is coarse-grained away and the correlations between the epidemiological states of individuals are disregarded.
In other words, these models assume that interactions occur in a \emph{mean-field} manner where individual details are replaced by averages. 
While these assumptions provide a convenient starting point, they are seldom met in real-world populations. In reality, interactions often occur repeatedly between the same pairs of individuals and are heavily influenced by social and spatial constraints. 
Such characteristics are better captured by modeling the contact structure as a static network that encodes these constraints~\cite{keeling1999effects, keeling2005networks, pastor-satorras2015epidemic}. In a static network model, the set of individuals one interacts with is assumed to be finite and fixed.

In this study, we use network epidemic models to reexamine how immunity induced by a past epidemic affects the outcome of future epidemics. Building on earlier studies that demonstrate the role of network structure in disease-induced herd immunity~\cite{newman2005threshold, ferrari2006network, bansal2012impact,  mann2021twopathogen, mann2021symbiotic, dilauro2021impact, mann2022strain}, we aim to unpack the mechanisms that shape herd immunity induced by disease
and to highlight the fundamental difference between the network model and mean-field model of herd immunity.

\begin{figure*}[tb!]
\centering
\includegraphics[width=0.75\linewidth]{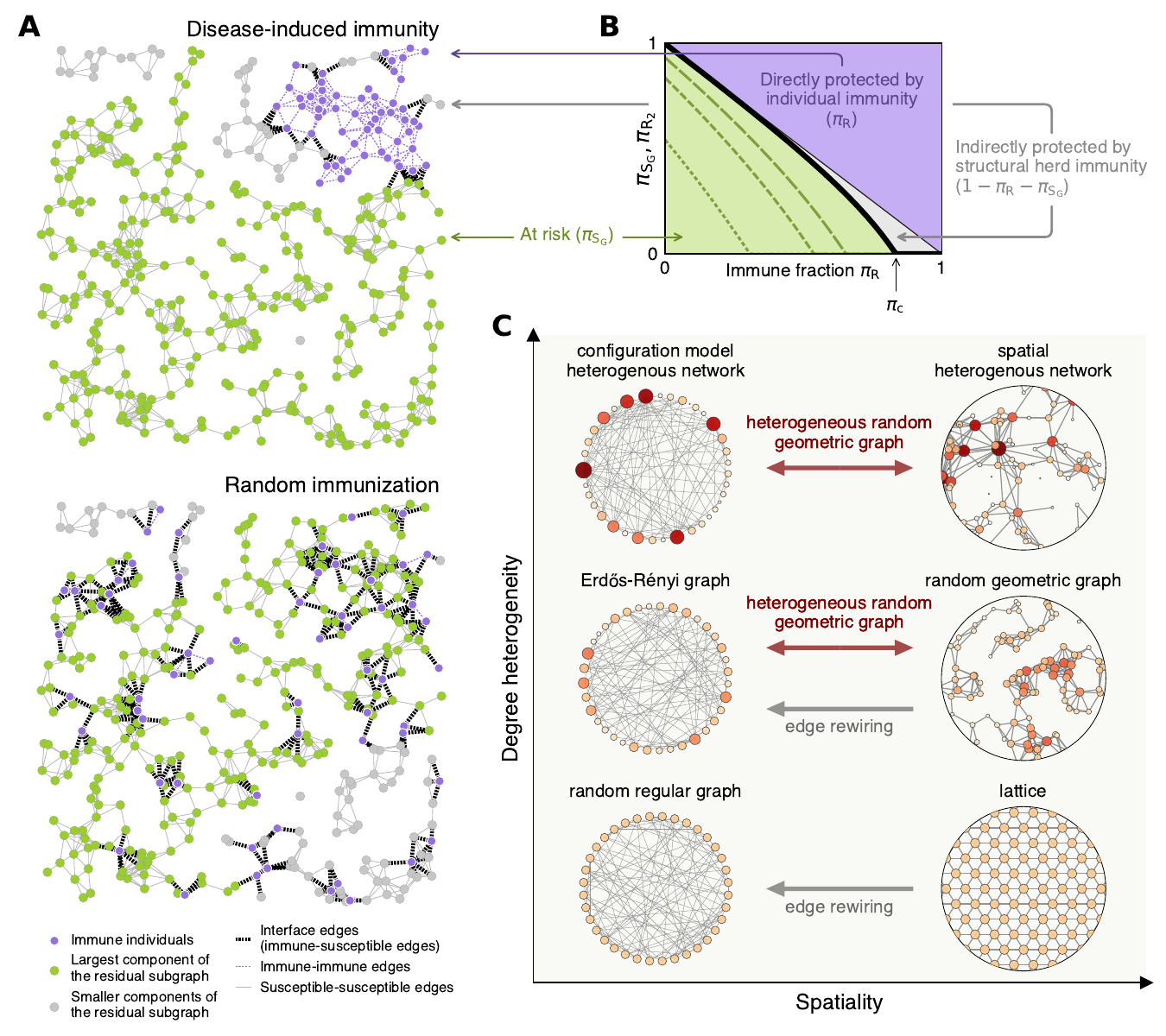}
    \caption{\textbf{Models.} \textbf{(A)} Comparison between the distributions of the same number of immune individuals (nodes) induced by disease spreading and by random immunization in a random geometric graph. Immune nodes are colored purple, while susceptible nodes are light green if they belong to the largest component of the residual subgraph and gray otherwise. The edges not included in the residual subgraph are represented by broken lines, with thin purple edges connecting two immune nodes and thick black edges connecting immune nodes and residual nodes. \textbf{(B)} Schematic illustration of structural herd immunity characterized by the giant residual component size $\pi_\mathrm{S_G}$ as a function of the immune fraction $\pi_\mathrm{R}$ (thick solid line). The gray area corresponds to the population indirectly protected by structural herd immunity. The green dashed curves are examples of sizes of post-immunity epidemics, with increasing transmissibility from left to right. The intersections of these curves with the horizontal axis indicate thresholds to complete herd immunity; the limiting value $\pi_\mathrm{c}$ marks the structural herd immunity threshold. \textbf{(C)} Network models used in this study, positioned according to their level of degree heterogeneity (vertical axis) and spatiality (horizontal axis).}
\label{fig:schematic}
\end{figure*}

Our key discovery is that in network models, disease-induced herd immunity is driven not only by the disproportionate distribution of immunity among socially active individuals---as already identified in mean-field models---but also by another, counteracting mechanism inherent to network models. In an outbreak originating from a single source (an initially infected individual), the set of individuals who become infected and subsequently immune is necessarily topologically contiguous in the network. We refer to this as \emph{localization of immunity}: since every immune node after an outbreak is adjacent to at least one other immune node, immunity is strongly correlated between neighbors and topologically clustered in the network compared to randomly distributed immunity; see Fig.~\ref{fig:schematic}A for a schematic illustration. Consequently, the \emph{interface} between immune and susceptible nodes is small: there are fewer edges between immune and susceptible nodes than there would be without such localization. This means that susceptible nodes are less protected because fewer of their neighbors are immune, which allows chains of infection that would otherwise be blocked to reach them. They also have more susceptible neighbors to infect, should they become infectious. Such mixing heterogeneities between susceptible and immune subpopulations resemble those discussed in the context of vaccination and other interventions~\cite{burgio2021homophily, rizi2022epidemic,burgio2022homophily, hiraoka2022herd, watanabe2022impact,rizi2024homophily}. 

We find that the localization of immunity has a significant impact on herd immunity even in maximally random networks. 
In homogeneous and/or spatially embedded networks, the effect of immunity localization on herd immunity is even more pronounced and can be strong enough to offset the advantage of disease-induced immunity over randomly distributed immunity.
Notably, the localization of immunity cannot be accounted for by mean-field models as typically used in the literature. As a result, mean-field models may overestimate the strength of disease-induced herd immunity.

The rest of the paper is organized as follows. First, we formalize how we quantify the effect of herd immunity in the network setting. We then show that unlike in the mean-field view, in network models disease-induced herd immunity differs from the effect of randomly distributed immunity. In particular, the net effect of disease-induced herd immunity is determined by the competition between two mechanisms: the biased distribution of immunity toward highly connected nodes and the localization of immunity within the network topology. After illustrating the effect of each mechanism in configuration model networks with different levels of heterogeneity in node degree (number of adjacent nodes), we quantify the effect of localization as a function of the level of spatiality of the network. Finally, we illustrate the model-specifity of herd immunity by showing that, in epidemic models informed by empirical population structure, a population that the mean-field model considers as completely protected can still be invaded by the disease in the network model.

\section*{Results}
In the following, we use {\emph{herd immunity}} as a broad term to describe any indirect protection of susceptible individuals against infection by the presence of immune individuals. This includes but is not limited to {\emph{complete herd immunity}} associated with a \emph{herd immunity threshold}, where any sustained transmission of the disease is prevented. Throughout, we assume immunity to be permanent and complete, fully preventing infection and transmission.

We use the canonical susceptible-infected-recovered (SIR) model to describe the dynamics of non-recurrent epidemics. Individuals and the contacts between them are represented as nodes and edges in an undirected contact network of size $N$. We assume that the contact network is large enough to consider the limit $N \to \infty$. We also assume that the contact network is quenched, i.e., remains static throughout the epidemic timescale. See Methods for details on the epidemic model, simulation methods, and analytical calculations.

We are interested in comparing two different scenarios in which immunity is introduced into a fully susceptible population. The first scenario is \emph{disease-induced immunity}, where individuals gain immunity through contracting the disease. An epidemic spreads from a source node until it eventually dies out, rendering the nodes that have experienced infection permanently and completely immune to future reinfection. In the second scenario, \emph{random immunization}, susceptible individuals are selected uniformly at random from the network and then permanently and completely immunized. This scenario embodies the outcome of homogeneous vaccination, while it can also be interpreted as the mean-field view of immunity acquired through any means, as there is no correlation between the immunity of neighboring nodes.

Regardless of how immunity is induced, its protective effect at the population level is measured by how much future outbreaks are reduced in size and in likelihood of occurrence.
In the case of disease-induced immunity, such post-immunity outbreaks can be caused by a more infectious variant of the original pathogen that induced immunity. Consider now the network of nodes that remain susceptible after the immunity-inducing epidemic or after random immunization. When a fraction $\pi_\mathrm{R}$ of nodes are immune, the proportion of this \emph{residual subgraph}~\cite{newman2005threshold, ferrari2006network, bansal2010shifting, funk2010interacting, karrer2011competing, hasegawa2011robustness, bansal2012impact,  hasegawa2012robustness, miller2013cocirculation, mann2021twopathogen, mann2021symbiotic, dilauro2021impact, mann2022strain} is $1 - \pi_\mathrm{R}$, providing an upper bound for a post-immunity outbreak due to direct protection. 

The actual size of the post-immunity epidemic depends on the transmissibility of the disease. However, even if the disease is infectious enough to be always transmitted from an infected individual to a susceptible neighbor, an epidemic caused by a single introduction cannot grow larger than the largest connected component of the residual subgraph. When the size of this largest connected component scales with $N$, we call it the \emph{giant residual component} and denote its relative size by $\pi_\mathrm{S_G}$. Naturally, $\pi_\mathrm{S_G} \leq 1 - \pi_\mathrm{R}$, where the equality holds only when the residual subgraph is connected. 

Within the class of network models we consider in this work, the giant residual component is the only connected component in the residual subgraph whose size scales with $N$. For large $N$, all smaller connected components in the residual subgraph become negligible in size. Consequently, nodes in these small components are almost certainly shielded from future epidemics of any transmissibility despite not being immune because their risk of exposure is vanishingly small. We call this type of indirect protection \emph{structural herd immunity}: there are susceptible nodes that are isolated by the immune nodes in the residual contact network. 

The effect of structural herd immunity can be quantified by the sum of the sizes of the small components or, equivalently, by the difference between the size of the residual subgraph and its giant component, $1 - \pi_\mathrm{R} - \pi_\mathrm{S_G}$ (Fig.~\ref{fig:schematic}B). 
If there is no giant component in the residual subgraph (i.e., $\pi_\mathrm{S_G} \to 0$ when $N \to \infty$), all residual nodes are indirectly protected, and complete herd immunity is achieved regardless of the infectiousness of the disease. The value of $\pi_\mathrm{R}$ at which the giant component disappears, denoted by $\pi_\mathrm{c}$, signifies the structural herd immunity threshold.

The fraction of nodes $\pi_\mathrm{S_G}$ in the giant residual component, which represents the maximum possible epidemic size, can vary depending on the specific distribution of immune nodes on the network even if their fraction $\pi_\mathrm{R}$ is the same. In particular, we focus on two aspects of how disease-induced immunity is distributed in the network: its bias toward high-degree nodes and its localization. We quantify the first with the ratio $\langle k \rangle_\mathrm{R} / \langle k \rangle$ between the mean degree of immune nodes and the mean degree of the entire network. The second aspect, localization of immunity, refers to how adjacent immune nodes are in the network---in a maximally localized configuration, there are as many edges between the immune nodes as possible and in a non-localized configuration, immune nodes are located randomly in the network. In the following, we measure the localization of immunity straightforwardly by the share of interface edges between the immune and residual subgraphs, $\rho_\mathrm{SR}=E_\mathrm{SR}/E$, where $E_\mathrm{SR}$ is the number of edges between susceptible and immune nodes and $E$ is the total number of edges in the network. This measure is linearly related to the correlation coefficient between the epidemiological states of adjacent nodes (see SI Appendix~A). 
 When immune nodes are more localized, i.e., likely to be adjacent to each other in the network, there are fewer edges at the interface between the two subgraphs. Conversely, in a non-localized configuration, the interface is larger and as the susceptible nodes at the interface have at least one immune neighbor, there are fewer edges to carry a subsequent infection (see Fig.~\ref{fig:schematic}A). 

Epidemic dynamics is largely influenced by individual variance in contact and transmission patterns~\cite{lloyd-smith2005superspreading, mossong2008social, hebert-dufresne2020beyond}, which translates to the heterogeneity of node degrees in the contact network. The paradigmatic network model used to express degree heterogeneity is called the configuration model, where the distribution of node degrees solely determines the network structure~\cite{newman2018networks}. In a configuration model network, the structure is locally tree-like, meaning that the likelihood of a node being part of a finite-length cycle diminishes as the network size increases. This feature often simplifies analytical calculations and makes the model more tractable. 

However, real-world contact networks through which diseases are transmitted are hardly tree-like; rather, they are characterized by the abundance of short cycles. This is because contact and transmission between individuals can only occur when individuals are physically close to each other. If two individuals have a common neighbor in the contact network, they are likely to be near each other, which implies a high probability that they are also connected. As a result, many triangles and short cycles are formed. We will refer to the propensity of individuals to be in contact with other spatially proximate individuals as \emph{spatiality}.

In this work, we explore a wide range of network structures that differ in terms of \emph{degree heterogeneity} and \emph{spatiality}. Degree heterogeneity is characterized by the variance of the degree distribution. For spatial features, we consider that the nodes are endowed with fixed coordinates in a two-dimensional space. Then, spatiality can be measured by the average length of edges (i.e., the average distance between adjacent nodes) in this two-dimensional space. For example, in random geometric graphs, edge lengths are short because only local contacts are allowed, representing maximal spatiality, while in Erd\H{o}s-R\'{e}nyi random graphs, nodes are in contact independently of their spatial positions, resulting in longer edges on average despite the same Poisson degree distribution. Figure~\ref{fig:schematic}C illustrates these two features. See Methods for details on the network models used in this study.

\subsection*{Localization of disease-induced immunity significantly weakens herd immunity}
We first focus on the structural herd immunity in configuration model networks, where the degree distribution is the only defining feature. We study, in increasing order of degree heterogeneity, random regular graphs (RRGs), Erd\H{o}s-R\'{e}nyi random graphs (ERGs), and configuration model networks with negative binomial and power law (scale-free) degree distributions. 
For each of the network ensembles, we first calculate, both analytically and numerically, the expected size $\pi_\mathrm{R}$ of a large epidemic (an outbreak that infects a finite fraction of the population) as a function of transmission rate $\beta$. 
Then, we compute the size of the giant residual component, $\pi_\mathrm{S_G}$, for three cases: after removing the nodes that are naturally infected in the epidemic (disease-induced immunity), after removing the same number of nodes but randomly (random immunization), and after removing the same number of nodes with the same degrees but randomly. We refer to the third case as \emph{proportional immunization}. The analytical calculations are performed using an averaged message-passing framework~\cite{newman2002spread, newman2005threshold, newman2023message}; see SI Appendix~F for details. 

\begin{figure*}[tb!]
\centering
\includegraphics[width=0.95\linewidth]{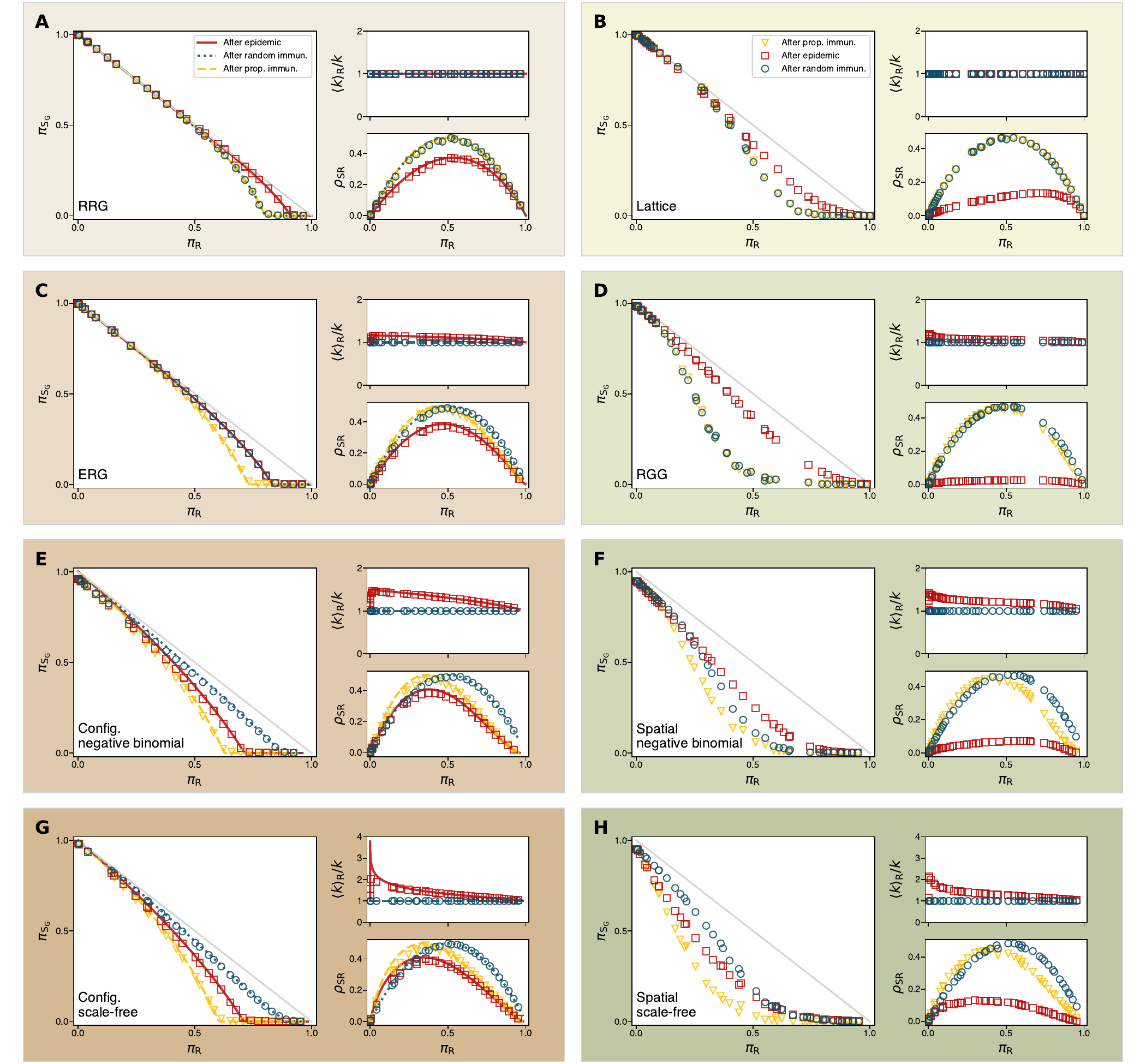}
\caption{\textbf{Comparison of the strength of  herd immunity for different network models.} The configuration model networks are on the right, spatial networks on the left. Each row corresponds to a different degree distribution. For all networks, the number of nodes is $N = 10^4$ and the average degree is $\langle k \rangle = 6$. 
The negative binomial degree distribution (third row) is parameterized by the dispersion parameter $r = 3$. The scale-free degree distribution (fourth row) has a power-law tail with exponent $\alpha = 3.1$. The spatial heterogeneous networks (F, H) are generated with the temperature parameter $\tau = 0.05$. For each set of panels, the left panel shows the size $\pi_\mathrm{S_G}$ of the giant residual component, the top right panel shows the mean degree of immune nodes normalized by the mean degree of the entire network,  $\langle k \rangle_\mathrm{R} / \langle k \rangle$, and the bottom right panel shows the fraction of edges between the immune and residual subgraphs, $\rho_\mathrm{SR}$. 
Symbols denote numerical results averaged over 50 different realizations, and lines represent analytical predictions by the averaged message-passing approach. The colors indicate natural infection (red), random immunization (blue), and proportional immunization (yellow). See Methods and SI Appendixes~E and F for details of numerical and analytical methods used. }
\label{fig:main_results}
\end{figure*}

The left column of Fig.~\ref{fig:main_results} summarizes our results. For RRGs, we observe that, strikingly, structural herd immunity induced by disease is weaker than that of random immunization; the giant residual component is always larger in the case of disease-induced immunity (Fig.~\ref{fig:main_results}A). As RRGs have no degree heterogeneity and their structure is entirely random, the strength of herd immunity is entirely dictated by the localization of immunity in the wake of the outbreak. The contiguous nature of the subgraph covered by disease-induced immunity is clearly visible in the smaller size $\rho_\mathrm{SR}$ of the interface between the immune and residual subgraphs. Conversely, for random immunization, the larger interface is associated with a high level of structural herd immunity, with a residual subgraph whose giant component is smaller. 

When the contact network is an ERG where node degrees are moderately heterogeneous, disease-induced immunity and random immunization result in a giant residual component of the same size. Although structural herd immunity is strengthened by the fact that the outbreak disproportionately infects high-degree nodes, it is weakened to an equal extent by the localization of immunity after the outbreak. Both effects are visible in Fig.~\ref{fig:main_results}C: the average degree ${\langle k \rangle}_\mathrm{R}$ of immune nodes is higher for disease-induced immunity, but the size of the susceptible-immune interface $\rho_\mathrm{SR}$ is smaller. The equal magnitude of these two effects can be shown analytically; we prove in SI Appendix~G that the Poisson distribution is, in fact, the unique degree distribution of the configuration model in which the two effects exactly offset each other. The giant residual component size under proportional immunization implies that the structural herd immunity would be stronger if only the effect of preference for high-degree nodes were considered: immunizing high-degree nodes wherever they are positioned in the network results in a more splintered residual subgraph. This leads to the important conclusion that if the contribution of localization is neglected, the strength of disease-induced herd immunity is overestimated. 

As the degree heterogeneity of the contact network increases, the advantage of disease-induced immunity in exploiting degree heterogeneity and residing disproportionately among high-degree nodes outweighs the localization effect, as seen in the results for configuration model networks with a negative binomial degree distribution with dispersion parameter $r = 3$ (Fig.~\ref{fig:main_results}E). Importantly, disease-induced herd immunity is still significantly weaker than what would be expected from the degrees of immune nodes alone. This is clearly visible in the difference in the size of the giant residual component for disease-induced immunity and proportional immunization, as well as in the average degrees of immunized nodes and interface sizes. Qualitatively similar results are obtained for scale-free networks, corroborating our findings (Fig.~\ref{fig:main_results}G). See SI Appendix~K for the robustness of our results against changes in population size and average degree.

To conclude, the net effect of disease-induced herd immunity is determined by the competition between the biased distribution of immunity toward high-degree nodes and the localization of immunity in the network. We observe that localization significantly weakens structural herd immunity in networks with any degree distribution: greater localization of immune nodes leaves the residual subgraph more intact. While the degree heterogeneity of the contact network amplifies the high-degree bias of disease-induced immunity and strengthens herd immunity, it is always counteracted by the effect of localization. Without this countering effect, structural herd immunity would be even stronger, as exemplified by the smaller giant residual component after proportional immunization. 

\subsection*{Spatiality enhances localization, which further weakens disease-induced herd immunity}
Unlike the configuration model networks studied so far, real-world contact networks are spatially constrained and are, therefore, effectively low-dimensional. As the ratio of surface area to volume is smaller in lower dimensions, one can expect that the susceptible-immune interface under disease-induced immunity is smaller and that the effect of localization is more pronounced in networks embedded in low-dimensional space. Accordingly, we expect disease-induced herd immunity to be weaker in spatial networks. To study this, we numerically investigate the strength of structural herd immunity in regular lattices, random geometric graphs (RGGs), and heterogeneous random geometric graphs (HRGGs; see Methods for details).

We observe that, for regular lattices, disease-induced immunity leads to a larger giant residual component than random or proportional immunization (Fig.~\ref{fig:main_results}B), similar to the case for RRGs. The smaller size of the interface, $\rho_\mathrm{SR}$, implies that the effect of localization of disease-induced immunity is stronger in a regular lattice than in an RRG due to its spatiality. 

For RGGs, the gap between disease-induced immunity and random immunization is even larger, confirming the above hypothesis and implying a greater advantage of random immunization over disease-induced immunity in efficiently shrinking the giant residual component (Fig.~\ref{fig:main_results}D). 
In an RGG, the immune nodes under disease-induced immunity have a very small interface with the residual nodes, indicating that they are highly localized. This leaves the residual subgraph susceptible to a future outbreak, as only a few residual nodes have immune neighbors that would break chains of transmission and reduce the size of the largest residual component.
Although disease-induced immunity can exploit the modest heterogeneity of the Poisson degree distribution, the impact of localization is much more pronounced, overriding the effect of high-degree bias of immunity. 
This is in contrast to the case of ERGs, the configuration model counterpart of RGGs, where the effects of the two mechanisms exactly cancel each other out. 
Note that in RGGs, the two mechanisms are intertwined; high-degree bias amplifies localization because of degree correlations.

The impact of spatiality is particularly evident for networks with higher degree heterogeneity. For a HRGG with a negative binomial degree distribution, structural herd immunity induced by natural infection can be weaker than that achieved by random immunization (Fig.~\ref{fig:main_results}F). Juxtaposed against the configuration model counterpart, where the opposite result is found, this highlights that the spatiality of the contact network can boost the effect of localization to the extent that it overcomes the counteracting effects of biased distribution of immunity toward high-degree nodes, thus reversing the outcome. In scale-free HRGGs, characterized by an even higher degree heterogeneity, disease-induced immunity still proves more efficient than random immunization in dismantling the giant residual component (Fig.~\ref{fig:main_results}H). Even then, there is a significant gap between $\pi_\mathrm{S_G}$ for disease-induced immunity and proportional immunization, highlighting the effect of immunity localization to attenuate structural herd immunity. SI Appendix~K confirms the robustness of our results to changes in population size and average degree.

We have so far established that the spatiality of the contact network diminishes the disease-induced herd immunity by amplifying immunity localization. We next ask: How strongly does the network need to be embedded in space for the effect of high-degree bias of immunity to be outweighed by the effect of immunity localization?
To this end, we compare the effectiveness of disease-induced immunity and random immunization across the spatiality spectrum. We use the edge rewiring process and the HRGG model with varying temperatures to cover the spectrum of spatiality continuously. 
As a  measure of spatiality, we use the normalized mean edge length $\langle d \rangle / d^*$ as described in Methods. In the configuration model, where the edges are completely random, we have that $\langle d \rangle / d^* = 1$; edges are shorter for networks that are more strongly constrained in space.
To estimate the effectiveness of structural herd immunity with a single number, we use the structural herd immunity threshold and measure the minimum fraction of nodes that need to be immune to eliminate the giant residual component: $\pi_\mathrm{c} = \min \{\pi_\mathrm{R} \mid \pi_\mathrm{S_G} = 0\}$. In other words, even a disease with an infinitely large transmission rate cannot invade the population if $\pi_\mathrm{R} \geq \pi_\mathrm{c}$. Thus, $\pi_\mathrm{c}$ represents the worst-case bound for the herd immunity threshold, i.e., the structural herd immunity threshold. Here, we numerically identify $\pi_\mathrm{c}$ as the minimum value of $\pi_\mathrm{R}$ that leads to $\pi_\mathrm{S_G} \leq 0.01$. We let $\pi_\mathrm{c}^\mathrm{DII}$ and $\pi_\mathrm{c}^\mathrm{RI}$ denote the threshold under disease-induced immunity and random immunization, respectively.

\begin{figure}[tb!]
\centering
    \includegraphics[width=\linewidth]{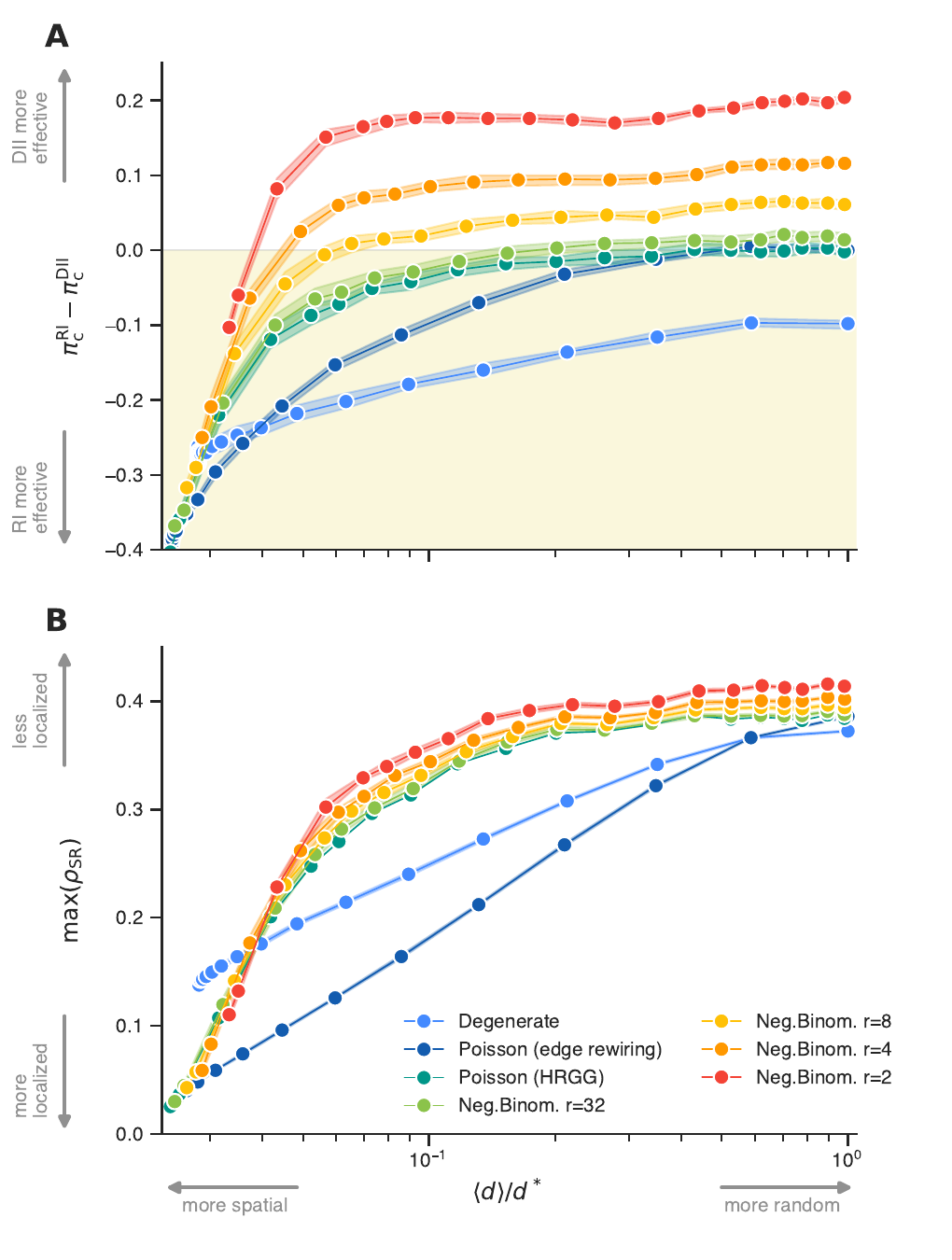}
    \caption{\textbf{Disease-induced herd immunity as a function of spatiality}. Spatiality is measured by the normalized mean edge length $\langle d \rangle / d^*$. \textbf{(A)} Difference in the structural herd immunity thresholds $\pi_\mathrm{c}^\mathrm{RI} - \pi_\mathrm{c}^\mathrm{DII}$. Different degree distributions are indicated by line colors. The shaded area along each line represents the $95\%$ confidence interval. Positive values indicate that natural infection induces a stronger herd immunity than random immunization, while negative values suggest the opposite. \textbf{(B)} The maximum (peak value) of $\rho_\mathrm{SR}$ as a function of $\pi_\mathrm{R}$ for disease-induced immunity. Smaller values imply stronger localization of immunity. Note that the value of $\pi_\mathrm{R}$ at which $\rho_\mathrm{SR}$ is maximized generally does not coincide with $\pi_\mathrm{c}$. }
    \label{fig:diff_transition_points}
\end{figure}

Figure~\ref{fig:diff_transition_points}A shows the difference between the structural herd immunity thresholds of random immunization and disease-induced immunity, $\pi_\mathrm{c}^\mathrm{RI} - \pi_\mathrm{c}^\mathrm{DII}$, as a function of normalized mean edge length $\langle d \rangle / d^*$ for different degree distributions. Positive values indicate that $\pi_\mathrm{c}^\mathrm{DII}$ is smaller, i.e., disease-induced immunity results in a lower threshold than random immunization, while negative values imply the opposite. 
For all the degree distributions, the difference between the thresholds increases with spatiality. In other words, the spatiality of the contact network decreases the relative advantage of disease-induced immunity. 
Figure~\ref{fig:diff_transition_points}B and SI Appendix~H show that, when the degree distribution is held constant, increasing the spatiality of the network generally shrinks the interface, implying that disease-induced immunity becomes increasingly localized. This increase in localization corresponds with the decrease in the strength of structural herd immunity from natural infection shown in Fig.~\ref{fig:diff_transition_points}A. Therefore, the more spatial the networks are, the more localized disease-induced immunity becomes, and consequently, the weaker the herd immunity acquired through natural infection.

As we have already seen, for Poisson degree distributions, the two immunity scenarios induce an equally strong herd immunity without spatiality (i.e., at $\langle d \rangle / d^* = 1$). 
For more heterogeneous networks, there is a crossover from the regime where disease-induced immunity is more effective ($ \pi_\mathrm{c}^\mathrm{DII} < \pi_\mathrm{c}^\mathrm{RI}$) to the regime where random immunization becomes more advantageous ($ \pi_\mathrm{c}^\mathrm{DII} > \pi_\mathrm{c}^\mathrm{RI}$) as the network becomes more spatial. As the degree distribution becomes broader, as in negative binomial distributions with increasing variance (i.e., decreasing dispersion parameter), the crossover point shifts toward the spatial end, indicating that a higher level of spatiality is required to counterbalance the effect of degree heterogeneity (see also SI Appendix~I). This highlights the competition between degree heterogeneity and spatiality of the contact network in determining the strength of disease-induced herd immunity. 

\subsection*{Comparing mean-field and network models with population structure}

Our analysis above suggests that disease-induced herd immunity in network models is driven by high-degree bias and localization of immunity, while mean-field models only account for the former. This difference between models has real-world consequences for estimating the effect of disease-induced herd immunity. In particular, mean-field models may overestimate the strength of herd immunity and provide an overly optimistic outlook. We demonstrate this by showing, based on empirical contact patterns, that even when the population reaches the condition of complete herd immunity according to a mean-field model, the disease can still reinvade the population in a similarly parameterized network model.

We compare two age-stratified models informed by empirical data: a mean-field model and a network model of SIR dynamics, both with the same age structure and age-specific contact patterns. The age structure is captured by vector $\mathbf{p}$, whose entries are the proportion $p_i$ of the population in each age group $i$. The contact patterns are represented by the contact matrix $M$ whose element $M_{ij}$ represents the average number of contacts an individual in age group $i$ makes with individuals in age group $j$. For the contacts between groups to be symmetric, we impose on the contact matrix $p_i M_{ij} = p_j M_{ji}$.

The mean-field model is defined by the following set of differential equations:
\begin{equation}
    \begin{aligned}
        \dot{s}_i &= - \lambda_i s_i, \\
        \dot{y}_i &= \lambda_i s_i - \gamma y_i, \\
        \dot{r}_i &= \gamma y_i.
    \end{aligned}
    \label{eq:structured_pop_model}
\end{equation}
Here, $s_i, y_i, r_i$ denotes the proportion of susceptible, infected, and recovered individuals in age group $i$, respectively, and satisfies $s_i + y_i + r_i = 1$; $\gamma$ is the rate at which infected individuals recover; $\lambda_i$ is the force of infection to which an individual in age group $i$ is subject, which is calculated as
\begin{equation}
    \lambda_i = \beta_\mathrm{MF} \sum_j M_{ij} y_j,
\end{equation}
where $\beta_\mathrm{MF}$ denotes the transmission rate of the mean-field model. In this model, the basic reproduction number can be computed as~\cite{diekmann1990definition}
\begin{equation}
    R_0 = \frac{\beta_\mathrm{MF}}{\gamma}\, \Lambda(M),
\end{equation}
where $\Lambda(\cdot)$ denotes the spectral radius. By linearizing Eq.~\eqref{eq:structured_pop_model} around $s_i$ and $y_i \to 0$, we obtain the condition that disease with transmission rate $\beta_\mathrm{MF}$ cannot invade the population:
\begin{equation}
    R_0 \,\frac{\Lambda\big(\operatorname{diag}(\mathbf{s}) M\big)}{\Lambda(M)}\leq 1,
    \label{eq:herd_immunity_threshold_mf}
\end{equation}
where $\mathbf{s}$ denotes the vector of which $i$th element is $s_i$. 
Thus, this condition represents complete herd immunity of the mean-field model, with the left-hand side representing the effective reproduction number. Note that Eq.~\eqref{eq:herd_immunity_threshold_mf} is a condition for the vector $\mathbf{s}$ and does not represent a single scalar value of the herd immunity threshold. There are many different $\mathbf{s}$ that make both sides of Eq.~\eqref{eq:herd_immunity_threshold_mf} equal, and the herd immunity threshold as the total proportion of susceptible individuals in the population, given by $\mathbf{p}\cdot\mathbf{s}$, can vary depending on the specifics of the system and dynamics.

\begin{figure*}[tb!]
\centering
    \includegraphics[width=0.95\linewidth]{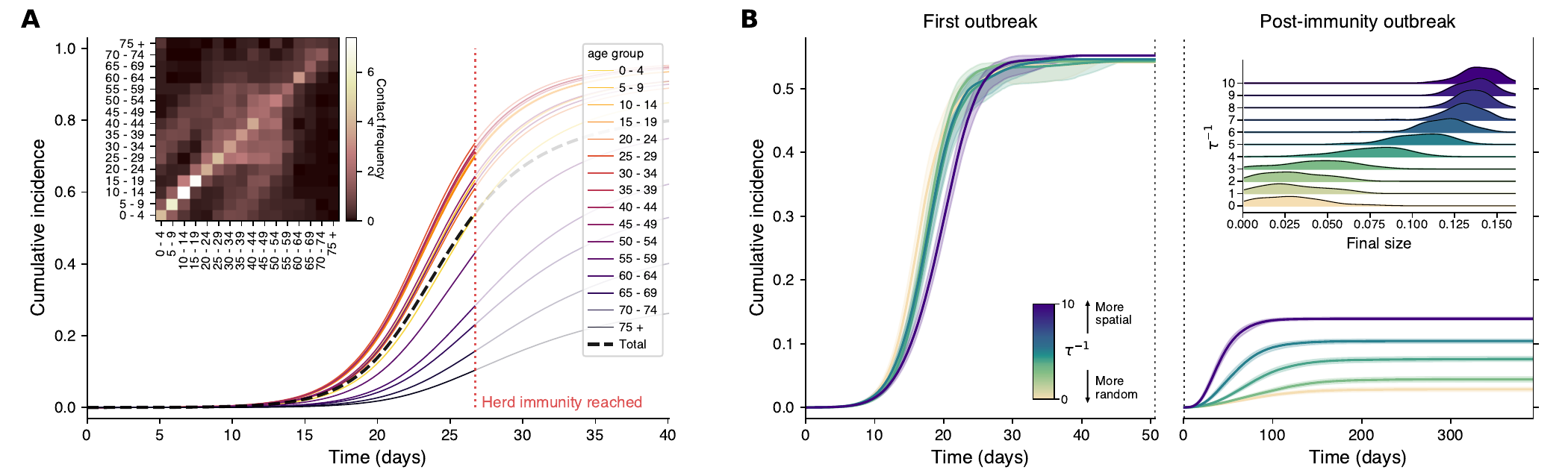}
    \caption{\textbf{Comparing mean-field and network epidemic models of the empirical population of Finland.} \textbf{(A)} Mean-field model simulation of an epidemic in an initially susceptible population. The black dashed line represents the cumulative incidence in the entire population, while the solid curves represent the cumulative incidence in each age group, indicated by different colors. The red dotted vertical line indicates the time at which herd immunity is reached. Inset: Contact matrix $M$ for Finland's population. \textbf{(B)} Cumulative incidence in network model simulations of the first epidemic (left) and after disease reintroduction (right). Solid lines show fixed-time averages over 100 realizations, and shaded areas represent $95\%$ confidence intervals. The colors of the curves represent spatial SBM networks with different temperatures: $\tau^{-1} = 0, 3, 4, 5, 10$ from light to dark colors, respectively. The inset in the right panel shows the kernel density estimate of the final sizes of post-immunity outbreaks in networks with different temperatures.}
    \label{fig:comparison_mf_net}
\end{figure*}

Figure~\ref{fig:comparison_mf_net}A shows a simulated epidemic with a mean-field model of the population of Finland in 2007.  The age structure~\cite{unstat} and contact matrix~\cite{mossong2008social, prem2017projecting} are stratified into 16 age groups. The basic reproduction number is set to $R_0 = 3$ and the mean recovery time is set to $\gamma^{-1} = 5$ days. When the population reaches the herd immunity threshold defined by Eq.~\eqref{eq:herd_immunity_threshold_mf}, $54.1\%$ of the total population has contracted the disease. However, due to the heterogeneity of contact patterns within the population, there is a large variance in attack rate between age groups, ranging from $10.4\%$ among those over 75 to $73.6\%$ among 25--29-year-olds.

We formulate the corresponding network model as follows. The population of $N$ individuals is divided into age groups, each with $n_i \simeq N p_i$ individuals. 
The contact network is generated by a stochastic block model (SBM). The standard SBM assumes that the probability that individuals are linked to each other depends only on the groups they belong to. It is defined by the number of individuals $n_i$ in each group $i$ and the edge probability $B_{ij}$ between an individual in group $i$ and an individual in group $j$. The edge probability is related to the contact matrix as
\begin{equation}
    B_{ij} = a M_{ij} / n_j, 
\end{equation}
where $a$ is the coefficient that controls the total number of edges in the network. 
In addition to the age-stratified contact patterns, we further incorporate spatiality by considering a mixture of the SBM and the HRGG model parameterized by temperature $\tau$; see SI Appendix~D for details.

Using the same age structure and contact matrix of Finland, we construct the contact network with $N=2 \times 10^4$ and $a=1$. For this setup, the normalized mean edge length $\langle d \rangle / d^*$ decreases nonlinearly with inverse temperature from $\langle d \rangle / d^* = 1$ for $\tau^{-1} = 0$ to $\langle d \rangle / d^* = 0.31$ for $\tau^{-1} = 3$ and $\langle d \rangle / d^* = 0.16$ for $\tau^{-1} = 10$. We let an epidemic of $R_0 = 3$ evolve in the network until it reaches the herd immunity threshold defined by Eq.~\eqref{eq:herd_immunity_threshold_mf}, at which point we halt the transmission process and let every infected individual recover. 
Once the population is free of infection, we reintroduce the disease into the population by infecting an individual who remained susceptible during the first epidemic. If the population has already reached herd immunity during the first outbreak, the disease will be quickly eliminated, and there will be no epidemic resurgence that infects a finite fraction of the population. To highlight the maximum risk, we sample post-immunity outbreaks of significant sizes, if they occur at all. 

Figure~\ref{fig:comparison_mf_net}B shows the cumulative incidence during the first epidemic and after the reintroduction of the disease according to the network model. The size of the first epidemic is hardly affected by spatiality, varying from an average of $54.2\%$ for $\tau^{-1}=0$ to $55.2\%$ for $\tau^{-1}=10$, which is also consistent with the results of the mean-field model. Upon reintroduction of the disease, simulations for standard SBM networks ($\tau^{-1} = 0$) indicate the occurrence of a relatively small epidemic, affecting an additional $2.8\%$ on average. When the network is spatially embedded, we observe a larger resurgence, infecting up to $13.8\%$ of the population. This suggests that the population is still in at risk, even though the number of immune individuals should be sufficient to achieve herd immunity according to the mean-field model. In other words, the mean-field model underestimates the herd immunity threshold of networked populations in such a scenario, especially when they are spatially embedded. We discuss the significance of our results for different population sizes in SI Appendix~K and provide an explicit calculation of the network model threshold in SI Appendix~L.

\section*{Discussion}
We have studied the effectiveness of disease-induced immunity in network epidemic models.
We have found that disease-induced herd immunity is driven by a ``tug-of-war'' between two antagonistic factors, both stemming from the distribution of immune nodes in the network. On the one hand, an epidemic tends to spread and induce immunity disproportionately among highly connected nodes, which strengthens herd immunity. On the other hand, it leads to a localized distribution of immunity in a topologically contiguous section of the network; this has the opposite effect, that is, it weakens herd immunity, drawing parallels to the effects of inhomogeneous vaccine coverage~\cite{hiraoka2022herd}. Importantly, the effect of localization cannot be captured by mean-field epidemic models. 

We have shown that the strength of the effects of the two mechanisms (high-degree bias and immunity localization) on disease-induced herd immunity is determined by network structure.
In regular networks, where the degrees are identical and only localization is at play, disease-induced immunity provides weaker protection than random immunization. The impact of high-degree bias of immunity increases as the degree distribution becomes more heterogeneous. 
In Erd\H{o}s-R\'{e}nyi graphs that display a moderate amount of degree heterogeneity, the two mechanisms are equally influential, with their effects canceling each other out exactly. In more degree-heterogeneous configuration model networks, disease-induced immunity is more effective than random immunization in conferring protection. However, even in this case, we found that immunity localization attenuates collective protection to a lower level than expected from the degree distribution of immune nodes alone.

When spatiality is added to the picture, the effect of immunity localization is boosted. Notably, even in a degree-heterogeneous network where the distribution of disease-induced immunity is strongly biased toward high-degree nodes, its advantage over random immunization may be offset and overturned because the strong localization weakens the amount of structural herd immunity. 
We have shown that increasing spatiality generally makes disease-induced herd immunity weaker compared to random immunization. 
The competition between spatiality and degree heterogeneity of the network is evident from the fact that the former needs to be compensated for by the latter for disease-induced immunity to be more effective than random immunization. In summary, our results suggest that the two underlying mechanisms of disease-induced herd immunity are each influenced by one of the two features of network topology. While degree heterogeneity promotes the effect of the biased distribution of immunity toward high-degree nodes, spatiality enhances the impact of immunity localization. 

The difference in the mechanisms of disease-induced herd immunity, namely the presence/absence of the effect of immunity localization, can lead to a discrepancy in the estimated strength of herd immunity between mean-field and network models. We have shown that this inconsistency can arise in stylized models of a real-world population. 
Even when the mean-field model deems that the population has reached complete herd immunity through natural infection, the disease can reinvade the population in a network model. In other words, the mean-field model estimates a lower threshold and a stronger herd immunity than the network model. Note that this discrepancy is not due to the spatial structure of the network; even when the network model is informed with the same contact data as the mean-field model, its herd immunity threshold is higher.

The tension between the two models stems from the fact that the mean-field model does not account for dynamical correlations between the epidemiological states of individuals. It implicitly assumes that the population is mixed much faster than the epidemic dynamics. The network model, on the other hand, represents the slow-mixing regime; the contact partners of each individual remain the same throughout the epidemic~\cite{volz2007susceptible}. This gives rise to dynamical correlations that manifest themselves in the localized distribution of immune nodes after the spreading of the disease.

In studying network epidemic models, we have based our argument on the notion of \emph{structural} herd immunity embodied by the residual subgraph and its giant component. Although the notion of giant residual component has been used in previous network literature~\cite{newman2005threshold, ferrari2006network, bansal2012impact}, it is not a common practice in epidemiology, where herd immunity is usually characterized against a disease with a specific transmissibility. There are two different ways to interpret what the giant residual component represents. On the one hand, it represents the individuals who will with certainty be infected in a post-immunity epidemic caused by a highly infectious pathogen (i.e., when the transmissibility is infinite). On the other hand, it represents the population that is still at risk of becoming infected in a post-immunity epidemic of lower transmissibility. When the pathogen is infectious enough to spread in the giant residual component but not infectious enough to infect it entirely, any individual in it can still be infected in the post-immunity epidemic: exactly who will be infected is determined by the stochasticity of seeding and transmission. In contrast, individuals in the smaller components of the residual subgraph are exempt from the risk of infection. This is seen in how the structural constraints affect the final sizes of post-immunity epidemics with specific transmissibility (SI Appendix~J): the results are remarkably consistent with the findings for structural herd immunity.

In a mean-field model, be it a fully-mixed mass-action model or a stratified population model parameterized by a contact matrix, an implicit assumption is that every individual interacts with every other individual. This can be mapped to a fully connected contact network (a complete graph). When a subset of nodes is immune and removed from this contact network, the residual subgraph is naturally a complete graph and forms a single connected component. In other words, the residual subgraph is identical to its giant component. This means that (i) if the disease is transmissible enough, every individual in the residual subgraph will be infected in a post-immunity epidemic, and (ii) some susceptible individuals may not get infected in a post-immunity epidemic of a less infectious disease because of herd immunity, but no one is free from the risk of infection.

We remark that epidemic localization in a network has been discussed in a different context, namely, for the susceptible-infected-susceptible model of recurrent infection~\cite{goltsev2012localization, pastor-satorras2018eigenvector, st-onge2021master, st-onge2022influential}. In these works, the term ``localization'' implies that infection tends to reside disproportionately on nodes with certain connectivity patterns, such as nodes with high degrees, in inner $k$-cores, and in dense subgraphs. In our work, we focus on non-recurrent diseases and define localization in terms of strong correlations between the immunity of adjacent nodes as a result of  infection spreading over a contiguous region of the network. Despite the different implications of the term, both lines of work recognize the importance of the interplay between network structure and epidemic dynamics.

The concept of disease-induced immunity, particularly its threshold, gained significant attention during the early stages of the COVID-19 pandemic before vaccines became available. Mean-field model predictions suggested that the threshold was lower than expected, leading to policy discussions in some countries on the possibility of protecting high-risk groups by allowing those at lower risk of severe illness to acquire immunity through infection~\cite{walker2020impact, orlowski2020four, perra2021nonpharmaceutical, brusselaers2022evaluation, zenone2022analyzing}. In addition to several issues raised by this approach, including ethical concerns, the assumption that at-risk individuals can be identified, and the assumption of permanent and complete immunity~\cite{brett2020transmission, buss2021threequarters, sridhar2021herd}, our analysis suggests that the optimistic projections were, in part, a consequence of the peculiarities inherent in the mean-field models used for estimation.

We emphasize that our aim is not to claim that network models are an accurate depiction of reality. Rather, our results indicate that there is a strong model dependence in the estimation of the strength of herd immunity, which calls for caution in interpreting the results of any mathematical model of infectious disease and warrants careful consideration of its underlying assumptions. The mean-field modeling approach, albeit widely used in epidemic modeling, implicitly assumes a particular set of idealized assumptions that may bias its estimation of disease-induced herd immunity. Outcomes of stylized network models that account for repeated and spatially-constrained contacts highlight the need to reassess conclusions drawn from mean-field models in light of more realistic interaction structures. 
We stress that our network models are also highly simplified and their outcomes are to be taken as theoretical. In this work, we have deliberately disregarded several characteristics of contact patterns in real-world populations in order to single out the effects of two prominent structural features: degree heterogeneity and spatiality. 

An advantage of the network approach is that it allows us to build a more individualistic and structural understanding of herd immunity, rather than quantifying it only in terms of thresholds and final epidemic sizes. From this perspective, the role of correlations and inhomogeneities in the network microstructure as well as the presence of mesoscopic structures~\cite{stegehuis2016epidemic, hebert-dufresne2019smeared, morita2021solvable, ball2023impact} on herd immunity remain to be understood.
Furthermore, contacts in real-world populations are not static but associated with specific times, durations, and frequencies. In particular, a large fraction of human interactions occur recurrently with a small set of others (e.g., family, friends, colleagues) while there may be many random encounters of limited duration in public places. In other words, real-world contact networks lie somewhere on the spectrum between the slow and fast mixing regimes. This calls for improved models of contact networks beyond the static, binary ones we have used in our work. Extending our framework for analyzing herd immunity to account for temporality of interactions and heterogeneity in contact frequency and duration is an important goal for future research.

\section*{Methods}
\subsection*{Epidemic model}
The susceptible-infected-recovered (SIR) model on a static network is defined as follows. 
The dynamical state of each node is either susceptible, infected, or removed, and this state is updated in continuous time. Transmission occurs between each connected pair of an infected node and a susceptible node independently at rate $\beta$, after which the susceptible node becomes infected. Each infected node transitions to the recovered (immune) state at rate $\gamma$. The nodes in the immune state can no longer become infected or transmit the disease, and are thus effectively removed from the system. In this work, we set $\gamma = 1$ unless specified otherwise.

In our numerical simulations, we compute the outcome of the SIR model by mapping it to an epidemic percolation network~\cite{kenah2007second}, a directed network that exactly encodes the stochastic epidemic dynamics on the original network. This mapping significantly simplifies the numerical analysis; for instance, all the nodes that will be infected in an outbreak can be obtained as the descendants of the initially infected nodes in the epidemic percolation network. The details of the epidemic percolation network framework can be found in SI Appendix~E.

We derive the message-passing formalism for bond percolation~\cite{newman2023message} to analytically calculate relevant quantities, such as giant residual component $\pi_\mathrm{S_G}$, average immune node degree $\langle k \rangle_\mathrm{R}$, and the boundary size between susceptible and immune nodes, $\rho_\mathrm{SR}$. For configuration model networks, this method is equivalent to the probability generating function method for solving bond percolation~\cite{newman2002spread, newman2005threshold}. We discuss the details of the message-passing formalism in SI Appendix~F.

\subsection*{Network structures}
In this study, we use different network models to control two key topological properties: degree heterogeneity and spatiality. 
Degree heterogeneity refers to the variance in node connectivity. Node degrees are as homogeneous as possible when all nodes have the same number of neighbors, i.e., the degree distribution is degenerate, as in random regular graphs (RRGs) and regular lattices. In Erd\H{o}s-R\'{e}nyi graphs (ERGs) and random geometric graphs (RGGs), the degrees are moderately heterogeneous and follow Poisson distributions, where the variance is equal to the mean. At the more heterogeneous end of the spectrum, the network is characterized by the presence of nodes with significantly more connections than average. We use negative binomial distributions (parameterized by dispersion parameter $r$; a smaller $r$ implies a more heterogeneous degree distribution) and power law distributions (parameterized by exponent $- \alpha$; a smaller $\alpha$ implies a more heterogeneous degree distribution) to represent such high heterogeneity. For all the network models, we fix the mean degree $\langle k \rangle = 6$. 

Spatiality refers to the extent to which the underlying geometric configuration of nodes within a low-dimensional metric space determines the connectivity between them. Here, each node occupies a position in the underlying space. In a highly spatial network, the nodes are more likely to be linked to each other if they are closer to each other in space. This is the case for regular lattices and RGGs. At the non-spatial end of the spectrum is the family of configuration network models, such as RRGs and ERGs, representing maximum randomness under the given degree distribution. 

We adopt two different approaches to interpolate continuously between spatial and random networks. We use an edge randomization scheme to tune the spatiality for relatively homogeneous networks with degenerate and Poisson degree distributions. Starting with an instance of a network model with the highest spatiality, i.e., a lattice or an RGG, we rewire a fraction $\phi$ of all edges by exchanging the endpoints of two randomly selected edges~\cite{fosdick2018configuring}. This process, commonly known as the \emph{double edge swap}, preserves the degree of each node and allows us to adjust the spatiality without altering the original degree distribution. By completely randomizing edges (i.e., $\phi = 1$), the double edge swap operation transforms a lattice into an RRG and an RGG into an ERG. See SI Appendix~B for details of the edge rewiring process.

To explore the spatial network topologies with higher levels of degree heterogeneity, we adopt the heterogeneous random geometric graph (HRGG) model proposed by Bogu\~{n}\'{a} et al.~\cite{boguna2020small}, which allows us to control the degree distribution and spatiality independently.
In the HRGG model, each node is assigned an expected degree and a position in a metric space, which is assumed to be a two-dimensional unit square with periodic boundaries. 
Given the expected degrees and positions, the model generates random networks that satisfy the following properties: (i) the degree of each node is a Poisson random variable with expectation equal to the expected degree assigned to the node; (ii) the spatiality of the network is expressed as the propensity of nodes to form local connections in the underlying space, which is governed by an independent parameter called the \emph{temperature} $\tau > 0$. Low values of $\tau$ imply that nodes in proximity are more likely to be linked, and thus, the generated network is more strongly embedded in the space. On the other hand, in the limit of $\tau \to \infty$, the edges are agnostic to the positions of nodes in the underlying space, making the model equivalent to the configuration model. We use this model to generate networks with Poisson and negative binomial degree distributions with varying levels of spatiality. Note that this model cannot generate networks with a degree distribution more homogeneous than Poisson.
See SI Appendix~C for details about the HRGG model.

The two methods both induce randomness in the connection patterns between nodes embedded in space, but in different ways. We use the mean spatial length of edges to evaluate the spatiality of the generated networks in a unified manner. The length $d_{ij}$ of the edge between nodes $i$ and $j$ is the Euclidean distance between the positions of $i$ and $j$ in the underlying space, and $\langle d \rangle$ denotes the mean length of all edges. 
A completely rewired network with $\phi = 1$ and a ``hot'' HRGG with $\tau \to \infty$ are equivalent to the configuration model; in such cases, the mean edge length is equal to the expected distance $d^*$ between two random points:
\begin{equation}
\begin{aligned}
d^* & = 4 \int_0^{\frac{1}{2}} \int_0^{\frac{1}{2}} \sqrt{x^2 + y^2} \, dx dy \\
& = \frac{\sqrt{2} + \ln (\sqrt{2} + 1)}{6} \simeq 0.3826\dots .
\end{aligned}
\end{equation}
On the other hand, when $\phi = 0$ or $\tau \to 0$, each node is connected exclusively to other nodes in their proximity; therefore $\langle d \rangle \to 0$ in the large system size limit. Between these two extremes, $\langle d \rangle$ responds monotonically as a function of $\phi$ or $\tau$. We adopt $\langle d \rangle / d^*$ as the normalized measure that represents the randomness of a network with respect to the underlying space.

\begin{acknowledgments}
The authors wish to acknowledge Aalto University Science-IT project for generous computational resources.

\paragraph*{Funding:} 
MK acknowledges support from project 105572 NordicMathCovid, which is part of the Nordic Programme on Health and Welfare funded by NordForsk, and Research Council of Finland project 349366.

\paragraph*{Author contributions:}
T.H., A.K.R., M.K., and J.S. conceived the research. All authors contributed in developing the models and writing the manuscript. T.H. and Z.G. implemented the code and performed simulations. 

\paragraph*{Data and materials availability: } All the data used for the empirical population simulation are publicly available: The age structure data is available in the Demographic Statistics Database by the United Nations Statistics Division, \url{https://unstats.un.org/unsd/demographic-social/}. The contact matrix data is available in the Supporting Information of Ref.~\cite{prem2017projecting} with the identifier doi:10.1371/journal.pcbi.1005697.s002. The code used in this study is publicly available at \url{https://version.aalto.fi/gitlab/hiraokt1/herd_immunity/}.
\end{acknowledgments}


\begin{thebibliography}{88}%
\makeatletter
\providecommand \@ifxundefined [1]{%
 \@ifx{#1\undefined}
}%
\providecommand \@ifnum [1]{%
 \ifnum #1\expandafter \@firstoftwo
 \else \expandafter \@secondoftwo
 \fi
}%
\providecommand \@ifx [1]{%
 \ifx #1\expandafter \@firstoftwo
 \else \expandafter \@secondoftwo
 \fi
}%
\providecommand \natexlab [1]{#1}%
\providecommand \enquote  [1]{``#1''}%
\providecommand \bibnamefont  [1]{#1}%
\providecommand \bibfnamefont [1]{#1}%
\providecommand \citenamefont [1]{#1}%
\providecommand \href@noop [0]{\@secondoftwo}%
\providecommand \href [0]{\begingroup \@sanitize@url \@href}%
\providecommand \@href[1]{\@@startlink{#1}\@@href}%
\providecommand \@@href[1]{\endgroup#1\@@endlink}%
\providecommand \@sanitize@url [0]{\catcode `\\12\catcode `\$12\catcode
  `\&12\catcode `\#12\catcode `\^12\catcode `\_12\catcode `\%12\relax}%
\providecommand \@@startlink[1]{}%
\providecommand \@@endlink[0]{}%
\providecommand \url  [0]{\begingroup\@sanitize@url \@url }%
\providecommand \@url [1]{\endgroup\@href {#1}{\urlprefix }}%
\providecommand \urlprefix  [0]{URL }%
\providecommand \Eprint [0]{\href }%
\providecommand \doibase [0]{https://doi.org/}%
\providecommand \selectlanguage [0]{\@gobble}%
\providecommand \bibinfo  [0]{\@secondoftwo}%
\providecommand \bibfield  [0]{\@secondoftwo}%
\providecommand \translation [1]{[#1]}%
\providecommand \BibitemOpen [0]{}%
\providecommand \bibitemStop [0]{}%
\providecommand \bibitemNoStop [0]{.\EOS\space}%
\providecommand \EOS [0]{\spacefactor3000\relax}%
\providecommand \BibitemShut  [1]{\csname bibitem#1\endcsname}%
\let\auto@bib@innerbib\@empty
\bibitem [{\citenamefont {Anderson}\ and\ \citenamefont
  {May}(1991)}]{anderson1991infectious}%
  \BibitemOpen
  \bibfield  {author} {\bibinfo {author} {\bibfnamefont {R.~M.}\ \bibnamefont
  {Anderson}}\ and\ \bibinfo {author} {\bibfnamefont {R.~M.}\ \bibnamefont
  {May}},\ }\href {https://doi.org/10.1093/oso/9780198545996.001.0001} {\emph
  {\bibinfo {title} {Infectious Diseases of Humans: Dynamics and Control}}}\
  (\bibinfo  {publisher} {Oxford University Press},\ \bibinfo {year}
  {1991})\BibitemShut {NoStop}%
\bibitem [{\citenamefont {Orenstein}\ \emph {et~al.}(2000)\citenamefont
  {Orenstein}, \citenamefont {Strebel}, \citenamefont {Papania}, \citenamefont
  {Sutter}, \citenamefont {Bellini},\ and\ \citenamefont
  {Cochi}}]{orenstein2000measles}%
  \BibitemOpen
  \bibfield  {author} {\bibinfo {author} {\bibfnamefont {W.~A.}\ \bibnamefont
  {Orenstein}}, \bibinfo {author} {\bibfnamefont {P.~M.}\ \bibnamefont
  {Strebel}}, \bibinfo {author} {\bibfnamefont {M.}~\bibnamefont {Papania}},
  \bibinfo {author} {\bibfnamefont {R.~W.}\ \bibnamefont {Sutter}}, \bibinfo
  {author} {\bibfnamefont {W.~J.}\ \bibnamefont {Bellini}},\ and\ \bibinfo
  {author} {\bibfnamefont {S.~L.}\ \bibnamefont {Cochi}},\ }\bibfield  {title}
  {\bibinfo {title} {Measles eradication: is it in our future?},\ }\href
  {https://doi.org/10.2105/AJPH.90.10.1521} {\bibfield  {journal} {\bibinfo
  {journal} {American Journal of Public Health}\ }\textbf {\bibinfo {volume}
  {90}},\ \bibinfo {pages} {1521} (\bibinfo {year} {2000})}\BibitemShut
  {NoStop}%
\bibitem [{\citenamefont {Reichert}\ \emph {et~al.}(2001)\citenamefont
  {Reichert}, \citenamefont {Sugaya}, \citenamefont {Fedson}, \citenamefont
  {Glezen}, \citenamefont {Simonsen},\ and\ \citenamefont
  {Tashiro}}]{reichert2001japanese}%
  \BibitemOpen
  \bibfield  {author} {\bibinfo {author} {\bibfnamefont {T.~A.}\ \bibnamefont
  {Reichert}}, \bibinfo {author} {\bibfnamefont {N.}~\bibnamefont {Sugaya}},
  \bibinfo {author} {\bibfnamefont {D.~S.}\ \bibnamefont {Fedson}}, \bibinfo
  {author} {\bibfnamefont {W.~P.}\ \bibnamefont {Glezen}}, \bibinfo {author}
  {\bibfnamefont {L.}~\bibnamefont {Simonsen}},\ and\ \bibinfo {author}
  {\bibfnamefont {M.}~\bibnamefont {Tashiro}},\ }\bibfield  {title} {\bibinfo
  {title} {The {{Japanese}} experience with vaccinating schoolchildren against
  influenza},\ }\href {https://doi.org/10.1056/NEJM200103223441204} {\bibfield
  {journal} {\bibinfo  {journal} {New England Journal of Medicine}\ }\textbf
  {\bibinfo {volume} {344}},\ \bibinfo {pages} {889} (\bibinfo {year}
  {2001})}\BibitemShut {NoStop}%
\bibitem [{\citenamefont {Drolet}\ \emph {et~al.}(2015)\citenamefont {Drolet},
  \citenamefont {B{\'e}nard}, \citenamefont {Boily}, \citenamefont {Ali},
  \citenamefont {Baandrup}, \citenamefont {Bauer}, \citenamefont {Beddows},
  \citenamefont {Brisson}, \citenamefont {Brotherton}, \citenamefont
  {Cummings}, \citenamefont {Donovan}, \citenamefont {Fairley}, \citenamefont
  {Flagg}, \citenamefont {Johnson}, \citenamefont {Kahn}, \citenamefont
  {Kavanagh}, \citenamefont {Kjaer}, \citenamefont {Kliewer}, \citenamefont
  {{Lemieux-Mellouki}}, \citenamefont {Markowitz}, \citenamefont {Mboup},
  \citenamefont {Mesher}, \citenamefont {Niccolai}, \citenamefont {Oliphant},
  \citenamefont {Pollock}, \citenamefont {Soldan}, \citenamefont {Sonnenberg},
  \citenamefont {Tabrizi}, \citenamefont {Tanton},\ and\ \citenamefont
  {Brisson}}]{drolet2015populationlevel}%
  \BibitemOpen
  \bibfield  {author} {\bibinfo {author} {\bibfnamefont {M.}~\bibnamefont
  {Drolet}}, \bibinfo {author} {\bibfnamefont {{\'E}.}~\bibnamefont
  {B{\'e}nard}}, \bibinfo {author} {\bibfnamefont {M.-C.}\ \bibnamefont
  {Boily}}, \bibinfo {author} {\bibfnamefont {H.}~\bibnamefont {Ali}}, \bibinfo
  {author} {\bibfnamefont {L.}~\bibnamefont {Baandrup}}, \bibinfo {author}
  {\bibfnamefont {H.}~\bibnamefont {Bauer}}, \bibinfo {author} {\bibfnamefont
  {S.}~\bibnamefont {Beddows}}, \bibinfo {author} {\bibfnamefont
  {J.}~\bibnamefont {Brisson}}, \bibinfo {author} {\bibfnamefont {J.~M.~L.}\
  \bibnamefont {Brotherton}}, \bibinfo {author} {\bibfnamefont
  {T.}~\bibnamefont {Cummings}}, \bibinfo {author} {\bibfnamefont
  {B.}~\bibnamefont {Donovan}}, \bibinfo {author} {\bibfnamefont {C.~K.}\
  \bibnamefont {Fairley}}, \bibinfo {author} {\bibfnamefont {E.~W.}\
  \bibnamefont {Flagg}}, \bibinfo {author} {\bibfnamefont {A.~M.}\ \bibnamefont
  {Johnson}}, \bibinfo {author} {\bibfnamefont {J.~A.}\ \bibnamefont {Kahn}},
  \bibinfo {author} {\bibfnamefont {K.}~\bibnamefont {Kavanagh}}, \bibinfo
  {author} {\bibfnamefont {S.~K.}\ \bibnamefont {Kjaer}}, \bibinfo {author}
  {\bibfnamefont {E.~V.}\ \bibnamefont {Kliewer}}, \bibinfo {author}
  {\bibfnamefont {P.}~\bibnamefont {{Lemieux-Mellouki}}}, \bibinfo {author}
  {\bibfnamefont {L.}~\bibnamefont {Markowitz}}, \bibinfo {author}
  {\bibfnamefont {A.}~\bibnamefont {Mboup}}, \bibinfo {author} {\bibfnamefont
  {D.}~\bibnamefont {Mesher}}, \bibinfo {author} {\bibfnamefont
  {L.}~\bibnamefont {Niccolai}}, \bibinfo {author} {\bibfnamefont
  {J.}~\bibnamefont {Oliphant}}, \bibinfo {author} {\bibfnamefont {K.~G.}\
  \bibnamefont {Pollock}}, \bibinfo {author} {\bibfnamefont {K.}~\bibnamefont
  {Soldan}}, \bibinfo {author} {\bibfnamefont {P.}~\bibnamefont {Sonnenberg}},
  \bibinfo {author} {\bibfnamefont {S.~N.}\ \bibnamefont {Tabrizi}}, \bibinfo
  {author} {\bibfnamefont {C.}~\bibnamefont {Tanton}},\ and\ \bibinfo {author}
  {\bibfnamefont {M.}~\bibnamefont {Brisson}},\ }\bibfield  {title} {\bibinfo
  {title} {Population-level impact and herd effects following human
  papillomavirus vaccination programmes: a systematic review and
  meta-analysis},\ }\href {https://doi.org/10.1016/S1473-3099(14)71073-4}
  {\bibfield  {journal} {\bibinfo  {journal} {The Lancet Infectious Diseases}\
  }\textbf {\bibinfo {volume} {15}},\ \bibinfo {pages} {565} (\bibinfo {year}
  {2015})}\BibitemShut {NoStop}%
\bibitem [{\citenamefont {Toor}\ \emph {et~al.}(2021)\citenamefont {Toor},
  \citenamefont {{Echeverria-Londono}}, \citenamefont {Li}, \citenamefont
  {Abbas}, \citenamefont {Carter}, \citenamefont {Clapham}, \citenamefont
  {Clark}, \citenamefont {De~Villiers}, \citenamefont {Eilertson},
  \citenamefont {Ferrari}, \citenamefont {Gamkrelidze}, \citenamefont
  {Hallett}, \citenamefont {Hinsley}, \citenamefont {Hogan}, \citenamefont
  {Huber}, \citenamefont {Jackson}, \citenamefont {Jean}, \citenamefont {Jit},
  \citenamefont {Karachaliou}, \citenamefont {Klepac}, \citenamefont {Kraay},
  \citenamefont {Lessler}, \citenamefont {Li}, \citenamefont {Lopman},
  \citenamefont {Mengistu}, \citenamefont {Metcalf}, \citenamefont {Moore},
  \citenamefont {Nayagam}, \citenamefont {Papadopoulos}, \citenamefont
  {Perkins}, \citenamefont {Portnoy}, \citenamefont {Razavi}, \citenamefont
  {{Razavi-Shearer}}, \citenamefont {Resch}, \citenamefont {Sanderson},
  \citenamefont {Sweet}, \citenamefont {Tam}, \citenamefont {Tanvir},
  \citenamefont {Tran~Minh}, \citenamefont {Trotter}, \citenamefont {Truelove},
  \citenamefont {Vynnycky}, \citenamefont {Walker}, \citenamefont {Winter},
  \citenamefont {Woodruff}, \citenamefont {Ferguson},\ and\ \citenamefont
  {Gaythorpe}}]{toor2021lives}%
  \BibitemOpen
  \bibfield  {author} {\bibinfo {author} {\bibfnamefont {J.}~\bibnamefont
  {Toor}}, \bibinfo {author} {\bibfnamefont {S.}~\bibnamefont
  {{Echeverria-Londono}}}, \bibinfo {author} {\bibfnamefont {X.}~\bibnamefont
  {Li}}, \bibinfo {author} {\bibfnamefont {K.}~\bibnamefont {Abbas}}, \bibinfo
  {author} {\bibfnamefont {E.~D.}\ \bibnamefont {Carter}}, \bibinfo {author}
  {\bibfnamefont {H.~E.}\ \bibnamefont {Clapham}}, \bibinfo {author}
  {\bibfnamefont {A.}~\bibnamefont {Clark}}, \bibinfo {author} {\bibfnamefont
  {M.~J.}\ \bibnamefont {De~Villiers}}, \bibinfo {author} {\bibfnamefont
  {K.}~\bibnamefont {Eilertson}}, \bibinfo {author} {\bibfnamefont
  {M.}~\bibnamefont {Ferrari}}, \bibinfo {author} {\bibfnamefont
  {I.}~\bibnamefont {Gamkrelidze}}, \bibinfo {author} {\bibfnamefont {T.~B.}\
  \bibnamefont {Hallett}}, \bibinfo {author} {\bibfnamefont {W.~R.}\
  \bibnamefont {Hinsley}}, \bibinfo {author} {\bibfnamefont {D.}~\bibnamefont
  {Hogan}}, \bibinfo {author} {\bibfnamefont {J.~H.}\ \bibnamefont {Huber}},
  \bibinfo {author} {\bibfnamefont {M.~L.}\ \bibnamefont {Jackson}}, \bibinfo
  {author} {\bibfnamefont {K.}~\bibnamefont {Jean}}, \bibinfo {author}
  {\bibfnamefont {M.}~\bibnamefont {Jit}}, \bibinfo {author} {\bibfnamefont
  {A.}~\bibnamefont {Karachaliou}}, \bibinfo {author} {\bibfnamefont
  {P.}~\bibnamefont {Klepac}}, \bibinfo {author} {\bibfnamefont
  {A.}~\bibnamefont {Kraay}}, \bibinfo {author} {\bibfnamefont
  {J.}~\bibnamefont {Lessler}}, \bibinfo {author} {\bibfnamefont
  {X.}~\bibnamefont {Li}}, \bibinfo {author} {\bibfnamefont {B.~A.}\
  \bibnamefont {Lopman}}, \bibinfo {author} {\bibfnamefont {T.}~\bibnamefont
  {Mengistu}}, \bibinfo {author} {\bibfnamefont {C.~J.~E.}\ \bibnamefont
  {Metcalf}}, \bibinfo {author} {\bibfnamefont {S.~M.}\ \bibnamefont {Moore}},
  \bibinfo {author} {\bibfnamefont {S.}~\bibnamefont {Nayagam}}, \bibinfo
  {author} {\bibfnamefont {T.}~\bibnamefont {Papadopoulos}}, \bibinfo {author}
  {\bibfnamefont {T.~A.}\ \bibnamefont {Perkins}}, \bibinfo {author}
  {\bibfnamefont {A.}~\bibnamefont {Portnoy}}, \bibinfo {author} {\bibfnamefont
  {H.}~\bibnamefont {Razavi}}, \bibinfo {author} {\bibfnamefont
  {D.}~\bibnamefont {{Razavi-Shearer}}}, \bibinfo {author} {\bibfnamefont
  {S.}~\bibnamefont {Resch}}, \bibinfo {author} {\bibfnamefont
  {C.}~\bibnamefont {Sanderson}}, \bibinfo {author} {\bibfnamefont
  {S.}~\bibnamefont {Sweet}}, \bibinfo {author} {\bibfnamefont
  {Y.}~\bibnamefont {Tam}}, \bibinfo {author} {\bibfnamefont {H.}~\bibnamefont
  {Tanvir}}, \bibinfo {author} {\bibfnamefont {Q.}~\bibnamefont {Tran~Minh}},
  \bibinfo {author} {\bibfnamefont {C.~L.}\ \bibnamefont {Trotter}}, \bibinfo
  {author} {\bibfnamefont {S.~A.}\ \bibnamefont {Truelove}}, \bibinfo {author}
  {\bibfnamefont {E.}~\bibnamefont {Vynnycky}}, \bibinfo {author}
  {\bibfnamefont {N.}~\bibnamefont {Walker}}, \bibinfo {author} {\bibfnamefont
  {A.}~\bibnamefont {Winter}}, \bibinfo {author} {\bibfnamefont
  {K.}~\bibnamefont {Woodruff}}, \bibinfo {author} {\bibfnamefont {N.~M.}\
  \bibnamefont {Ferguson}},\ and\ \bibinfo {author} {\bibfnamefont {K.~A.}\
  \bibnamefont {Gaythorpe}},\ }\bibfield  {title} {\bibinfo {title} {Lives
  saved with vaccination for 10 pathogens across 112 countries in a
  pre-{{COVID-19}} world},\ }\href {https://doi.org/10.7554/eLife.67635}
  {\bibfield  {journal} {\bibinfo  {journal} {eLife}\ }\textbf {\bibinfo
  {volume} {10}},\ \bibinfo {pages} {e67635} (\bibinfo {year}
  {2021})}\BibitemShut {NoStop}%
\bibitem [{\citenamefont {Shattock}\ \emph {et~al.}(2024)\citenamefont
  {Shattock}, \citenamefont {Johnson}, \citenamefont {Sim}, \citenamefont
  {Carter}, \citenamefont {Lambach}, \citenamefont {Hutubessy}, \citenamefont
  {Thompson}, \citenamefont {Badizadegan}, \citenamefont {Lambert},
  \citenamefont {Ferrari}, \citenamefont {Jit}, \citenamefont {Fu},
  \citenamefont {Silal}, \citenamefont {Hounsell}, \citenamefont {White},
  \citenamefont {Mosser}, \citenamefont {Gaythorpe}, \citenamefont {Trotter},
  \citenamefont {Lindstrand}, \citenamefont {O'Brien},\ and\ \citenamefont
  {{Bar-Zeev}}}]{shattock2024contribution}%
  \BibitemOpen
  \bibfield  {author} {\bibinfo {author} {\bibfnamefont {A.~J.}\ \bibnamefont
  {Shattock}}, \bibinfo {author} {\bibfnamefont {H.~C.}\ \bibnamefont
  {Johnson}}, \bibinfo {author} {\bibfnamefont {S.~Y.}\ \bibnamefont {Sim}},
  \bibinfo {author} {\bibfnamefont {A.}~\bibnamefont {Carter}}, \bibinfo
  {author} {\bibfnamefont {P.}~\bibnamefont {Lambach}}, \bibinfo {author}
  {\bibfnamefont {R.~C.~W.}\ \bibnamefont {Hutubessy}}, \bibinfo {author}
  {\bibfnamefont {K.~M.}\ \bibnamefont {Thompson}}, \bibinfo {author}
  {\bibfnamefont {K.}~\bibnamefont {Badizadegan}}, \bibinfo {author}
  {\bibfnamefont {B.}~\bibnamefont {Lambert}}, \bibinfo {author} {\bibfnamefont
  {M.~J.}\ \bibnamefont {Ferrari}}, \bibinfo {author} {\bibfnamefont
  {M.}~\bibnamefont {Jit}}, \bibinfo {author} {\bibfnamefont {H.}~\bibnamefont
  {Fu}}, \bibinfo {author} {\bibfnamefont {S.~P.}\ \bibnamefont {Silal}},
  \bibinfo {author} {\bibfnamefont {R.~A.}\ \bibnamefont {Hounsell}}, \bibinfo
  {author} {\bibfnamefont {R.~G.}\ \bibnamefont {White}}, \bibinfo {author}
  {\bibfnamefont {J.~F.}\ \bibnamefont {Mosser}}, \bibinfo {author}
  {\bibfnamefont {K.~A.~M.}\ \bibnamefont {Gaythorpe}}, \bibinfo {author}
  {\bibfnamefont {C.~L.}\ \bibnamefont {Trotter}}, \bibinfo {author}
  {\bibfnamefont {A.}~\bibnamefont {Lindstrand}}, \bibinfo {author}
  {\bibfnamefont {K.~L.}\ \bibnamefont {O'Brien}},\ and\ \bibinfo {author}
  {\bibfnamefont {N.}~\bibnamefont {{Bar-Zeev}}},\ }\bibfield  {title}
  {\bibinfo {title} {Contribution of vaccination to improved survival and
  health: modelling 50 years of the {{Expanded Programme}} on
  {{Immunization}}},\ }\href {https://doi.org/10.1016/S0140-6736(24)00850-X}
  {\bibfield  {journal} {\bibinfo  {journal} {The Lancet}\ }\textbf {\bibinfo
  {volume} {403}},\ \bibinfo {pages} {2307} (\bibinfo {year}
  {2024})}\BibitemShut {NoStop}%
\bibitem [{\citenamefont {Anderson}\ and\ \citenamefont
  {May}(1985)}]{anderson1985vaccination}%
  \BibitemOpen
  \bibfield  {author} {\bibinfo {author} {\bibfnamefont {R.~M.}\ \bibnamefont
  {Anderson}}\ and\ \bibinfo {author} {\bibfnamefont {R.~M.}\ \bibnamefont
  {May}},\ }\bibfield  {title} {\bibinfo {title} {Vaccination and herd immunity
  to infectious diseases},\ }\href {https://doi.org/10.1038/318323a0}
  {\bibfield  {journal} {\bibinfo  {journal} {Nature}\ }\textbf {\bibinfo
  {volume} {318}},\ \bibinfo {pages} {323} (\bibinfo {year}
  {1985})}\BibitemShut {NoStop}%
\bibitem [{\citenamefont {Anderson}\ and\ \citenamefont
  {May}(1990)}]{anderson1990immunisation}%
  \BibitemOpen
  \bibfield  {author} {\bibinfo {author} {\bibfnamefont {R.}~\bibnamefont
  {Anderson}}\ and\ \bibinfo {author} {\bibfnamefont {R.}~\bibnamefont {May}},\
  }\bibfield  {title} {\bibinfo {title} {Immunisation and herd immunity},\
  }\href {https://doi.org/10.1016/0140-6736(90)90420-A} {\bibfield  {journal}
  {\bibinfo  {journal} {The Lancet}\ }\textbf {\bibinfo {volume} {335}},\
  \bibinfo {pages} {641} (\bibinfo {year} {1990})}\BibitemShut {NoStop}%
\bibitem [{\citenamefont {Anderson}(1992)}]{anderson1992concept}%
  \BibitemOpen
  \bibfield  {author} {\bibinfo {author} {\bibfnamefont {R.~M.}\ \bibnamefont
  {Anderson}},\ }\bibfield  {title} {\bibinfo {title} {The concept of herd
  immunity and the design of community-based immunization programmes},\ }\href
  {https://doi.org/10.1016/0264-410X(92)90327-G} {\bibfield  {journal}
  {\bibinfo  {journal} {Vaccine}\ }\textbf {\bibinfo {volume} {10}},\ \bibinfo
  {pages} {928} (\bibinfo {year} {1992})}\BibitemShut {NoStop}%
\bibitem [{\citenamefont {Fine}\ \emph {et~al.}(2011)\citenamefont {Fine},
  \citenamefont {Eames},\ and\ \citenamefont {Heymann}}]{fine2011herd}%
  \BibitemOpen
  \bibfield  {author} {\bibinfo {author} {\bibfnamefont {P.}~\bibnamefont
  {Fine}}, \bibinfo {author} {\bibfnamefont {K.}~\bibnamefont {Eames}},\ and\
  \bibinfo {author} {\bibfnamefont {D.~L.}\ \bibnamefont {Heymann}},\
  }\bibfield  {title} {\bibinfo {title} {``herd immunity'': A rough guide},\
  }\href {https://doi.org/10.1093/cid/cir007} {\bibfield  {journal} {\bibinfo
  {journal} {Clinical Infectious Diseases}\ }\textbf {\bibinfo {volume} {52}},\
  \bibinfo {pages} {911} (\bibinfo {year} {2011})}\BibitemShut {NoStop}%
\bibitem [{\citenamefont {Fine}(1993)}]{fine1993herd}%
  \BibitemOpen
  \bibfield  {author} {\bibinfo {author} {\bibfnamefont {P.~E.~M.}\
  \bibnamefont {Fine}},\ }\bibfield  {title} {\bibinfo {title} {{Herd Immunity:
  History, Theory, Practice}},\ }\href
  {https://doi.org/10.1093/oxfordjournals.epirev.a036121} {\bibfield  {journal}
  {\bibinfo  {journal} {Epidemiologic Reviews}\ }\textbf {\bibinfo {volume}
  {15}},\ \bibinfo {pages} {265} (\bibinfo {year} {1993})}\BibitemShut
  {NoStop}%
\bibitem [{\citenamefont {Jones}\ and\ \citenamefont
  {Helmreich}(2020)}]{jones2020history}%
  \BibitemOpen
  \bibfield  {author} {\bibinfo {author} {\bibfnamefont {D.}~\bibnamefont
  {Jones}}\ and\ \bibinfo {author} {\bibfnamefont {S.}~\bibnamefont
  {Helmreich}},\ }\bibfield  {title} {\bibinfo {title} {A history of herd
  immunity},\ }\href
  {https://doi.org/https://doi.org/10.1016/S0140-6736(20)31924-3} {\bibfield
  {journal} {\bibinfo  {journal} {The Lancet}\ }\textbf {\bibinfo {volume}
  {396}},\ \bibinfo {pages} {810} (\bibinfo {year} {2020})}\BibitemShut
  {NoStop}%
\bibitem [{\citenamefont {Morens}\ \emph {et~al.}(2022)\citenamefont {Morens},
  \citenamefont {Folkers},\ and\ \citenamefont {Fauci}}]{morens2022concept}%
  \BibitemOpen
  \bibfield  {author} {\bibinfo {author} {\bibfnamefont {D.~M.}\ \bibnamefont
  {Morens}}, \bibinfo {author} {\bibfnamefont {G.~K.}\ \bibnamefont
  {Folkers}},\ and\ \bibinfo {author} {\bibfnamefont {A.~S.}\ \bibnamefont
  {Fauci}},\ }\bibfield  {title} {\bibinfo {title} {The concept of classical
  herd immunity may not apply to {{COVID-19}}},\ }\href
  {https://doi.org/10.1093/infdis/jiac109} {\bibfield  {journal} {\bibinfo
  {journal} {The Journal of Infectious Diseases}\ }\textbf {\bibinfo {volume}
  {226}},\ \bibinfo {pages} {195} (\bibinfo {year} {2022})}\BibitemShut
  {NoStop}%
\bibitem [{\citenamefont {Anderson}\ and\ \citenamefont
  {May}(1982)}]{anderson1982directly}%
  \BibitemOpen
  \bibfield  {author} {\bibinfo {author} {\bibfnamefont {R.~M.}\ \bibnamefont
  {Anderson}}\ and\ \bibinfo {author} {\bibfnamefont {R.~M.}\ \bibnamefont
  {May}},\ }\bibfield  {title} {\bibinfo {title} {Directly transmitted
  infections diseases: Control by vaccination},\ }\href
  {https://doi.org/10.1126/science.7063839} {\bibfield  {journal} {\bibinfo
  {journal} {Science}\ }\textbf {\bibinfo {volume} {215}},\ \bibinfo {pages}
  {1053} (\bibinfo {year} {1982})}\BibitemShut {NoStop}%
\bibitem [{\citenamefont {Fox}\ \emph {et~al.}(1971)\citenamefont {Fox},
  \citenamefont {Elveback}, \citenamefont {Scott}, \citenamefont {Gatewood},\
  and\ \citenamefont {Ackerman}}]{fox1971herd}%
  \BibitemOpen
  \bibfield  {author} {\bibinfo {author} {\bibfnamefont {J.~P.}\ \bibnamefont
  {Fox}}, \bibinfo {author} {\bibfnamefont {L.}~\bibnamefont {Elveback}},
  \bibinfo {author} {\bibfnamefont {W.}~\bibnamefont {Scott}}, \bibinfo
  {author} {\bibfnamefont {L.}~\bibnamefont {Gatewood}},\ and\ \bibinfo
  {author} {\bibfnamefont {E.}~\bibnamefont {Ackerman}},\ }\bibfield  {title}
  {\bibinfo {title} {{Herd Immunity: Basic Concept and Relevance to Public
  Health Immunization Practices}},\ }\href
  {https://doi.org/10.1093/oxfordjournals.aje.a121310} {\bibfield  {journal}
  {\bibinfo  {journal} {American Journal of Epidemiology}\ }\textbf {\bibinfo
  {volume} {94}},\ \bibinfo {pages} {179} (\bibinfo {year} {1971})}\BibitemShut
  {NoStop}%
\bibitem [{\citenamefont {Schenzle}(1984)}]{schenzle1984agestructured}%
  \BibitemOpen
  \bibfield  {author} {\bibinfo {author} {\bibfnamefont {D.}~\bibnamefont
  {Schenzle}},\ }\bibfield  {title} {\bibinfo {title} {An age-structured model
  of pre- and post-vaccination measles transmission},\ }\href
  {https://doi.org/10.1093/imammb/1.2.169} {\bibfield  {journal} {\bibinfo
  {journal} {Mathematical Medicine and Biology}\ }\textbf {\bibinfo {volume}
  {1}},\ \bibinfo {pages} {169} (\bibinfo {year} {1984})}\BibitemShut {NoStop}%
\bibitem [{\citenamefont {May}\ and\ \citenamefont
  {Anderson}(1984)}]{may1984spatial}%
  \BibitemOpen
  \bibfield  {author} {\bibinfo {author} {\bibfnamefont {R.~M.}\ \bibnamefont
  {May}}\ and\ \bibinfo {author} {\bibfnamefont {R.~M.}\ \bibnamefont
  {Anderson}},\ }\bibfield  {title} {\bibinfo {title} {Spatial heterogeneity
  and the design of immunization programs},\ }\href
  {https://doi.org/10.1016/0025-5564(84)90063-4} {\bibfield  {journal}
  {\bibinfo  {journal} {Mathematical Biosciences}\ }\textbf {\bibinfo {volume}
  {72}},\ \bibinfo {pages} {83} (\bibinfo {year} {1984})}\BibitemShut {NoStop}%
\bibitem [{\citenamefont {Anderson}\ and\ \citenamefont
  {May}(1984)}]{anderson1984spatial}%
  \BibitemOpen
  \bibfield  {author} {\bibinfo {author} {\bibfnamefont {R.~M.}\ \bibnamefont
  {Anderson}}\ and\ \bibinfo {author} {\bibfnamefont {R.~M.}\ \bibnamefont
  {May}},\ }\bibfield  {title} {\bibinfo {title} {Spatial, temporal, and
  genetic heterogeneity in host populations and the design of immunization
  programmes},\ }\href {https://doi.org/10.1093/imammb/1.3.233} {\bibfield
  {journal} {\bibinfo  {journal} {Mathematical Medicine and Biology}\ }\textbf
  {\bibinfo {volume} {1}},\ \bibinfo {pages} {233} (\bibinfo {year}
  {1984})}\BibitemShut {NoStop}%
\bibitem [{\citenamefont {Hethcote}\ and\ \citenamefont {{Van
  Ark}}(1987)}]{hethcote1987epidemiological}%
  \BibitemOpen
  \bibfield  {author} {\bibinfo {author} {\bibfnamefont {H.~W.}\ \bibnamefont
  {Hethcote}}\ and\ \bibinfo {author} {\bibfnamefont {J.~W.}\ \bibnamefont
  {{Van Ark}}},\ }\bibfield  {title} {\bibinfo {title} {{Epidemiological models
  for heterogeneous populations: proportionate mixing, parameter estimation,
  and immunization programs}},\ }\href
  {https://doi.org/10.1016/0025-5564(87)90044-7} {\bibfield  {journal}
  {\bibinfo  {journal} {Mathematical Biosciences}\ }\textbf {\bibinfo {volume}
  {84}},\ \bibinfo {pages} {85} (\bibinfo {year} {1987})}\BibitemShut {NoStop}%
\bibitem [{\citenamefont {Kermack}\ and\ \citenamefont
  {McKendrick}(1927)}]{kermack1927contribution}%
  \BibitemOpen
  \bibfield  {author} {\bibinfo {author} {\bibfnamefont {W.~O.}\ \bibnamefont
  {Kermack}}\ and\ \bibinfo {author} {\bibfnamefont {A.~G.}\ \bibnamefont
  {McKendrick}},\ }\bibfield  {title} {\bibinfo {title} {A contribution to the
  mathematical theory of epidemics},\ }\href
  {https://doi.org/10.1098/rspa.1927.0118} {\bibfield  {journal} {\bibinfo
  {journal} {Proceedings of the Royal Society of London. Series A, Containing
  Papers of a Mathematical and Physical Character}\ }\textbf {\bibinfo {volume}
  {115}},\ \bibinfo {pages} {700} (\bibinfo {year} {1927})}\BibitemShut
  {NoStop}%
\bibitem [{\citenamefont {Britton}\ \emph {et~al.}(2020)\citenamefont
  {Britton}, \citenamefont {Ball},\ and\ \citenamefont
  {Trapman}}]{britton2020mathematical}%
  \BibitemOpen
  \bibfield  {author} {\bibinfo {author} {\bibfnamefont {T.}~\bibnamefont
  {Britton}}, \bibinfo {author} {\bibfnamefont {F.}~\bibnamefont {Ball}},\ and\
  \bibinfo {author} {\bibfnamefont {P.}~\bibnamefont {Trapman}},\ }\bibfield
  {title} {\bibinfo {title} {A mathematical model reveals the influence of
  population heterogeneity on herd immunity to {{SARS-CoV-2}}},\ }\href
  {https://doi.org/10.1126/science.abc6810} {\bibfield  {journal} {\bibinfo
  {journal} {Science}\ }\textbf {\bibinfo {volume} {369}},\ \bibinfo {pages}
  {846} (\bibinfo {year} {2020})}\BibitemShut {NoStop}%
\bibitem [{\citenamefont {Neipel}\ \emph {et~al.}(2020)\citenamefont {Neipel},
  \citenamefont {Bauermann}, \citenamefont {Bo}, \citenamefont {Harmon},\ and\
  \citenamefont {J{\"u}licher}}]{neipel2020powerlaw}%
  \BibitemOpen
  \bibfield  {author} {\bibinfo {author} {\bibfnamefont {J.}~\bibnamefont
  {Neipel}}, \bibinfo {author} {\bibfnamefont {J.}~\bibnamefont {Bauermann}},
  \bibinfo {author} {\bibfnamefont {S.}~\bibnamefont {Bo}}, \bibinfo {author}
  {\bibfnamefont {T.}~\bibnamefont {Harmon}},\ and\ \bibinfo {author}
  {\bibfnamefont {F.}~\bibnamefont {J{\"u}licher}},\ }\bibfield  {title}
  {\bibinfo {title} {Power-law population heterogeneity governs epidemic
  waves},\ }\href {https://doi.org/10.1371/journal.pone.0239678} {\bibfield
  {journal} {\bibinfo  {journal} {PLOS ONE}\ }\textbf {\bibinfo {volume}
  {15}},\ \bibinfo {pages} {e0239678} (\bibinfo {year} {2020})}\BibitemShut
  {NoStop}%
\bibitem [{\citenamefont {Lu}\ \emph {et~al.}(2021)\citenamefont {Lu},
  \citenamefont {Aleta}, \citenamefont {Ajelli}, \citenamefont
  {{Pastor-Satorras}}, \citenamefont {Vespignani},\ and\ \citenamefont
  {Moreno}}]{lu2021datadriven}%
  \BibitemOpen
  \bibfield  {author} {\bibinfo {author} {\bibfnamefont {D.}~\bibnamefont
  {Lu}}, \bibinfo {author} {\bibfnamefont {A.}~\bibnamefont {Aleta}}, \bibinfo
  {author} {\bibfnamefont {M.}~\bibnamefont {Ajelli}}, \bibinfo {author}
  {\bibfnamefont {R.}~\bibnamefont {{Pastor-Satorras}}}, \bibinfo {author}
  {\bibfnamefont {A.}~\bibnamefont {Vespignani}},\ and\ \bibinfo {author}
  {\bibfnamefont {Y.}~\bibnamefont {Moreno}},\ }\bibfield  {title} {\bibinfo
  {title} {Data-driven estimate of {{SARS-CoV-2}} herd immunity threshold in
  populations with individual contact pattern variations},\ }\bibfield
  {journal} {\bibinfo  {journal} {medRxiv}\ }\href
  {https://doi.org/10.1101/2021.03.19.21253974} {10.1101/2021.03.19.21253974}
  (\bibinfo {year} {2021})\BibitemShut {NoStop}%
\bibitem [{\citenamefont {Gomes}\ \emph {et~al.}(2022)\citenamefont {Gomes},
  \citenamefont {Ferreira}, \citenamefont {Corder}, \citenamefont {King},
  \citenamefont {{Souto-Maior}}, \citenamefont {{Penha-Gon{\c c}alves}},
  \citenamefont {Gon{\c c}alves}, \citenamefont {Chikina}, \citenamefont
  {Pegden},\ and\ \citenamefont {Aguas}}]{gomes2022individual}%
  \BibitemOpen
  \bibfield  {author} {\bibinfo {author} {\bibfnamefont {M.~G.~M.}\
  \bibnamefont {Gomes}}, \bibinfo {author} {\bibfnamefont {M.~U.}\ \bibnamefont
  {Ferreira}}, \bibinfo {author} {\bibfnamefont {R.~M.}\ \bibnamefont
  {Corder}}, \bibinfo {author} {\bibfnamefont {J.~G.}\ \bibnamefont {King}},
  \bibinfo {author} {\bibfnamefont {C.}~\bibnamefont {{Souto-Maior}}}, \bibinfo
  {author} {\bibfnamefont {C.}~\bibnamefont {{Penha-Gon{\c c}alves}}}, \bibinfo
  {author} {\bibfnamefont {G.}~\bibnamefont {Gon{\c c}alves}}, \bibinfo
  {author} {\bibfnamefont {M.}~\bibnamefont {Chikina}}, \bibinfo {author}
  {\bibfnamefont {W.}~\bibnamefont {Pegden}},\ and\ \bibinfo {author}
  {\bibfnamefont {R.}~\bibnamefont {Aguas}},\ }\bibfield  {title} {\bibinfo
  {title} {Individual variation in susceptibility or exposure to {{SARS-CoV-2}}
  lowers the herd immunity threshold},\ }\href
  {https://doi.org/10.1016/j.jtbi.2022.111063} {\bibfield  {journal} {\bibinfo
  {journal} {Journal of Theoretical Biology}\ }\textbf {\bibinfo {volume}
  {540}},\ \bibinfo {pages} {111063} (\bibinfo {year} {2022})}\BibitemShut
  {NoStop}%
\bibitem [{\citenamefont {Aguas}\ \emph {et~al.}(2022)\citenamefont {Aguas},
  \citenamefont {Gon{\c c}alves}, \citenamefont {Ferreira},\ and\ \citenamefont
  {M.~Gomes}}]{aguas2022herd}%
  \BibitemOpen
  \bibfield  {author} {\bibinfo {author} {\bibfnamefont {R.}~\bibnamefont
  {Aguas}}, \bibinfo {author} {\bibfnamefont {G.}~\bibnamefont {Gon{\c
  c}alves}}, \bibinfo {author} {\bibfnamefont {M.~U.}\ \bibnamefont
  {Ferreira}},\ and\ \bibinfo {author} {\bibfnamefont {M.~G.}\ \bibnamefont
  {M.~Gomes}},\ }\bibfield  {title} {\bibinfo {title} {Herd immunity thresholds
  for {{SARS-CoV-2}} estimated from unfolding epidemics},\ }\bibfield
  {journal} {\bibinfo  {journal} {medRxiv}\ }\href
  {https://doi.org/10.1101/2020.07.23.20160762} {10.1101/2020.07.23.20160762}
  (\bibinfo {year} {2022})\BibitemShut {NoStop}%
\bibitem [{\citenamefont {Murayama}\ \emph {et~al.}(2024)\citenamefont
  {Murayama}, \citenamefont {Pearson}, \citenamefont {Abbott}, \citenamefont
  {Miura}, \citenamefont {Jung}, \citenamefont {Fearon}, \citenamefont {Funk},\
  and\ \citenamefont {Endo}}]{murayama2024accumulation}%
  \BibitemOpen
  \bibfield  {author} {\bibinfo {author} {\bibfnamefont {H.}~\bibnamefont
  {Murayama}}, \bibinfo {author} {\bibfnamefont {C.~A.~B.}\ \bibnamefont
  {Pearson}}, \bibinfo {author} {\bibfnamefont {S.}~\bibnamefont {Abbott}},
  \bibinfo {author} {\bibfnamefont {F.}~\bibnamefont {Miura}}, \bibinfo
  {author} {\bibfnamefont {S.-m.}\ \bibnamefont {Jung}}, \bibinfo {author}
  {\bibfnamefont {E.}~\bibnamefont {Fearon}}, \bibinfo {author} {\bibfnamefont
  {S.}~\bibnamefont {Funk}},\ and\ \bibinfo {author} {\bibfnamefont
  {A.}~\bibnamefont {Endo}},\ }\bibfield  {title} {\bibinfo {title}
  {Accumulation of immunity in heavy-tailed sexual contact networks shapes mpox
  outbreak sizes},\ }\href {https://doi.org/10.1093/infdis/jiad254} {\bibfield
  {journal} {\bibinfo  {journal} {The Journal of Infectious Diseases}\ }\textbf
  {\bibinfo {volume} {229}},\ \bibinfo {pages} {59} (\bibinfo {year}
  {2024})}\BibitemShut {NoStop}%
\bibitem [{\citenamefont {Xiridou}\ \emph {et~al.}(2024)\citenamefont
  {Xiridou}, \citenamefont {Miura}, \citenamefont {Adam}, \citenamefont {Op~de
  Coul}, \citenamefont {de~Wit},\ and\ \citenamefont
  {Wallinga}}]{xiridou2024fading}%
  \BibitemOpen
  \bibfield  {author} {\bibinfo {author} {\bibfnamefont {M.}~\bibnamefont
  {Xiridou}}, \bibinfo {author} {\bibfnamefont {F.}~\bibnamefont {Miura}},
  \bibinfo {author} {\bibfnamefont {P.}~\bibnamefont {Adam}}, \bibinfo {author}
  {\bibfnamefont {E.}~\bibnamefont {Op~de Coul}}, \bibinfo {author}
  {\bibfnamefont {J.}~\bibnamefont {de~Wit}},\ and\ \bibinfo {author}
  {\bibfnamefont {J.}~\bibnamefont {Wallinga}},\ }\bibfield  {title} {\bibinfo
  {title} {The fading of the mpox outbreak among men who have sex with men: A
  mathematical modelling study},\ }\href
  {https://doi.org/10.1093/infdis/jiad414} {\bibfield  {journal} {\bibinfo
  {journal} {The Journal of Infectious Diseases}\ }\textbf {\bibinfo {volume}
  {230}},\ \bibinfo {pages} {e121} (\bibinfo {year} {2024})}\BibitemShut
  {NoStop}%
\bibitem [{\citenamefont {Diekmann}\ \emph {et~al.}(1990)\citenamefont
  {Diekmann}, \citenamefont {Heesterbeek},\ and\ \citenamefont
  {Metz}}]{diekmann1990definition}%
  \BibitemOpen
  \bibfield  {author} {\bibinfo {author} {\bibfnamefont {O.}~\bibnamefont
  {Diekmann}}, \bibinfo {author} {\bibfnamefont {J.}~\bibnamefont
  {Heesterbeek}},\ and\ \bibinfo {author} {\bibfnamefont {J.}~\bibnamefont
  {Metz}},\ }\bibfield  {title} {\bibinfo {title} {On the definition and the
  computation of the basic reproduction ratio {{$R_0$}} in models for
  infectious diseases in heterogeneous populations},\ }\href
  {https://doi.org/10.1007/BF00178324} {\bibfield  {journal} {\bibinfo
  {journal} {Journal of Mathematical Biology}\ }\textbf {\bibinfo {volume}
  {28}},\ \bibinfo {pages} {365} (\bibinfo {year} {1990})}\BibitemShut
  {NoStop}%
\bibitem [{\citenamefont {Sattenspiel}\ and\ \citenamefont
  {Dietz}(1995)}]{sattenspiel1995structured}%
  \BibitemOpen
  \bibfield  {author} {\bibinfo {author} {\bibfnamefont {L.}~\bibnamefont
  {Sattenspiel}}\ and\ \bibinfo {author} {\bibfnamefont {K.}~\bibnamefont
  {Dietz}},\ }\bibfield  {title} {\bibinfo {title} {A structured epidemic model
  incorporating geographic mobility among regions},\ }\href
  {https://doi.org/10.1016/0025-5564(94)00068-B} {\bibfield  {journal}
  {\bibinfo  {journal} {Mathematical Biosciences}\ }\textbf {\bibinfo {volume}
  {128}},\ \bibinfo {pages} {71} (\bibinfo {year} {1995})}\BibitemShut
  {NoStop}%
\bibitem [{\citenamefont {Lloyd}\ and\ \citenamefont
  {May}(1996)}]{lloyd1996spatial}%
  \BibitemOpen
  \bibfield  {author} {\bibinfo {author} {\bibfnamefont {A.~L.}\ \bibnamefont
  {Lloyd}}\ and\ \bibinfo {author} {\bibfnamefont {R.~M.}\ \bibnamefont
  {May}},\ }\bibfield  {title} {\bibinfo {title} {Spatial heterogeneity in
  epidemic models},\ }\href {https://doi.org/10.1006/jtbi.1996.0042} {\bibfield
   {journal} {\bibinfo  {journal} {Journal of Theoretical Biology}\ }\textbf
  {\bibinfo {volume} {179}},\ \bibinfo {pages} {1} (\bibinfo {year}
  {1996})}\BibitemShut {NoStop}%
\bibitem [{\citenamefont {Ball}\ \emph {et~al.}(1997)\citenamefont {Ball},
  \citenamefont {Mollison},\ and\ \citenamefont
  {Scalia-Tomba}}]{ball1997epidemics}%
  \BibitemOpen
  \bibfield  {author} {\bibinfo {author} {\bibfnamefont {F.}~\bibnamefont
  {Ball}}, \bibinfo {author} {\bibfnamefont {D.}~\bibnamefont {Mollison}},\
  and\ \bibinfo {author} {\bibfnamefont {G.}~\bibnamefont {Scalia-Tomba}},\
  }\bibfield  {title} {\bibinfo {title} {{Epidemics with two levels of
  mixing}},\ }\href {https://doi.org/10.1214/aoap/1034625252} {\bibfield
  {journal} {\bibinfo  {journal} {The Annals of Applied Probability}\ }\textbf
  {\bibinfo {volume} {7}},\ \bibinfo {pages} {46} (\bibinfo {year}
  {1997})}\BibitemShut {NoStop}%
\bibitem [{\citenamefont {Riley}(2007)}]{riley2007largescale}%
  \BibitemOpen
  \bibfield  {author} {\bibinfo {author} {\bibfnamefont {S.}~\bibnamefont
  {Riley}},\ }\bibfield  {title} {\bibinfo {title} {Large-scale
  spatial-transmission models of infectious disease},\ }\href
  {https://doi.org/10.1126/science.1134695} {\bibfield  {journal} {\bibinfo
  {journal} {Science}\ }\textbf {\bibinfo {volume} {316}},\ \bibinfo {pages}
  {1298} (\bibinfo {year} {2007})}\BibitemShut {NoStop}%
\bibitem [{\citenamefont {Colizza}\ and\ \citenamefont
  {Vespignani}(2008)}]{colizza2008epidemic}%
  \BibitemOpen
  \bibfield  {author} {\bibinfo {author} {\bibfnamefont {V.}~\bibnamefont
  {Colizza}}\ and\ \bibinfo {author} {\bibfnamefont {A.}~\bibnamefont
  {Vespignani}},\ }\bibfield  {title} {\bibinfo {title} {Epidemic modeling in
  metapopulation systems with heterogeneous coupling pattern: Theory and
  simulations},\ }\href {https://doi.org/10.1016/j.jtbi.2007.11.028} {\bibfield
   {journal} {\bibinfo  {journal} {Journal of Theoretical Biology}\ }\textbf
  {\bibinfo {volume} {251}},\ \bibinfo {pages} {450} (\bibinfo {year}
  {2008})}\BibitemShut {NoStop}%
\bibitem [{\citenamefont {House}\ and\ \citenamefont
  {Keeling}(2008)}]{house2008deterministic}%
  \BibitemOpen
  \bibfield  {author} {\bibinfo {author} {\bibfnamefont {T.}~\bibnamefont
  {House}}\ and\ \bibinfo {author} {\bibfnamefont {M.~J.}\ \bibnamefont
  {Keeling}},\ }\bibfield  {title} {\bibinfo {title} {Deterministic epidemic
  models with explicit household structure},\ }\href
  {https://doi.org/10.1016/j.mbs.2008.01.011} {\bibfield  {journal} {\bibinfo
  {journal} {Mathematical Biosciences}\ }\textbf {\bibinfo {volume} {213}},\
  \bibinfo {pages} {29} (\bibinfo {year} {2008})}\BibitemShut {NoStop}%
\bibitem [{\citenamefont {Ajelli}\ \emph {et~al.}(2010)\citenamefont {Ajelli},
  \citenamefont {Gon{\c c}alves}, \citenamefont {Balcan}, \citenamefont
  {Colizza}, \citenamefont {Hu}, \citenamefont {Ramasco}, \citenamefont
  {Merler},\ and\ \citenamefont {Vespignani}}]{ajelli2010comparing}%
  \BibitemOpen
  \bibfield  {author} {\bibinfo {author} {\bibfnamefont {M.}~\bibnamefont
  {Ajelli}}, \bibinfo {author} {\bibfnamefont {B.}~\bibnamefont {Gon{\c
  c}alves}}, \bibinfo {author} {\bibfnamefont {D.}~\bibnamefont {Balcan}},
  \bibinfo {author} {\bibfnamefont {V.}~\bibnamefont {Colizza}}, \bibinfo
  {author} {\bibfnamefont {H.}~\bibnamefont {Hu}}, \bibinfo {author}
  {\bibfnamefont {J.~J.}\ \bibnamefont {Ramasco}}, \bibinfo {author}
  {\bibfnamefont {S.}~\bibnamefont {Merler}},\ and\ \bibinfo {author}
  {\bibfnamefont {A.}~\bibnamefont {Vespignani}},\ }\bibfield  {title}
  {\bibinfo {title} {Comparing large-scale computational approaches to epidemic
  modeling: Agent-based versus structured metapopulation models},\ }\href
  {https://doi.org/10.1186/1471-2334-10-190} {\bibfield  {journal} {\bibinfo
  {journal} {BMC Infectious Diseases}\ }\textbf {\bibinfo {volume} {10}},\
  \bibinfo {pages} {190} (\bibinfo {year} {2010})}\BibitemShut {NoStop}%
\bibitem [{\citenamefont {Belik}\ \emph {et~al.}(2011)\citenamefont {Belik},
  \citenamefont {Geisel},\ and\ \citenamefont {Brockmann}}]{belik2011natural}%
  \BibitemOpen
  \bibfield  {author} {\bibinfo {author} {\bibfnamefont {V.}~\bibnamefont
  {Belik}}, \bibinfo {author} {\bibfnamefont {T.}~\bibnamefont {Geisel}},\ and\
  \bibinfo {author} {\bibfnamefont {D.}~\bibnamefont {Brockmann}},\ }\bibfield
  {title} {\bibinfo {title} {Natural human mobility patterns and spatial spread
  of infectious diseases},\ }\href {https://doi.org/10.1103/PhysRevX.1.011001}
  {\bibfield  {journal} {\bibinfo  {journal} {Physical Review X}\ }\textbf
  {\bibinfo {volume} {1}},\ \bibinfo {pages} {011001} (\bibinfo {year}
  {2011})}\BibitemShut {NoStop}%
\bibitem [{\citenamefont {Zachreson}\ \emph {et~al.}(2022)\citenamefont
  {Zachreson}, \citenamefont {Chang}, \citenamefont {Harding},\ and\
  \citenamefont {Prokopenko}}]{zachreson2022effects}%
  \BibitemOpen
  \bibfield  {author} {\bibinfo {author} {\bibfnamefont {C.}~\bibnamefont
  {Zachreson}}, \bibinfo {author} {\bibfnamefont {S.}~\bibnamefont {Chang}},
  \bibinfo {author} {\bibfnamefont {N.}~\bibnamefont {Harding}},\ and\ \bibinfo
  {author} {\bibfnamefont {M.}~\bibnamefont {Prokopenko}},\ }\bibfield  {title}
  {\bibinfo {title} {The effects of local homogeneity assumptions in
  metapopulation models of infectious disease},\ }\href
  {https://doi.org/10.1098/rsos.211919} {\bibfield  {journal} {\bibinfo
  {journal} {Royal Society Open Science}\ }\textbf {\bibinfo {volume} {9}},\
  \bibinfo {pages} {211919} (\bibinfo {year} {2022})}\BibitemShut {NoStop}%
\bibitem [{\citenamefont {Keeling}(1999)}]{keeling1999effects}%
  \BibitemOpen
  \bibfield  {author} {\bibinfo {author} {\bibfnamefont {M.~J.}\ \bibnamefont
  {Keeling}},\ }\bibfield  {title} {\bibinfo {title} {The effects of local
  spatial structure on epidemiological invasions},\ }\href
  {https://doi.org/10.1098/rspb.1999.0716} {\bibfield  {journal} {\bibinfo
  {journal} {Proceedings of the Royal Society of London. Series B: Biological
  Sciences}\ }\textbf {\bibinfo {volume} {266}},\ \bibinfo {pages} {859}
  (\bibinfo {year} {1999})}\BibitemShut {NoStop}%
\bibitem [{\citenamefont {Keeling}\ and\ \citenamefont
  {Eames}(2005)}]{keeling2005networks}%
  \BibitemOpen
  \bibfield  {author} {\bibinfo {author} {\bibfnamefont {M.~J.}\ \bibnamefont
  {Keeling}}\ and\ \bibinfo {author} {\bibfnamefont {K.~T.}\ \bibnamefont
  {Eames}},\ }\bibfield  {title} {\bibinfo {title} {Networks and epidemic
  models},\ }\href {https://doi.org/10.1098/rsif.2005.0051} {\bibfield
  {journal} {\bibinfo  {journal} {Journal of The Royal Society Interface}\
  }\textbf {\bibinfo {volume} {2}},\ \bibinfo {pages} {295} (\bibinfo {year}
  {2005})}\BibitemShut {NoStop}%
\bibitem [{\citenamefont {{Pastor-Satorras}}\ \emph {et~al.}(2015)\citenamefont
  {{Pastor-Satorras}}, \citenamefont {Castellano}, \citenamefont
  {Van~Mieghem},\ and\ \citenamefont
  {Vespignani}}]{pastor-satorras2015epidemic}%
  \BibitemOpen
  \bibfield  {author} {\bibinfo {author} {\bibfnamefont {R.}~\bibnamefont
  {{Pastor-Satorras}}}, \bibinfo {author} {\bibfnamefont {C.}~\bibnamefont
  {Castellano}}, \bibinfo {author} {\bibfnamefont {P.}~\bibnamefont
  {Van~Mieghem}},\ and\ \bibinfo {author} {\bibfnamefont {A.}~\bibnamefont
  {Vespignani}},\ }\bibfield  {title} {\bibinfo {title} {Epidemic processes in
  complex networks},\ }\href {https://doi.org/10.1103/RevModPhys.87.925}
  {\bibfield  {journal} {\bibinfo  {journal} {Reviews of Modern Physics}\
  }\textbf {\bibinfo {volume} {87}},\ \bibinfo {pages} {925} (\bibinfo {year}
  {2015})}\BibitemShut {NoStop}%
\bibitem [{\citenamefont {Newman}(2005)}]{newman2005threshold}%
  \BibitemOpen
  \bibfield  {author} {\bibinfo {author} {\bibfnamefont {M.~E.~J.}\
  \bibnamefont {Newman}},\ }\bibfield  {title} {\bibinfo {title} {Threshold
  effects for two pathogens spreading on a network},\ }\href
  {https://doi.org/10.1103/PhysRevLett.95.108701} {\bibfield  {journal}
  {\bibinfo  {journal} {Physical Review Letters}\ }\textbf {\bibinfo {volume}
  {95}},\ \bibinfo {pages} {108701} (\bibinfo {year} {2005})}\BibitemShut
  {NoStop}%
\bibitem [{\citenamefont {Ferrari}\ \emph {et~al.}(2006)\citenamefont
  {Ferrari}, \citenamefont {Bansal}, \citenamefont {Meyers},\ and\
  \citenamefont {Bj{\o}rnstad}}]{ferrari2006network}%
  \BibitemOpen
  \bibfield  {author} {\bibinfo {author} {\bibfnamefont {M.~J.}\ \bibnamefont
  {Ferrari}}, \bibinfo {author} {\bibfnamefont {S.}~\bibnamefont {Bansal}},
  \bibinfo {author} {\bibfnamefont {L.~A.}\ \bibnamefont {Meyers}},\ and\
  \bibinfo {author} {\bibfnamefont {O.~N.}\ \bibnamefont {Bj{\o}rnstad}},\
  }\bibfield  {title} {\bibinfo {title} {Network frailty and the geometry of
  herd immunity},\ }\href {https://doi.org/10.1098/rspb.2006.3636} {\bibfield
  {journal} {\bibinfo  {journal} {Proceedings of the Royal Society B:
  Biological Sciences}\ }\textbf {\bibinfo {volume} {273}},\ \bibinfo {pages}
  {2743} (\bibinfo {year} {2006})}\BibitemShut {NoStop}%
\bibitem [{\citenamefont {Bansal}\ and\ \citenamefont
  {Meyers}(2012)}]{bansal2012impact}%
  \BibitemOpen
  \bibfield  {author} {\bibinfo {author} {\bibfnamefont {S.}~\bibnamefont
  {Bansal}}\ and\ \bibinfo {author} {\bibfnamefont {L.~A.}\ \bibnamefont
  {Meyers}},\ }\bibfield  {title} {\bibinfo {title} {The impact of past
  epidemics on future disease dynamics},\ }\href
  {https://doi.org/10.1016/j.jtbi.2012.06.012} {\bibfield  {journal} {\bibinfo
  {journal} {Journal of Theoretical Biology}\ }\textbf {\bibinfo {volume}
  {309}},\ \bibinfo {pages} {176} (\bibinfo {year} {2012})}\BibitemShut
  {NoStop}%
\bibitem [{\citenamefont {Mann}\ \emph
  {et~al.}(2021{\natexlab{a}})\citenamefont {Mann}, \citenamefont {Smith},
  \citenamefont {Mitchell},\ and\ \citenamefont
  {Dobson}}]{mann2021twopathogen}%
  \BibitemOpen
  \bibfield  {author} {\bibinfo {author} {\bibfnamefont {P.}~\bibnamefont
  {Mann}}, \bibinfo {author} {\bibfnamefont {V.~A.}\ \bibnamefont {Smith}},
  \bibinfo {author} {\bibfnamefont {J.~B.~O.}\ \bibnamefont {Mitchell}},\ and\
  \bibinfo {author} {\bibfnamefont {S.}~\bibnamefont {Dobson}},\ }\bibfield
  {title} {\bibinfo {title} {Two-pathogen model with competition on clustered
  networks},\ }\href {https://doi.org/10.1103/PhysRevE.103.062308} {\bibfield
  {journal} {\bibinfo  {journal} {Physical Review E}\ }\textbf {\bibinfo
  {volume} {103}},\ \bibinfo {pages} {062308} (\bibinfo {year}
  {2021}{\natexlab{a}})}\BibitemShut {NoStop}%
\bibitem [{\citenamefont {Mann}\ \emph
  {et~al.}(2021{\natexlab{b}})\citenamefont {Mann}, \citenamefont {Smith},
  \citenamefont {Mitchell},\ and\ \citenamefont {Dobson}}]{mann2021symbiotic}%
  \BibitemOpen
  \bibfield  {author} {\bibinfo {author} {\bibfnamefont {P.}~\bibnamefont
  {Mann}}, \bibinfo {author} {\bibfnamefont {V.~A.}\ \bibnamefont {Smith}},
  \bibinfo {author} {\bibfnamefont {J.~B.~O.}\ \bibnamefont {Mitchell}},\ and\
  \bibinfo {author} {\bibfnamefont {S.}~\bibnamefont {Dobson}},\ }\bibfield
  {title} {\bibinfo {title} {Symbiotic and antagonistic disease dynamics on
  networks using bond percolation},\ }\href
  {https://doi.org/10.1103/PhysRevE.104.024303} {\bibfield  {journal} {\bibinfo
   {journal} {Physical Review E}\ }\textbf {\bibinfo {volume} {104}},\ \bibinfo
  {pages} {024303} (\bibinfo {year} {2021}{\natexlab{b}})}\BibitemShut
  {NoStop}%
\bibitem [{\citenamefont {Di~Lauro}\ \emph {et~al.}(2021)\citenamefont
  {Di~Lauro}, \citenamefont {Berthouze}, \citenamefont {Dorey}, \citenamefont
  {Miller},\ and\ \citenamefont {Kiss}}]{dilauro2021impact}%
  \BibitemOpen
  \bibfield  {author} {\bibinfo {author} {\bibfnamefont {F.}~\bibnamefont
  {Di~Lauro}}, \bibinfo {author} {\bibfnamefont {L.}~\bibnamefont {Berthouze}},
  \bibinfo {author} {\bibfnamefont {M.~D.}\ \bibnamefont {Dorey}}, \bibinfo
  {author} {\bibfnamefont {J.~C.}\ \bibnamefont {Miller}},\ and\ \bibinfo
  {author} {\bibfnamefont {I.~Z.}\ \bibnamefont {Kiss}},\ }\bibfield  {title}
  {\bibinfo {title} {The impact of contact structure and mixing on control
  measures and disease-induced herd immunity in epidemic models: A mean-field
  model perspective},\ }\href {https://doi.org/10.1007/s11538-021-00947-8}
  {\bibfield  {journal} {\bibinfo  {journal} {Bulletin of Mathematical
  Biology}\ }\textbf {\bibinfo {volume} {83}},\ \bibinfo {pages} {117}
  (\bibinfo {year} {2021})}\BibitemShut {NoStop}%
\bibitem [{\citenamefont {Mann}\ \emph {et~al.}(2022)\citenamefont {Mann},
  \citenamefont {Smith}, \citenamefont {Mitchell},\ and\ \citenamefont
  {Dobson}}]{mann2022strain}%
  \BibitemOpen
  \bibfield  {author} {\bibinfo {author} {\bibfnamefont {P.}~\bibnamefont
  {Mann}}, \bibinfo {author} {\bibfnamefont {V.~A.}\ \bibnamefont {Smith}},
  \bibinfo {author} {\bibfnamefont {J.~B.~O.}\ \bibnamefont {Mitchell}},\ and\
  \bibinfo {author} {\bibfnamefont {S.}~\bibnamefont {Dobson}},\ }\bibfield
  {title} {\bibinfo {title} {$n$-strain epidemic model using bond
  percolation},\ }\href {https://doi.org/10.1103/PhysRevE.106.014304}
  {\bibfield  {journal} {\bibinfo  {journal} {Physical Review E}\ }\textbf
  {\bibinfo {volume} {106}},\ \bibinfo {pages} {014304} (\bibinfo {year}
  {2022})}\BibitemShut {NoStop}%
\bibitem [{\citenamefont {Burgio}\ \emph {et~al.}(2021)\citenamefont {Burgio},
  \citenamefont {Steinegger}, \citenamefont {Rapisardi},\ and\ \citenamefont
  {Arenas}}]{burgio2021homophily}%
  \BibitemOpen
  \bibfield  {author} {\bibinfo {author} {\bibfnamefont {G.}~\bibnamefont
  {Burgio}}, \bibinfo {author} {\bibfnamefont {B.}~\bibnamefont {Steinegger}},
  \bibinfo {author} {\bibfnamefont {G.}~\bibnamefont {Rapisardi}},\ and\
  \bibinfo {author} {\bibfnamefont {A.}~\bibnamefont {Arenas}},\ }\bibfield
  {title} {\bibinfo {title} {Homophily in the adoption of digital proximity
  tracing apps shapes the evolution of epidemics},\ }\href
  {https://doi.org/10.1103/PhysRevResearch.3.033128} {\bibfield  {journal}
  {\bibinfo  {journal} {Physical Review Research}\ }\textbf {\bibinfo {volume}
  {3}},\ \bibinfo {pages} {033128} (\bibinfo {year} {2021})}\BibitemShut
  {NoStop}%
\bibitem [{\citenamefont {Rizi}\ \emph {et~al.}(2022)\citenamefont {Rizi},
  \citenamefont {Faqeeh}, \citenamefont {{Badie-Modiri}},\ and\ \citenamefont
  {Kivel{\"a}}}]{rizi2022epidemic}%
  \BibitemOpen
  \bibfield  {author} {\bibinfo {author} {\bibfnamefont {A.~K.}\ \bibnamefont
  {Rizi}}, \bibinfo {author} {\bibfnamefont {A.}~\bibnamefont {Faqeeh}},
  \bibinfo {author} {\bibfnamefont {A.}~\bibnamefont {{Badie-Modiri}}},\ and\
  \bibinfo {author} {\bibfnamefont {M.}~\bibnamefont {Kivel{\"a}}},\ }\bibfield
   {title} {\bibinfo {title} {Epidemic spreading and digital contact tracing:
  Effects of heterogeneous mixing and quarantine failures},\ }\href
  {https://doi.org/10.1103/PhysRevE.105.044313} {\bibfield  {journal} {\bibinfo
   {journal} {Physical Review E}\ }\textbf {\bibinfo {volume} {105}},\ \bibinfo
  {pages} {044313} (\bibinfo {year} {2022})}\BibitemShut {NoStop}%
\bibitem [{\citenamefont {Burgio}\ \emph {et~al.}(2022)\citenamefont {Burgio},
  \citenamefont {Steinegger},\ and\ \citenamefont
  {Arenas}}]{burgio2022homophily}%
  \BibitemOpen
  \bibfield  {author} {\bibinfo {author} {\bibfnamefont {G.}~\bibnamefont
  {Burgio}}, \bibinfo {author} {\bibfnamefont {B.}~\bibnamefont {Steinegger}},\
  and\ \bibinfo {author} {\bibfnamefont {A.}~\bibnamefont {Arenas}},\
  }\bibfield  {title} {\bibinfo {title} {Homophily impacts the success of
  vaccine roll-outs},\ }\href {https://doi.org/10.1038/s42005-022-00849-8}
  {\bibfield  {journal} {\bibinfo  {journal} {Communications Physics}\ }\textbf
  {\bibinfo {volume} {5}},\ \bibinfo {pages} {70} (\bibinfo {year}
  {2022})}\BibitemShut {NoStop}%
\bibitem [{\citenamefont {Hiraoka}\ \emph {et~al.}(2022)\citenamefont
  {Hiraoka}, \citenamefont {Rizi}, \citenamefont {Kivel{\"a}},\ and\
  \citenamefont {Saram{\"a}ki}}]{hiraoka2022herd}%
  \BibitemOpen
  \bibfield  {author} {\bibinfo {author} {\bibfnamefont {T.}~\bibnamefont
  {Hiraoka}}, \bibinfo {author} {\bibfnamefont {A.~K.}\ \bibnamefont {Rizi}},
  \bibinfo {author} {\bibfnamefont {M.}~\bibnamefont {Kivel{\"a}}},\ and\
  \bibinfo {author} {\bibfnamefont {J.}~\bibnamefont {Saram{\"a}ki}},\
  }\bibfield  {title} {\bibinfo {title} {Herd immunity and epidemic size in
  networks with vaccination homophily},\ }\href
  {https://doi.org/10.1103/PhysRevE.105.L052301} {\bibfield  {journal}
  {\bibinfo  {journal} {Physical Review E}\ }\textbf {\bibinfo {volume}
  {105}},\ \bibinfo {pages} {L052301} (\bibinfo {year} {2022})}\BibitemShut
  {NoStop}%
\bibitem [{\citenamefont {Watanabe}\ and\ \citenamefont
  {Hasegawa}(2022)}]{watanabe2022impact}%
  \BibitemOpen
  \bibfield  {author} {\bibinfo {author} {\bibfnamefont {H.}~\bibnamefont
  {Watanabe}}\ and\ \bibinfo {author} {\bibfnamefont {T.}~\bibnamefont
  {Hasegawa}},\ }\bibfield  {title} {\bibinfo {title} {Impact of assortative
  mixing by mask-wearing on the propagation of epidemics in networks},\ }\href
  {https://doi.org/10.1016/j.physa.2022.127760} {\bibfield  {journal} {\bibinfo
   {journal} {Physica A: Statistical Mechanics and its Applications}\ }\textbf
  {\bibinfo {volume} {603}},\ \bibinfo {pages} {127760} (\bibinfo {year}
  {2022})}\BibitemShut {NoStop}%
\bibitem [{\citenamefont {K.~Rizi}\ \emph {et~al.}(2024)\citenamefont
  {K.~Rizi}, \citenamefont {Michielan}, \citenamefont {Stegehuis},\ and\
  \citenamefont {Kivel{\"a}}}]{rizi2024homophily}%
  \BibitemOpen
  \bibfield  {author} {\bibinfo {author} {\bibfnamefont {A.}~\bibnamefont
  {K.~Rizi}}, \bibinfo {author} {\bibfnamefont {R.}~\bibnamefont {Michielan}},
  \bibinfo {author} {\bibfnamefont {C.}~\bibnamefont {Stegehuis}},\ and\
  \bibinfo {author} {\bibfnamefont {M.}~\bibnamefont {Kivel{\"a}}},\ }\bibfield
   {title} {\bibinfo {title} {Homophily within and across groups},\ }\href@noop
  {} {\bibfield  {journal} {\bibinfo  {journal} {arXiv:2412.07901}\ } (\bibinfo
  {year} {2024})}\BibitemShut {NoStop}%
\bibitem [{\citenamefont {Bansal}\ \emph {et~al.}(2010)\citenamefont {Bansal},
  \citenamefont {Pourbohloul}, \citenamefont {Hupert}, \citenamefont
  {Grenfell},\ and\ \citenamefont {Meyers}}]{bansal2010shifting}%
  \BibitemOpen
  \bibfield  {author} {\bibinfo {author} {\bibfnamefont {S.}~\bibnamefont
  {Bansal}}, \bibinfo {author} {\bibfnamefont {B.}~\bibnamefont {Pourbohloul}},
  \bibinfo {author} {\bibfnamefont {N.}~\bibnamefont {Hupert}}, \bibinfo
  {author} {\bibfnamefont {B.}~\bibnamefont {Grenfell}},\ and\ \bibinfo
  {author} {\bibfnamefont {L.~A.}\ \bibnamefont {Meyers}},\ }\bibfield  {title}
  {\bibinfo {title} {The shifting demographic landscape of pandemic
  influenza},\ }\href {https://doi.org/10.1371/journal.pone.0009360} {\bibfield
   {journal} {\bibinfo  {journal} {PLoS ONE}\ }\textbf {\bibinfo {volume}
  {5}},\ \bibinfo {pages} {e9360} (\bibinfo {year} {2010})}\BibitemShut
  {NoStop}%
\bibitem [{\citenamefont {Funk}\ and\ \citenamefont
  {Jansen}(2010)}]{funk2010interacting}%
  \BibitemOpen
  \bibfield  {author} {\bibinfo {author} {\bibfnamefont {S.}~\bibnamefont
  {Funk}}\ and\ \bibinfo {author} {\bibfnamefont {V.~A.~A.}\ \bibnamefont
  {Jansen}},\ }\bibfield  {title} {\bibinfo {title} {Interacting epidemics on
  overlay networks},\ }\href {https://doi.org/10.1103/PhysRevE.81.036118}
  {\bibfield  {journal} {\bibinfo  {journal} {Physical Review E}\ }\textbf
  {\bibinfo {volume} {81}},\ \bibinfo {pages} {036118} (\bibinfo {year}
  {2010})}\BibitemShut {NoStop}%
\bibitem [{\citenamefont {Karrer}\ and\ \citenamefont
  {Newman}(2011)}]{karrer2011competing}%
  \BibitemOpen
  \bibfield  {author} {\bibinfo {author} {\bibfnamefont {B.}~\bibnamefont
  {Karrer}}\ and\ \bibinfo {author} {\bibfnamefont {M.~E.~J.}\ \bibnamefont
  {Newman}},\ }\bibfield  {title} {\bibinfo {title} {Competing epidemics on
  complex networks},\ }\href {https://doi.org/10.1103/PhysRevE.84.036106}
  {\bibfield  {journal} {\bibinfo  {journal} {Physical Review E}\ }\textbf
  {\bibinfo {volume} {84}},\ \bibinfo {pages} {036106} (\bibinfo {year}
  {2011})}\BibitemShut {NoStop}%
\bibitem [{\citenamefont {Hasegawa}\ and\ \citenamefont
  {Masuda}(2011)}]{hasegawa2011robustness}%
  \BibitemOpen
  \bibfield  {author} {\bibinfo {author} {\bibfnamefont {T.}~\bibnamefont
  {Hasegawa}}\ and\ \bibinfo {author} {\bibfnamefont {N.}~\bibnamefont
  {Masuda}},\ }\bibfield  {title} {\bibinfo {title} {Robustness of networks
  against propagating attacks under vaccination strategies},\ }\href
  {https://doi.org/10.1088/1742-5468/2011/09/P09014} {\bibfield  {journal}
  {\bibinfo  {journal} {Journal of Statistical Mechanics: Theory and
  Experiment}\ }\textbf {\bibinfo {volume} {2011}},\ \bibinfo {pages} {P09014}
  (\bibinfo {year} {2011})}\BibitemShut {NoStop}%
\bibitem [{\citenamefont {Hasegawa}\ \emph {et~al.}(2012)\citenamefont
  {Hasegawa}, \citenamefont {Konno},\ and\ \citenamefont
  {Nemoto}}]{hasegawa2012robustness}%
  \BibitemOpen
  \bibfield  {author} {\bibinfo {author} {\bibfnamefont {T.}~\bibnamefont
  {Hasegawa}}, \bibinfo {author} {\bibfnamefont {K.}~\bibnamefont {Konno}},\
  and\ \bibinfo {author} {\bibfnamefont {K.}~\bibnamefont {Nemoto}},\
  }\bibfield  {title} {\bibinfo {title} {Robustness of correlated networks
  against propagating attacks},\ }\href
  {https://doi.org/10.1140/epjb/e2012-30290-0} {\bibfield  {journal} {\bibinfo
  {journal} {The European Physical Journal B}\ }\textbf {\bibinfo {volume}
  {85}},\ \bibinfo {pages} {262} (\bibinfo {year} {2012})}\BibitemShut
  {NoStop}%
\bibitem [{\citenamefont {Miller}(2013)}]{miller2013cocirculation}%
  \BibitemOpen
  \bibfield  {author} {\bibinfo {author} {\bibfnamefont {J.~C.}\ \bibnamefont
  {Miller}},\ }\bibfield  {title} {\bibinfo {title} {Cocirculation of
  infectious diseases on networks},\ }\href
  {https://doi.org/10.1103/PhysRevE.87.060801} {\bibfield  {journal} {\bibinfo
  {journal} {Physical Review E}\ }\textbf {\bibinfo {volume} {87}},\ \bibinfo
  {pages} {060801} (\bibinfo {year} {2013})}\BibitemShut {NoStop}%
\bibitem [{\citenamefont {{Lloyd-Smith}}\ \emph {et~al.}(2005)\citenamefont
  {{Lloyd-Smith}}, \citenamefont {Schreiber}, \citenamefont {Kopp},\ and\
  \citenamefont {Getz}}]{lloyd-smith2005superspreading}%
  \BibitemOpen
  \bibfield  {author} {\bibinfo {author} {\bibfnamefont {J.~O.}\ \bibnamefont
  {{Lloyd-Smith}}}, \bibinfo {author} {\bibfnamefont {S.~J.}\ \bibnamefont
  {Schreiber}}, \bibinfo {author} {\bibfnamefont {P.~E.}\ \bibnamefont
  {Kopp}},\ and\ \bibinfo {author} {\bibfnamefont {W.~M.}\ \bibnamefont
  {Getz}},\ }\bibfield  {title} {\bibinfo {title} {Superspreading and the
  effect of individual variation on disease emergence},\ }\href
  {https://doi.org/10.1038/nature04153} {\bibfield  {journal} {\bibinfo
  {journal} {Nature}\ }\textbf {\bibinfo {volume} {438}},\ \bibinfo {pages}
  {355} (\bibinfo {year} {2005})}\BibitemShut {NoStop}%
\bibitem [{\citenamefont {Mossong}\ \emph {et~al.}(2008)\citenamefont
  {Mossong}, \citenamefont {Hens}, \citenamefont {Jit}, \citenamefont
  {Beutels}, \citenamefont {Auranen}, \citenamefont {Mikolajczyk},
  \citenamefont {Massari}, \citenamefont {Salmaso}, \citenamefont {Tomba},
  \citenamefont {Wallinga}, \citenamefont {Heijne}, \citenamefont
  {{Sadkowska-Todys}}, \citenamefont {Rosinska},\ and\ \citenamefont
  {Edmunds}}]{mossong2008social}%
  \BibitemOpen
  \bibfield  {author} {\bibinfo {author} {\bibfnamefont {J.}~\bibnamefont
  {Mossong}}, \bibinfo {author} {\bibfnamefont {N.}~\bibnamefont {Hens}},
  \bibinfo {author} {\bibfnamefont {M.}~\bibnamefont {Jit}}, \bibinfo {author}
  {\bibfnamefont {P.}~\bibnamefont {Beutels}}, \bibinfo {author} {\bibfnamefont
  {K.}~\bibnamefont {Auranen}}, \bibinfo {author} {\bibfnamefont
  {R.}~\bibnamefont {Mikolajczyk}}, \bibinfo {author} {\bibfnamefont
  {M.}~\bibnamefont {Massari}}, \bibinfo {author} {\bibfnamefont
  {S.}~\bibnamefont {Salmaso}}, \bibinfo {author} {\bibfnamefont {G.~S.}\
  \bibnamefont {Tomba}}, \bibinfo {author} {\bibfnamefont {J.}~\bibnamefont
  {Wallinga}}, \bibinfo {author} {\bibfnamefont {J.}~\bibnamefont {Heijne}},
  \bibinfo {author} {\bibfnamefont {M.}~\bibnamefont {{Sadkowska-Todys}}},
  \bibinfo {author} {\bibfnamefont {M.}~\bibnamefont {Rosinska}},\ and\
  \bibinfo {author} {\bibfnamefont {W.~J.}\ \bibnamefont {Edmunds}},\
  }\bibfield  {title} {\bibinfo {title} {Social contacts and mixing patterns
  relevant to the spread of infectious diseases},\ }\href
  {https://doi.org/10.1371/journal.pmed.0050074} {\bibfield  {journal}
  {\bibinfo  {journal} {PLoS Medicine}\ }\textbf {\bibinfo {volume} {5}},\
  \bibinfo {pages} {e74} (\bibinfo {year} {2008})}\BibitemShut {NoStop}%
\bibitem [{\citenamefont {{H{\'e}bert-Dufresne}}\ \emph
  {et~al.}(2020)\citenamefont {{H{\'e}bert-Dufresne}}, \citenamefont
  {Althouse}, \citenamefont {Scarpino},\ and\ \citenamefont
  {Allard}}]{hebert-dufresne2020beyond}%
  \BibitemOpen
  \bibfield  {author} {\bibinfo {author} {\bibfnamefont {L.}~\bibnamefont
  {{H{\'e}bert-Dufresne}}}, \bibinfo {author} {\bibfnamefont {B.~M.}\
  \bibnamefont {Althouse}}, \bibinfo {author} {\bibfnamefont {S.~V.}\
  \bibnamefont {Scarpino}},\ and\ \bibinfo {author} {\bibfnamefont
  {A.}~\bibnamefont {Allard}},\ }\bibfield  {title} {\bibinfo {title} {Beyond
  {{$R_0$}} : heterogeneity in secondary infections and probabilistic epidemic
  forecasting},\ }\href {https://doi.org/10.1098/rsif.2020.0393} {\bibfield
  {journal} {\bibinfo  {journal} {Journal of The Royal Society Interface}\
  }\textbf {\bibinfo {volume} {17}},\ \bibinfo {pages} {20200393} (\bibinfo
  {year} {2020})}\BibitemShut {NoStop}%
\bibitem [{\citenamefont {Newman}(2018)}]{newman2018networks}%
  \BibitemOpen
  \bibfield  {author} {\bibinfo {author} {\bibfnamefont {M.}~\bibnamefont
  {Newman}},\ }\href {https://doi.org/10.1093/oso/9780198805090.001.0001}
  {\emph {\bibinfo {title} {Networks}}}\ (\bibinfo  {publisher} {Oxford
  University Press},\ \bibinfo {year} {2018})\BibitemShut {NoStop}%
\bibitem [{\citenamefont {Newman}(2002)}]{newman2002spread}%
  \BibitemOpen
  \bibfield  {author} {\bibinfo {author} {\bibfnamefont {M.~E.~J.}\
  \bibnamefont {Newman}},\ }\bibfield  {title} {\bibinfo {title} {Spread of
  epidemic disease on networks},\ }\href
  {https://doi.org/10.1103/PhysRevE.66.016128} {\bibfield  {journal} {\bibinfo
  {journal} {Physical Review E}\ }\textbf {\bibinfo {volume} {66}},\ \bibinfo
  {pages} {016128} (\bibinfo {year} {2002})}\BibitemShut {NoStop}%
\bibitem [{\citenamefont {Newman}(2023)}]{newman2023message}%
  \BibitemOpen
  \bibfield  {author} {\bibinfo {author} {\bibfnamefont {M.~E.~J.}\
  \bibnamefont {Newman}},\ }\bibfield  {title} {\bibinfo {title} {Message
  passing methods on complex networks},\ }\href
  {https://doi.org/10.1098/rspa.2022.0774} {\bibfield  {journal} {\bibinfo
  {journal} {Proceedings of the Royal Society A: Mathematical, Physical and
  Engineering Sciences}\ }\textbf {\bibinfo {volume} {479}},\ \bibinfo {pages}
  {20220774} (\bibinfo {year} {2023})}\BibitemShut {NoStop}%
\bibitem [{\citenamefont {{United Nations Statistics Division}}()}]{unstat}%
  \BibitemOpen
  \bibfield  {author} {\bibinfo {author} {\bibnamefont {{United Nations
  Statistics Division}}},\ }\href@noop {} {\bibinfo {title} {Population by age,
  sex and urban/rural residence}},\ \bibinfo {note} {retrieved from Demographic
  Statistics Database \url{https://unstats.un.org/unsd/demographic-social/};
  Data accessed on 2024-03-15}\BibitemShut {NoStop}%
\bibitem [{\citenamefont {Prem}\ \emph {et~al.}(2017)\citenamefont {Prem},
  \citenamefont {Cook},\ and\ \citenamefont {Jit}}]{prem2017projecting}%
  \BibitemOpen
  \bibfield  {author} {\bibinfo {author} {\bibfnamefont {K.}~\bibnamefont
  {Prem}}, \bibinfo {author} {\bibfnamefont {A.~R.}\ \bibnamefont {Cook}},\
  and\ \bibinfo {author} {\bibfnamefont {M.}~\bibnamefont {Jit}},\ }\bibfield
  {title} {\bibinfo {title} {Projecting social contact matrices in 152
  countries using contact surveys and demographic data},\ }\href
  {https://doi.org/10.1371/journal.pcbi.1005697} {\bibfield  {journal}
  {\bibinfo  {journal} {PLOS Computational Biology}\ }\textbf {\bibinfo
  {volume} {13}},\ \bibinfo {pages} {e1005697} (\bibinfo {year}
  {2017})}\BibitemShut {NoStop}%
\bibitem [{\citenamefont {Volz}\ and\ \citenamefont
  {Meyers}(2007)}]{volz2007susceptible}%
  \BibitemOpen
  \bibfield  {author} {\bibinfo {author} {\bibfnamefont {E.}~\bibnamefont
  {Volz}}\ and\ \bibinfo {author} {\bibfnamefont {L.~A.}\ \bibnamefont
  {Meyers}},\ }\bibfield  {title} {\bibinfo {title}
  {Susceptible–infected–recovered epidemics in dynamic contact networks},\
  }\href {https://doi.org/10.1098/rspb.2007.1159} {\bibfield  {journal}
  {\bibinfo  {journal} {Proceedings of the Royal Society B: Biological
  Sciences}\ }\textbf {\bibinfo {volume} {274}},\ \bibinfo {pages} {2925}
  (\bibinfo {year} {2007})}\BibitemShut {NoStop}%
\bibitem [{\citenamefont {Goltsev}\ \emph {et~al.}(2012)\citenamefont
  {Goltsev}, \citenamefont {Dorogovtsev}, \citenamefont {Oliveira},\ and\
  \citenamefont {Mendes}}]{goltsev2012localization}%
  \BibitemOpen
  \bibfield  {author} {\bibinfo {author} {\bibfnamefont {A.~V.}\ \bibnamefont
  {Goltsev}}, \bibinfo {author} {\bibfnamefont {S.~N.}\ \bibnamefont
  {Dorogovtsev}}, \bibinfo {author} {\bibfnamefont {J.~G.}\ \bibnamefont
  {Oliveira}},\ and\ \bibinfo {author} {\bibfnamefont {J.~F.~F.}\ \bibnamefont
  {Mendes}},\ }\bibfield  {title} {\bibinfo {title} {Localization and spreading
  of diseases in complex networks},\ }\href
  {https://doi.org/10.1103/PhysRevLett.109.128702} {\bibfield  {journal}
  {\bibinfo  {journal} {Physical Review Letters}\ }\textbf {\bibinfo {volume}
  {109}},\ \bibinfo {pages} {128702} (\bibinfo {year} {2012})}\BibitemShut
  {NoStop}%
\bibitem [{\citenamefont {{Pastor-Satorras}}\ and\ \citenamefont
  {Castellano}(2018)}]{pastor-satorras2018eigenvector}%
  \BibitemOpen
  \bibfield  {author} {\bibinfo {author} {\bibfnamefont {R.}~\bibnamefont
  {{Pastor-Satorras}}}\ and\ \bibinfo {author} {\bibfnamefont {C.}~\bibnamefont
  {Castellano}},\ }\bibfield  {title} {\bibinfo {title} {Eigenvector
  localization in real networks and its implications for epidemic spreading},\
  }\href {https://doi.org/10.1007/s10955-018-1970-8} {\bibfield  {journal}
  {\bibinfo  {journal} {Journal of Statistical Physics}\ }\textbf {\bibinfo
  {volume} {173}},\ \bibinfo {pages} {1110} (\bibinfo {year}
  {2018})}\BibitemShut {NoStop}%
\bibitem [{\citenamefont {{St-Onge}}\ \emph {et~al.}(2021)\citenamefont
  {{St-Onge}}, \citenamefont {Thibeault}, \citenamefont {Allard}, \citenamefont
  {Dub{\'e}},\ and\ \citenamefont {{H{\'e}bert-Dufresne}}}]{st-onge2021master}%
  \BibitemOpen
  \bibfield  {author} {\bibinfo {author} {\bibfnamefont {G.}~\bibnamefont
  {{St-Onge}}}, \bibinfo {author} {\bibfnamefont {V.}~\bibnamefont
  {Thibeault}}, \bibinfo {author} {\bibfnamefont {A.}~\bibnamefont {Allard}},
  \bibinfo {author} {\bibfnamefont {L.~J.}\ \bibnamefont {Dub{\'e}}},\ and\
  \bibinfo {author} {\bibfnamefont {L.}~\bibnamefont {{H{\'e}bert-Dufresne}}},\
  }\bibfield  {title} {\bibinfo {title} {Master equation analysis of mesoscopic
  localization in contagion dynamics on higher-order networks},\ }\href
  {https://doi.org/10.1103/PhysRevE.103.032301} {\bibfield  {journal} {\bibinfo
   {journal} {Physical Review E}\ }\textbf {\bibinfo {volume} {103}},\ \bibinfo
  {pages} {032301} (\bibinfo {year} {2021})}\BibitemShut {NoStop}%
\bibitem [{\citenamefont {{St-Onge}}\ \emph {et~al.}(2022)\citenamefont
  {{St-Onge}}, \citenamefont {Iacopini}, \citenamefont {Latora}, \citenamefont
  {Barrat}, \citenamefont {Petri}, \citenamefont {Allard},\ and\ \citenamefont
  {{H{\'e}bert-Dufresne}}}]{st-onge2022influential}%
  \BibitemOpen
  \bibfield  {author} {\bibinfo {author} {\bibfnamefont {G.}~\bibnamefont
  {{St-Onge}}}, \bibinfo {author} {\bibfnamefont {I.}~\bibnamefont {Iacopini}},
  \bibinfo {author} {\bibfnamefont {V.}~\bibnamefont {Latora}}, \bibinfo
  {author} {\bibfnamefont {A.}~\bibnamefont {Barrat}}, \bibinfo {author}
  {\bibfnamefont {G.}~\bibnamefont {Petri}}, \bibinfo {author} {\bibfnamefont
  {A.}~\bibnamefont {Allard}},\ and\ \bibinfo {author} {\bibfnamefont
  {L.}~\bibnamefont {{H{\'e}bert-Dufresne}}},\ }\bibfield  {title} {\bibinfo
  {title} {Influential groups for seeding and sustaining nonlinear contagion in
  heterogeneous hypergraphs},\ }\href
  {https://doi.org/10.1038/s42005-021-00788-w} {\bibfield  {journal} {\bibinfo
  {journal} {Communications Physics}\ }\textbf {\bibinfo {volume} {5}},\
  \bibinfo {pages} {25} (\bibinfo {year} {2022})}\BibitemShut {NoStop}%
\bibitem [{\citenamefont {Walker}\ \emph {et~al.}(2020)\citenamefont {Walker},
  \citenamefont {Whittaker}, \citenamefont {Watson}, \citenamefont {Baguelin},
  \citenamefont {Winskill}, \citenamefont {Hamlet}, \citenamefont {Djafaara},
  \citenamefont {Cucunub{\'a}}, \citenamefont {Olivera~Mesa}, \citenamefont
  {Green}, \citenamefont {Thompson}, \citenamefont {Nayagam}, \citenamefont
  {Ainslie}, \citenamefont {Bhatia}, \citenamefont {Bhatt}, \citenamefont
  {Boonyasiri}, \citenamefont {Boyd}, \citenamefont {Brazeau}, \citenamefont
  {Cattarino}, \citenamefont {{Cuomo-Dannenburg}}, \citenamefont {Dighe},
  \citenamefont {Donnelly}, \citenamefont {Dorigatti}, \citenamefont
  {Van~Elsland}, \citenamefont {FitzJohn}, \citenamefont {Fu}, \citenamefont
  {Gaythorpe}, \citenamefont {Geidelberg}, \citenamefont {Grassly},
  \citenamefont {Haw}, \citenamefont {Hayes}, \citenamefont {Hinsley},
  \citenamefont {Imai}, \citenamefont {Jorgensen}, \citenamefont {Knock},
  \citenamefont {Laydon}, \citenamefont {Mishra}, \citenamefont
  {{Nedjati-Gilani}}, \citenamefont {Okell}, \citenamefont {Unwin},
  \citenamefont {Verity}, \citenamefont {Vollmer}, \citenamefont {Walters},
  \citenamefont {Wang}, \citenamefont {Wang}, \citenamefont {Xi}, \citenamefont
  {Lalloo}, \citenamefont {Ferguson},\ and\ \citenamefont
  {Ghani}}]{walker2020impact}%
  \BibitemOpen
  \bibfield  {author} {\bibinfo {author} {\bibfnamefont {P.~G.~T.}\
  \bibnamefont {Walker}}, \bibinfo {author} {\bibfnamefont {C.}~\bibnamefont
  {Whittaker}}, \bibinfo {author} {\bibfnamefont {O.~J.}\ \bibnamefont
  {Watson}}, \bibinfo {author} {\bibfnamefont {M.}~\bibnamefont {Baguelin}},
  \bibinfo {author} {\bibfnamefont {P.}~\bibnamefont {Winskill}}, \bibinfo
  {author} {\bibfnamefont {A.}~\bibnamefont {Hamlet}}, \bibinfo {author}
  {\bibfnamefont {B.~A.}\ \bibnamefont {Djafaara}}, \bibinfo {author}
  {\bibfnamefont {Z.}~\bibnamefont {Cucunub{\'a}}}, \bibinfo {author}
  {\bibfnamefont {D.}~\bibnamefont {Olivera~Mesa}}, \bibinfo {author}
  {\bibfnamefont {W.}~\bibnamefont {Green}}, \bibinfo {author} {\bibfnamefont
  {H.}~\bibnamefont {Thompson}}, \bibinfo {author} {\bibfnamefont
  {S.}~\bibnamefont {Nayagam}}, \bibinfo {author} {\bibfnamefont {K.~E.~C.}\
  \bibnamefont {Ainslie}}, \bibinfo {author} {\bibfnamefont {S.}~\bibnamefont
  {Bhatia}}, \bibinfo {author} {\bibfnamefont {S.}~\bibnamefont {Bhatt}},
  \bibinfo {author} {\bibfnamefont {A.}~\bibnamefont {Boonyasiri}}, \bibinfo
  {author} {\bibfnamefont {O.}~\bibnamefont {Boyd}}, \bibinfo {author}
  {\bibfnamefont {N.~F.}\ \bibnamefont {Brazeau}}, \bibinfo {author}
  {\bibfnamefont {L.}~\bibnamefont {Cattarino}}, \bibinfo {author}
  {\bibfnamefont {G.}~\bibnamefont {{Cuomo-Dannenburg}}}, \bibinfo {author}
  {\bibfnamefont {A.}~\bibnamefont {Dighe}}, \bibinfo {author} {\bibfnamefont
  {C.~A.}\ \bibnamefont {Donnelly}}, \bibinfo {author} {\bibfnamefont
  {I.}~\bibnamefont {Dorigatti}}, \bibinfo {author} {\bibfnamefont {S.~L.}\
  \bibnamefont {Van~Elsland}}, \bibinfo {author} {\bibfnamefont
  {R.}~\bibnamefont {FitzJohn}}, \bibinfo {author} {\bibfnamefont
  {H.}~\bibnamefont {Fu}}, \bibinfo {author} {\bibfnamefont {K.~A.~M.}\
  \bibnamefont {Gaythorpe}}, \bibinfo {author} {\bibfnamefont {L.}~\bibnamefont
  {Geidelberg}}, \bibinfo {author} {\bibfnamefont {N.}~\bibnamefont {Grassly}},
  \bibinfo {author} {\bibfnamefont {D.}~\bibnamefont {Haw}}, \bibinfo {author}
  {\bibfnamefont {S.}~\bibnamefont {Hayes}}, \bibinfo {author} {\bibfnamefont
  {W.}~\bibnamefont {Hinsley}}, \bibinfo {author} {\bibfnamefont
  {N.}~\bibnamefont {Imai}}, \bibinfo {author} {\bibfnamefont {D.}~\bibnamefont
  {Jorgensen}}, \bibinfo {author} {\bibfnamefont {E.}~\bibnamefont {Knock}},
  \bibinfo {author} {\bibfnamefont {D.}~\bibnamefont {Laydon}}, \bibinfo
  {author} {\bibfnamefont {S.}~\bibnamefont {Mishra}}, \bibinfo {author}
  {\bibfnamefont {G.}~\bibnamefont {{Nedjati-Gilani}}}, \bibinfo {author}
  {\bibfnamefont {L.~C.}\ \bibnamefont {Okell}}, \bibinfo {author}
  {\bibfnamefont {H.~J.}\ \bibnamefont {Unwin}}, \bibinfo {author}
  {\bibfnamefont {R.}~\bibnamefont {Verity}}, \bibinfo {author} {\bibfnamefont
  {M.}~\bibnamefont {Vollmer}}, \bibinfo {author} {\bibfnamefont {C.~E.}\
  \bibnamefont {Walters}}, \bibinfo {author} {\bibfnamefont {H.}~\bibnamefont
  {Wang}}, \bibinfo {author} {\bibfnamefont {Y.}~\bibnamefont {Wang}}, \bibinfo
  {author} {\bibfnamefont {X.}~\bibnamefont {Xi}}, \bibinfo {author}
  {\bibfnamefont {D.~G.}\ \bibnamefont {Lalloo}}, \bibinfo {author}
  {\bibfnamefont {N.~M.}\ \bibnamefont {Ferguson}},\ and\ \bibinfo {author}
  {\bibfnamefont {A.~C.}\ \bibnamefont {Ghani}},\ }\bibfield  {title} {\bibinfo
  {title} {The impact of {{COVID-19}} and strategies for mitigation and
  suppression in low- and middle-income countries},\ }\href
  {https://doi.org/10.1126/science.abc0035} {\bibfield  {journal} {\bibinfo
  {journal} {Science}\ }\textbf {\bibinfo {volume} {369}},\ \bibinfo {pages}
  {413} (\bibinfo {year} {2020})}\BibitemShut {NoStop}%
\bibitem [{\citenamefont {Orlowski}\ and\ \citenamefont
  {Goldsmith}(2020)}]{orlowski2020four}%
  \BibitemOpen
  \bibfield  {author} {\bibinfo {author} {\bibfnamefont {E.~J.~W.}\
  \bibnamefont {Orlowski}}\ and\ \bibinfo {author} {\bibfnamefont {D.~J.~A.}\
  \bibnamefont {Goldsmith}},\ }\bibfield  {title} {\bibinfo {title} {Four
  months into the {{COVID-19}} pandemic, {{Sweden's}} prized herd immunity is
  nowhere in sight},\ }\href {https://doi.org/10.1177/0141076820945282}
  {\bibfield  {journal} {\bibinfo  {journal} {Journal of the Royal Society of
  Medicine}\ }\textbf {\bibinfo {volume} {113}},\ \bibinfo {pages} {292}
  (\bibinfo {year} {2020})}\BibitemShut {NoStop}%
\bibitem [{\citenamefont {Perra}(2021)}]{perra2021nonpharmaceutical}%
  \BibitemOpen
  \bibfield  {author} {\bibinfo {author} {\bibfnamefont {N.}~\bibnamefont
  {Perra}},\ }\bibfield  {title} {\bibinfo {title} {Non-pharmaceutical
  interventions during the {{COVID-19}} pandemic: {{A}} review},\ }\href
  {https://doi.org/10.1016/j.physrep.2021.02.001} {\bibfield  {journal}
  {\bibinfo  {journal} {Physics Reports}\ }\textbf {\bibinfo {volume} {913}},\
  \bibinfo {pages} {1} (\bibinfo {year} {2021})}\BibitemShut {NoStop}%
\bibitem [{\citenamefont {Brusselaers}\ \emph {et~al.}(2022)\citenamefont
  {Brusselaers}, \citenamefont {Steadson}, \citenamefont {Bjorklund},
  \citenamefont {Breland}, \citenamefont {Stilhoff~S{\"o}rensen}, \citenamefont
  {Ewing}, \citenamefont {Bergmann},\ and\ \citenamefont
  {Steineck}}]{brusselaers2022evaluation}%
  \BibitemOpen
  \bibfield  {author} {\bibinfo {author} {\bibfnamefont {N.}~\bibnamefont
  {Brusselaers}}, \bibinfo {author} {\bibfnamefont {D.}~\bibnamefont
  {Steadson}}, \bibinfo {author} {\bibfnamefont {K.}~\bibnamefont {Bjorklund}},
  \bibinfo {author} {\bibfnamefont {S.}~\bibnamefont {Breland}}, \bibinfo
  {author} {\bibfnamefont {J.}~\bibnamefont {Stilhoff~S{\"o}rensen}}, \bibinfo
  {author} {\bibfnamefont {A.}~\bibnamefont {Ewing}}, \bibinfo {author}
  {\bibfnamefont {S.}~\bibnamefont {Bergmann}},\ and\ \bibinfo {author}
  {\bibfnamefont {G.}~\bibnamefont {Steineck}},\ }\bibfield  {title} {\bibinfo
  {title} {Evaluation of science advice during the {{COVID-19}} pandemic in
  {{Sweden}}},\ }\href {https://doi.org/10.1057/s41599-022-01097-5} {\bibfield
  {journal} {\bibinfo  {journal} {Humanities and Social Sciences
  Communications}\ }\textbf {\bibinfo {volume} {9}},\ \bibinfo {pages} {91}
  (\bibinfo {year} {2022})}\BibitemShut {NoStop}%
\bibitem [{\citenamefont {Zenone}\ \emph {et~al.}(2022)\citenamefont {Zenone},
  \citenamefont {Snyder}, \citenamefont {Marcon},\ and\ \citenamefont
  {Caulfield}}]{zenone2022analyzing}%
  \BibitemOpen
  \bibfield  {author} {\bibinfo {author} {\bibfnamefont {M.}~\bibnamefont
  {Zenone}}, \bibinfo {author} {\bibfnamefont {J.}~\bibnamefont {Snyder}},
  \bibinfo {author} {\bibfnamefont {A.}~\bibnamefont {Marcon}},\ and\ \bibinfo
  {author} {\bibfnamefont {T.}~\bibnamefont {Caulfield}},\ }\bibfield  {title}
  {\bibinfo {title} {Analyzing natural herd immunity media discourse in the
  {{United Kingdom}} and the {{United States}}},\ }\href
  {https://doi.org/10.1371/journal.pgph.0000078} {\bibfield  {journal}
  {\bibinfo  {journal} {PLOS Global Public Health}\ }\textbf {\bibinfo {volume}
  {2}},\ \bibinfo {pages} {e0000078} (\bibinfo {year} {2022})}\BibitemShut
  {NoStop}%
\bibitem [{\citenamefont {Brett}\ and\ \citenamefont
  {Rohani}(2020)}]{brett2020transmission}%
  \BibitemOpen
  \bibfield  {author} {\bibinfo {author} {\bibfnamefont {T.~S.}\ \bibnamefont
  {Brett}}\ and\ \bibinfo {author} {\bibfnamefont {P.}~\bibnamefont {Rohani}},\
  }\bibfield  {title} {\bibinfo {title} {Transmission dynamics reveal the
  impracticality of {{COVID-19}} herd immunity strategies},\ }\href
  {https://doi.org/10.1073/pnas.2008087117} {\bibfield  {journal} {\bibinfo
  {journal} {Proceedings of the National Academy of Sciences}\ }\textbf
  {\bibinfo {volume} {117}},\ \bibinfo {pages} {25897} (\bibinfo {year}
  {2020})}\BibitemShut {NoStop}%
\bibitem [{\citenamefont {Buss}\ \emph {et~al.}(2021)\citenamefont {Buss},
  \citenamefont {Prete}, \citenamefont {Abrahim}, \citenamefont {Mendrone},
  \citenamefont {Salomon}, \citenamefont {{De Almeida-Neto}}, \citenamefont
  {Fran{\c c}a}, \citenamefont {Belotti}, \citenamefont {Carvalho},
  \citenamefont {Costa}, \citenamefont {Crispim}, \citenamefont {Ferreira},
  \citenamefont {Fraiji}, \citenamefont {Gurzenda}, \citenamefont {Whittaker},
  \citenamefont {Kamaura}, \citenamefont {Takecian}, \citenamefont
  {Da~Silva~Peixoto}, \citenamefont {Oikawa}, \citenamefont {Nishiya},
  \citenamefont {Rocha}, \citenamefont {Salles}, \citenamefont
  {De~Souza~Santos}, \citenamefont {Da~Silva}, \citenamefont {Custer},
  \citenamefont {Parag}, \citenamefont {{Barral-Netto}}, \citenamefont
  {Kraemer}, \citenamefont {Pereira}, \citenamefont {Pybus}, \citenamefont
  {Busch}, \citenamefont {Castro}, \citenamefont {Dye}, \citenamefont
  {Nascimento}, \citenamefont {Faria},\ and\ \citenamefont
  {Sabino}}]{buss2021threequarters}%
  \BibitemOpen
  \bibfield  {author} {\bibinfo {author} {\bibfnamefont {L.~F.}\ \bibnamefont
  {Buss}}, \bibinfo {author} {\bibfnamefont {C.~A.}\ \bibnamefont {Prete}},
  \bibinfo {author} {\bibfnamefont {C.~M.~M.}\ \bibnamefont {Abrahim}},
  \bibinfo {author} {\bibfnamefont {A.}~\bibnamefont {Mendrone}}, \bibinfo
  {author} {\bibfnamefont {T.}~\bibnamefont {Salomon}}, \bibinfo {author}
  {\bibfnamefont {C.}~\bibnamefont {{De Almeida-Neto}}}, \bibinfo {author}
  {\bibfnamefont {R.~F.~O.}\ \bibnamefont {Fran{\c c}a}}, \bibinfo {author}
  {\bibfnamefont {M.~C.}\ \bibnamefont {Belotti}}, \bibinfo {author}
  {\bibfnamefont {M.~P. S.~S.}\ \bibnamefont {Carvalho}}, \bibinfo {author}
  {\bibfnamefont {A.~G.}\ \bibnamefont {Costa}}, \bibinfo {author}
  {\bibfnamefont {M.~A.~E.}\ \bibnamefont {Crispim}}, \bibinfo {author}
  {\bibfnamefont {S.~C.}\ \bibnamefont {Ferreira}}, \bibinfo {author}
  {\bibfnamefont {N.~A.}\ \bibnamefont {Fraiji}}, \bibinfo {author}
  {\bibfnamefont {S.}~\bibnamefont {Gurzenda}}, \bibinfo {author}
  {\bibfnamefont {C.}~\bibnamefont {Whittaker}}, \bibinfo {author}
  {\bibfnamefont {L.~T.}\ \bibnamefont {Kamaura}}, \bibinfo {author}
  {\bibfnamefont {P.~L.}\ \bibnamefont {Takecian}}, \bibinfo {author}
  {\bibfnamefont {P.}~\bibnamefont {Da~Silva~Peixoto}}, \bibinfo {author}
  {\bibfnamefont {M.~K.}\ \bibnamefont {Oikawa}}, \bibinfo {author}
  {\bibfnamefont {A.~S.}\ \bibnamefont {Nishiya}}, \bibinfo {author}
  {\bibfnamefont {V.}~\bibnamefont {Rocha}}, \bibinfo {author} {\bibfnamefont
  {N.~A.}\ \bibnamefont {Salles}}, \bibinfo {author} {\bibfnamefont {A.~A.}\
  \bibnamefont {De~Souza~Santos}}, \bibinfo {author} {\bibfnamefont {M.~A.}\
  \bibnamefont {Da~Silva}}, \bibinfo {author} {\bibfnamefont {B.}~\bibnamefont
  {Custer}}, \bibinfo {author} {\bibfnamefont {K.~V.}\ \bibnamefont {Parag}},
  \bibinfo {author} {\bibfnamefont {M.}~\bibnamefont {{Barral-Netto}}},
  \bibinfo {author} {\bibfnamefont {M.~U.~G.}\ \bibnamefont {Kraemer}},
  \bibinfo {author} {\bibfnamefont {R.~H.~M.}\ \bibnamefont {Pereira}},
  \bibinfo {author} {\bibfnamefont {O.~G.}\ \bibnamefont {Pybus}}, \bibinfo
  {author} {\bibfnamefont {M.~P.}\ \bibnamefont {Busch}}, \bibinfo {author}
  {\bibfnamefont {M.~C.}\ \bibnamefont {Castro}}, \bibinfo {author}
  {\bibfnamefont {C.}~\bibnamefont {Dye}}, \bibinfo {author} {\bibfnamefont
  {V.~H.}\ \bibnamefont {Nascimento}}, \bibinfo {author} {\bibfnamefont
  {N.~R.}\ \bibnamefont {Faria}},\ and\ \bibinfo {author} {\bibfnamefont
  {E.~C.}\ \bibnamefont {Sabino}},\ }\bibfield  {title} {\bibinfo {title}
  {Three-quarters attack rate of {{SARS-CoV-2}} in the {{Brazilian Amazon}}
  during a largely unmitigated epidemic},\ }\href
  {https://doi.org/10.1126/science.abe9728} {\bibfield  {journal} {\bibinfo
  {journal} {Science}\ }\textbf {\bibinfo {volume} {371}},\ \bibinfo {pages}
  {288} (\bibinfo {year} {2021})}\BibitemShut {NoStop}%
\bibitem [{\citenamefont {Sridhar}\ and\ \citenamefont
  {Gurdasani}(2021)}]{sridhar2021herd}%
  \BibitemOpen
  \bibfield  {author} {\bibinfo {author} {\bibfnamefont {D.}~\bibnamefont
  {Sridhar}}\ and\ \bibinfo {author} {\bibfnamefont {D.}~\bibnamefont
  {Gurdasani}},\ }\bibfield  {title} {\bibinfo {title} {Herd immunity by
  infection is not an option},\ }\href
  {https://doi.org/10.1126/science.abf7921} {\bibfield  {journal} {\bibinfo
  {journal} {Science}\ }\textbf {\bibinfo {volume} {371}},\ \bibinfo {pages}
  {230} (\bibinfo {year} {2021})}\BibitemShut {NoStop}%
\bibitem [{\citenamefont {Stegehuis}\ \emph {et~al.}(2016)\citenamefont
  {Stegehuis}, \citenamefont {Van Der~Hofstad},\ and\ \citenamefont
  {Van~Leeuwaarden}}]{stegehuis2016epidemic}%
  \BibitemOpen
  \bibfield  {author} {\bibinfo {author} {\bibfnamefont {C.}~\bibnamefont
  {Stegehuis}}, \bibinfo {author} {\bibfnamefont {R.}~\bibnamefont {Van
  Der~Hofstad}},\ and\ \bibinfo {author} {\bibfnamefont {J.~S.~H.}\
  \bibnamefont {Van~Leeuwaarden}},\ }\bibfield  {title} {\bibinfo {title}
  {Epidemic spreading on complex networks with community structures},\ }\href
  {https://doi.org/10.1038/srep29748} {\bibfield  {journal} {\bibinfo
  {journal} {Scientific Reports}\ }\textbf {\bibinfo {volume} {6}},\ \bibinfo
  {pages} {29748} (\bibinfo {year} {2016})}\BibitemShut {NoStop}%
\bibitem [{\citenamefont {{H{\'e}bert-Dufresne}}\ and\ \citenamefont
  {Allard}(2019)}]{hebert-dufresne2019smeared}%
  \BibitemOpen
  \bibfield  {author} {\bibinfo {author} {\bibfnamefont {L.}~\bibnamefont
  {{H{\'e}bert-Dufresne}}}\ and\ \bibinfo {author} {\bibfnamefont
  {A.}~\bibnamefont {Allard}},\ }\bibfield  {title} {\bibinfo {title} {Smeared
  phase transitions in percolation on real complex networks},\ }\href
  {https://doi.org/10.1103/PhysRevResearch.1.013009} {\bibfield  {journal}
  {\bibinfo  {journal} {Physical Review Research}\ }\textbf {\bibinfo {volume}
  {1}},\ \bibinfo {pages} {013009} (\bibinfo {year} {2019})}\BibitemShut
  {NoStop}%
\bibitem [{\citenamefont {Morita}(2021)}]{morita2021solvable}%
  \BibitemOpen
  \bibfield  {author} {\bibinfo {author} {\bibfnamefont {S.}~\bibnamefont
  {Morita}},\ }\bibfield  {title} {\bibinfo {title} {Solvable epidemic model on
  degree-correlated networks},\ }\href
  {https://doi.org/10.1016/j.physa.2020.125419} {\bibfield  {journal} {\bibinfo
   {journal} {Physica A: Statistical Mechanics and its Applications}\ }\textbf
  {\bibinfo {volume} {563}},\ \bibinfo {pages} {125419} (\bibinfo {year}
  {2021})}\BibitemShut {NoStop}%
\bibitem [{\citenamefont {Ball}\ \emph {et~al.}(2023)\citenamefont {Ball},
  \citenamefont {Critcher}, \citenamefont {Neal},\ and\ \citenamefont
  {Sirl}}]{ball2023impact}%
  \BibitemOpen
  \bibfield  {author} {\bibinfo {author} {\bibfnamefont {F.}~\bibnamefont
  {Ball}}, \bibinfo {author} {\bibfnamefont {L.}~\bibnamefont {Critcher}},
  \bibinfo {author} {\bibfnamefont {P.}~\bibnamefont {Neal}},\ and\ \bibinfo
  {author} {\bibfnamefont {D.}~\bibnamefont {Sirl}},\ }\bibfield  {title}
  {\bibinfo {title} {The impact of household structure on disease-induced herd
  immunity},\ }\href {https://doi.org/10.1007/s00285-023-02010-7} {\bibfield
  {journal} {\bibinfo  {journal} {Journal of Mathematical Biology}\ }\textbf
  {\bibinfo {volume} {87}},\ \bibinfo {pages} {83} (\bibinfo {year}
  {2023})}\BibitemShut {NoStop}%
\bibitem [{\citenamefont {Holme}\ and\ \citenamefont
  {Saram{\"a}ki}(2012)}]{holme2012temporal}%
  \BibitemOpen
  \bibfield  {author} {\bibinfo {author} {\bibfnamefont {P.}~\bibnamefont
  {Holme}}\ and\ \bibinfo {author} {\bibfnamefont {J.}~\bibnamefont
  {Saram{\"a}ki}},\ }\bibfield  {title} {\bibinfo {title} {Temporal networks},\
  }\href {https://doi.org/10.1016/j.physrep.2012.03.001} {\bibfield  {journal}
  {\bibinfo  {journal} {Physics Reports}\ }\textbf {\bibinfo {volume} {519}},\
  \bibinfo {pages} {97} (\bibinfo {year} {2012})}\BibitemShut {NoStop}%
\bibitem [{\citenamefont {Kenah}\ and\ \citenamefont
  {Robins}(2007)}]{kenah2007second}%
  \BibitemOpen
  \bibfield  {author} {\bibinfo {author} {\bibfnamefont {E.}~\bibnamefont
  {Kenah}}\ and\ \bibinfo {author} {\bibfnamefont {J.~M.}\ \bibnamefont
  {Robins}},\ }\bibfield  {title} {\bibinfo {title} {Second look at the spread
  of epidemics on networks},\ }\href@noop {} {\bibfield  {journal} {\bibinfo
  {journal} {Physical Review E}\ }\textbf {\bibinfo {volume} {76}},\ \bibinfo
  {pages} {036113} (\bibinfo {year} {2007})}\BibitemShut {NoStop}%
\bibitem [{\citenamefont {Fosdick}\ \emph {et~al.}(2018)\citenamefont
  {Fosdick}, \citenamefont {Larremore}, \citenamefont {Nishimura},\ and\
  \citenamefont {Ugander}}]{fosdick2018configuring}%
  \BibitemOpen
  \bibfield  {author} {\bibinfo {author} {\bibfnamefont {B.~K.}\ \bibnamefont
  {Fosdick}}, \bibinfo {author} {\bibfnamefont {D.~B.}\ \bibnamefont
  {Larremore}}, \bibinfo {author} {\bibfnamefont {J.}~\bibnamefont
  {Nishimura}},\ and\ \bibinfo {author} {\bibfnamefont {J.}~\bibnamefont
  {Ugander}},\ }\bibfield  {title} {\bibinfo {title} {Configuring random graph
  models with fixed degree sequences},\ }\href
  {https://doi.org/10.1137/16M1087175} {\bibfield  {journal} {\bibinfo
  {journal} {SIAM Review}\ }\textbf {\bibinfo {volume} {60}},\ \bibinfo {pages}
  {315} (\bibinfo {year} {2018})}\BibitemShut {NoStop}%
\bibitem [{\citenamefont {Bogu{\~n}{\'a}}\ \emph {et~al.}(2020)\citenamefont
  {Bogu{\~n}{\'a}}, \citenamefont {Krioukov}, \citenamefont {Almagro},\ and\
  \citenamefont {Serrano}}]{boguna2020small}%
  \BibitemOpen
  \bibfield  {author} {\bibinfo {author} {\bibfnamefont {M.}~\bibnamefont
  {Bogu{\~n}{\'a}}}, \bibinfo {author} {\bibfnamefont {D.}~\bibnamefont
  {Krioukov}}, \bibinfo {author} {\bibfnamefont {P.}~\bibnamefont {Almagro}},\
  and\ \bibinfo {author} {\bibfnamefont {M.~{\'A}.}\ \bibnamefont {Serrano}},\
  }\bibfield  {title} {\bibinfo {title} {Small worlds and clustering in spatial
  networks},\ }\href {https://doi.org/10.1103/PhysRevResearch.2.023040}
  {\bibfield  {journal} {\bibinfo  {journal} {Physical Review Research}\
  }\textbf {\bibinfo {volume} {2}},\ \bibinfo {pages} {023040} (\bibinfo {year}
  {2020})}\BibitemShut {NoStop}%
\end{thebibliography}
\end{document}


\title{Supplementary Information for \\``Strength and weakness of disease-induced herd immunity in networks''}
\author{Takayuki Hiraoka}
\email[Corresponding author. Email:]{takayuki.hiraoka@aalto.fi}
\author{Zahra Ghadiri}
\author{Abbas K.~Rizi}
\author{Mikko Kivel\"{a}}
\author{Jari Saram\"{a}ki}

\maketitle
\section{Correlation-based measure of immunity localization}
\label{section:correlation}
In the main text, we characterize the degree of immunity localization by the share of edges between immune and susceptible nodes, denoted by $\rho_\mathrm{SR}$. Here, we derive a measure of localization based on the correlation coefficient between epidemiological states of adjacent nodes, and show that it has a linear relationship with $\rho_\mathrm{SR}$. 

The localization is defined as the bias that the two nodes at the ends of an edge tend to be in the same epidemiological state (susceptible or immune) compared to the case where the node states are independent of network topology. Since the epidemiological state is binary, we can represent the state of node $i$ with $\sigma_i$, where $\sigma_i = +1$ if $i$ is susceptible and $\sigma_i = -1$ if it is immune (or the other way around; it does not affect the following discussion). For network $G = (V, E)$, the Pearson correlation coefficient between the states of two adjacent nodes is computed as $C = (\langle \sigma_i \sigma_j \rangle - \langle \sigma_i \rangle^2) / (\langle \sigma_i^2 \rangle - \langle \sigma_i \rangle^2)$, where
\begin{gather*}
\langle \sigma_i \sigma_j \rangle = \frac{1}{|E|} \sum_{(i, j) \in E} \sigma_i \sigma_j = 1 - 2 \rho_\mathrm{SR}, \\
\langle \sigma_i \rangle = \frac{1}{|V|} \sum_{i \in V} \sigma_i = 1 - 2 \pi_\mathrm{R},\\
\langle \sigma_i^2 \rangle = \frac{1}{|V|} \sum_{i \in V} \sigma_i^2 = 1.
\end{gather*}
Plugging these expressions into the original formula for the correlation coefficient, we have 
\begin{equation}
C = 1 - \frac{\rho_\mathrm{SR}}{2 \pi_\mathrm{R} (1 - \pi_\mathrm{R})}.
\end{equation}
This linear relationship between $C$ and $\pi_\mathrm{R}$ justifies our use of the share of interface edges as a measure of immunity localization.

\section{Edge rewiring}
\label{section:edge_rewiring}
The edge rewiring process is implemented according to the double edge swap procedure~\cite{fosdick2018configuring}. In this method, we select two edges $e_1=(u_1, v_1)$ and $e_2=(u_2, v_2)$ uniformly at random from all edges in the network. To avoid bias in the order of the endpoints, we may swap the endpoint with probability $1/2$, changing $(u, v)$ to $(v,u)$. Next, we exchange the endpoints of the selected edges to propose new edges $e_1^\mathrm{new}=(u_1, v_2)$ and $e_2^\mathrm{new}=(u_2, v_1)$. If $e_1^\mathrm{new}$ or $e_2^\mathrm{new}$ is a self-loop, or if either edge already exists in the network, we reject the proposed edges. Otherwise, we accept them and replace $e_1$ and $e_2$ with $e_1^\mathrm{new}$ and $e_2^\mathrm{new}$. This process is repeated until a fraction $\phi$ of all edges have been rewired. In the process of randomly selecting pairs of edges, we prioritize edges that have not yet been chosen. We do this by making a permutated list of edges and going through it to sequentially select two edges. If less than a fraction \(\phi\) of the edges have been rewired after a round of swaps, we continue the swapping process by permuting the list of edges again, this time including edges that were swapped in the previous round. This approach maintains the priority of unswapped edges while ensuring that each edge has an equal chance of being selected.

\section{Heterogeneous random geometric graphs}
\label{section:hrgg}
The heterogeneous random geometric graph proposed by Bogu\~{n}\'{a} et al.~\cite{boguna2020small} is a model for generating networks with heterogeneous degrees and tunable spatial embeddedness. In this model, each node $i$ in a network of size $n$ is assigned a point $x_i$ in the underlying manifold and a non-negative weight $w_i$, which is a function of its expected degree $\kappa_i$.  In addition, the model is globally parameterized by \emph{temperature} $\tau$, which governs the strength of spatial embeddedness, and a coefficient $\hat{\mu}$, controlling the expected number of edges (that is, relating weight $w$ to expected degree $\kappa$). The probability that a pair of nodes $i$ and $j$ are connected by an edge is given by a Fermi-Dirac distribution:
\begin{equation}
\begin{aligned}
p_{ij} &= p(|x_i - x_j|, w_i, w_j) \\
& = \frac{1}{\exp(\tau^{-1} \ln |x_i - x_j| + w_i + w_j) + 1} \\
&= \frac{1}{\{|x_i - x_j|\exp[\tau (w_i + w_j)]\}^{1/\tau} + 1},
\end{aligned}
\label{eq:connection_probability}
\end{equation}
where $|x - x'|$ denotes the Euclidean distance between points $x$ and $x'$.
In the following, we assume that the expected degree $\kappa_i$ of each node $i$ is independently drawn from a distribution with mean $\langle k \rangle$, and consequently $w_i$ is also an independent random variable distributed as $f(w)$. We also assume that the point of each node is distributed uniformly and independently on a $d$-dimensional unit hypercube with periodic boundary conditions (i.e., a $d$-dimensional unit hypertorus). 
Note that this differs from the original assumption in Bogu\~{n}\'{a} et al.~\cite{boguna2020small}, where the underlying space grows as the number of nodes $n$ increases so that the density of points in the space is kept to unity. For simplicity, we consider the case of $d=2$.

We assume, without loss of generality, that a node with weight $w$ is at the origin. In circular coordinates in $\mathbb{R}^2$, the volume element at radius $r$ is $r \mathrm{d}r \mathrm{d}\theta$ and the number of nodes in the annulus $[r, r+\mathrm{d}r]$ is $2 \pi n r \mathrm{d}r$. By integrating over the space, we get the expected degree $\kappa$ of the focal node as a function of $w$:
\begin{equation}
\kappa(w) = 2 \pi n \int \mathrm{d}w' \, f(w') \int_0^\infty \mathrm{d}r \, p(r, w, w') r, 
\label{eq:expected_degree}
\end{equation}

Equation~\eqref{eq:expected_degree} converges for $\tau < 1/d$ and diverges for $\tau \geq 1/d$. For $\tau < 1/d$, we can change the variable as $t = r \exp[\tau (w + w')]$ and obtain
\begin{equation}
\begin{aligned}
\kappa(w) &= 2 \pi n \int \mathrm{d}w' \, f(w') \exp\left[- 2 \tau (w + w')\right] \int_0^\infty \frac{t \mathrm{d}t}{t^{1/\tau} + 1} \\
&= 2 \pi n e^{- 2 \tau w} \left\langle e^{- 2 \tau w} \right\rangle \int_0^\infty \frac{t \mathrm{d}t}{t^{1/\tau} + 1} \\
&= 2 \pi n e^{- 2 \tau w} \left\langle e^{- 2 \tau w} \right\rangle \frac{\pi \tau}{\sin(2 \pi \tau)}.
\end{aligned}
\label{eq:expected_degree_beta_gt_d}
\end{equation}
Taking average over $w$, we find
\begin{equation}
\left\langle e^{- 2 \tau w} \right\rangle = \left[\frac{\sin(2 \pi \tau)}{2 \pi^2 n \tau} \langle k \rangle \right]^{1/2}.
\end{equation}
Plugging this back into Eq.~\eqref{eq:expected_degree_beta_gt_d} yields
\begin{gather}
w = - \frac{1}{2 \tau} \left(\ln \kappa + \frac{1}{2} \ln \hat{\mu} \right), \\
\hat{\mu} = \frac{\sin(2 \pi \tau)}{2 \pi^2 n \tau \langle k \rangle}.
\end{gather}
The connection probability is written, now as a function of $\kappa_i$ and $\kappa_j$, as
\begin{equation}
p_{ij} = \frac{1}{\left[\dfrac{|x_i - x_j|}{(\hat{\mu} \kappa_i \kappa_j)^{1/2}}\right]^{1/\tau} + 1}. 
\end{equation}
For an arbitrary dimension $d$, the above results are generalized as
\begin{gather}
w = - \frac{1}{d \tau} \left(\ln \kappa + \frac{1}{2} \ln \hat{\mu} \right), \\
\hat{\mu} = \frac{\Gamma(d/2) \sin(d \pi \tau)}{2 \pi^{1 + d/2} n \tau \langle k \rangle}, \\
p_{ij} = \frac{1}{\left[\dfrac{|x_i - x_j|}{(\hat{\mu} \kappa_i \kappa_j)^{1/d}}\right]^{1/\tau} + 1}.
\end{gather}

For $\tau \geq 1/d$, the range of integral in Eq.~\eqref{eq:expected_degree} should be limited inside the hypercube centered at the origin, i.e. $[-1/2, 1/2]^d$. For $r \leq 1/2$, the entire annulus is in the integrated region; for $1/2 < r \leq 1/\sqrt{2}$, however, fraction $\theta (r) = 1 - 4 \arccos(1/2r) / \pi$ of the annulus is in the region; for $r > 1/\sqrt{2}$, the annulus is completely outside the region. The expected degree is then given by
 \begin{widetext}
\begin{equation}
\kappa(w) = 2 \pi n \int \mathrm{d}w' \, f(w') \left[\int_0^{\frac{1}{2}} p(r, w, w') r \mathrm{d}r + \int_{\frac{1}{2}}^{\frac{1}{\sqrt{2}}} \theta(r) p(r, w, w') r \mathrm{d}r\right].
\end{equation}
 \end{widetext}
This cannot be easily integrated, but we can approximate it, for example, by neglecting $1$ in the denominator of Eq.~\eqref{eq:connection_probability} and integrate up to $1/2$ with respect to $r$:
\begin{equation}
\begin{aligned}
\kappa(w) &\approx 2 \pi n \int dw' \, f(w') \int_0^{1/2} dr \, e^{- (w + w')} r^{1 - \beta}  \\
& = \frac{2 \pi n e^{-w} \left\langle e^{-w} \right\rangle}{(2 - 1/\tau) 2^{2 - 1/\tau}}.
\end{aligned}
\end{equation}
Similarly to the case of $\tau < 1/d$, we find
\begin{equation}
\left\langle e^{-w} \right\rangle = \left[\frac{(2 - 1/\tau) 2^{2 - 1/\tau}}{2 \pi n} \langle k \rangle \right]^{1/2}, 
\end{equation}
which leads to
\begin{gather}
w = - \left(\ln \kappa + \frac{1}{2} \ln \hat{\mu} \right), \\
\hat{\mu} = \frac{(2 - 1/\tau) 2^{2 - 1/\tau}}{2 \pi n \langle k \rangle}.
\end{gather}
In most cases, this gives a good approximation for the value of $\hat{\mu}$; we can use this approximation as an initial guess and numerically optimize $\hat{\mu}$ to achieve the desired edge density.
The connection probability is then given by
\begin{equation}
p_{ij} = \frac{1}{\dfrac{{|x_i - x_j|}^{1/\tau}}{\hat{\mu} \kappa_i \kappa_j} + 1}.
\end{equation}
For arbitrary dimension $d$, only the formula for $\hat{\mu}$ needs to be changed:
\begin{equation}
\hat{\mu} = \frac{\Gamma(d/2) (d - 1/\tau) 2^{d - 1/\tau}}{2 \pi^{d/2} n \langle k \rangle}.
\end{equation}

\section{Spatial stochastic block model}
\label{section:spatial_sbm}
The stochastic block model (SBM) describes an ensemble of undirected networks generated as follows. Consider a complete graph of $N$ nodes, where each node belongs to one of $K$ groups. Let $N_i$ be the size of group $i$. An SBM network is obtained by filtering the edges of the complete graph according to a symmetric $K \times K$ matrix $B$, such that each of the $N_i N_j$ edges between groups $i$ and $j$ ($i \neq j$) is retained independently with probability $B_{ij}$, and each of the $\binom{N_i}{2}$ edges within group $i$ is independently kept with probability $B_{ii}$. As a result, the network becomes sparser, reducing from $\binom{N}{2}$ edges in the complete graph to an expected
\begin{equation}
m = \sum_{i = 1}^{K} \left[\binom{N_i}{2} B_{ii} + \sum_{j \neq i} N_i N_j B_{ij} \right]
\end{equation}
edges, corresponding to edge density $\delta = m / \binom{N}{2}$.

We combine the heterogeneous random geometric graph (HRGG) model and the SBM by letting the HRGG model undertake part of this edge filtering process. Specifically, we assign each node a random point in a metric space independently of the group it belongs to, and generate a HRGG where the expected degree of each node is $2 \eta m / N$, i.e., an expected $\eta m$ edges in total, with $\eta > 1$. This network is then further filtered by the SBM with an inflated block matrix
\begin{equation}
B' = \frac{1}{\eta \delta}B,
\end{equation}
so that the expected number of edges is equal to $m$. In this two-stage filtering process, the HRGG model reduces edges by a factor of $\eta \delta$ while the SBM reduces them by a factor of $1 / \eta$. The value of $\eta$ is bounded as 
\begin{equation}
\delta \max (B) \leq \eta \leq \frac{1}{\delta}.
\end{equation}
The first inequality derives from the requirement that each element of $B'$ should not exceed one and the second inequality ensures that the intermediate HRGG is not denser than the complete graph. The coefficient $\eta$ controls the relative contribution of SBM in the filtering process; we set $\eta = 1 / \sqrt{\delta}$ to balance the contribution of the two models. 

\section{Network epidemic simulations}
\label{section:epn}
We adopt the epidemic percolation network framework~\cite{kenah2007second} to implement the epidemic simulations of the network SIR model. Given an undirected contact network, we construct an effective weighted directed network that encodes stochastic transmission as follows:  (i) Generate a symmetric directed network by transforming each undirected edge in the contact network, represented by an unordered pair $\{i, j\}$, into two directed edges, represented by ordered pairs $i \to j$ and $j \to i$. (ii) Draw a random time $\Delta_i$ for each node $i$ to recover from infection from an exponential distribution with rate $\gamma$, that is, $\Delta_i \sim \operatorname{Exp}(\gamma)$. (iii) Draw a random time $t_{i \to j}$ for an infected node $i$, were it not to recover forever, to transmit the disease to a susceptible node $j$ from an exponential distribution with rate $\beta$, that is, $t_{i \to j} \sim \operatorname{Exp}(\beta)$. (iv) Each directed edge $i \to j$ is transmissible if node $i$ is infected, node $j$ is susceptible, and the transmission occurs before $i$ recovers. Retain edge $i \to j$ if $t_{i \to j} < \Delta_i$ and assign it a weight $t_{i \to j}$, or remove it otherwise. 

This weighted directed network accurately encodes stochastic realizations of the SIR dynamics~\cite{kenah2007second, karrer2010message}. The nodes that become infected during an outbreak originating from a source node can be identified as those that can be reached from the source node by traversing this effective directed network, i.e., the out-component of the source node. The time at which they become infected is given by the (weighted) shortest path length from the source node, which can be computed efficiently by Dijkstra's algorithm. If there is a giant strongly connected component (GSCC) in this effective network, an epidemic spreading to a finite fraction of the network will occur if the source node has a path to the GSCC, i.e., it is in the in-component of the GSCC.

\section{Message passing formalism}
\label{section:message_passing}
For locally tree-like networks, we can use the framework to map the epidemic dynamics to a bond percolation process~\cite{newman2002spread} to analytically calculate relevant quantities for herd immunity, such as the expected sizes of an epidemic and the giant component in the residual subgraph. We outline the message passing method for percolation~\cite{newman2023message}, which is generally valid for locally tree-like networks. Then, we derive the expression for the family of configuration model networks, where the degree distribution is the only determinant of the network structure.

\subsection*{Message passing framework for locally tree-like networks}
Let $\chi_i$ be the indicator function of node $i$'s presence in the residual subgraph. In other words, 
\begin{equation}
\chi_i = \begin{cases}
    1 & \text{if $i$ is not immune}, \\
    0 & \text{if $i$ is immune}.
\end{cases}
\end{equation}
Let us introduce two other indicator functions. First, we define $\zeta_i$ where $\zeta_i = 1$ if node $i$ is not immune \emph{and} not in the giant component of the residual subgraph, and $\zeta_i = 0$ otherwise. Second, we define $\zeta_{j \to i}$, which takes the value equal to $\zeta_j$ in a slightly modified network where node $i$ is removed. In other words,  $\zeta_{j \to i} = 1$ if node $j$ is not immune and not in the giant component of the residual subgraph if it were not for node $i$, otherwise $\zeta_{j \to i} = 0$.

Node $i$ in the residual subgraph is not part of its giant component if every adjacent node $j$ is either immune ($\chi_j = 0$) or not connected to the giant component in the absence of $i$ ($\zeta_{j \to i} = 1$), leading to 
\begin{multline}
P(\zeta_i = 1 \mid \chi_i = 1) \\ 
= \prod_{j \in \mathcal{N}_i} [P(\chi_j = 0 \mid \chi_i = 1) + P(\zeta_{j \to i} = 1 \mid \chi_i = 1)],
\end{multline}
where $\mathcal{N}_i$ denotes the set of nodes adjacent to $i$.
The sum and product rules dictate that the second term in the product of the rhs can be expressed as
\begin{multline}
P(\zeta_{j \to i} = 1 \mid \chi_i = 1) \\
= P(\zeta_{j \to i} = 1 \mid \chi_j = 1) P(\chi_j = 1 \mid \chi_i = 1).
\end{multline}
Note that $P(\zeta_{j \to i} = 1 \mid \chi_j = 0) = 0$ since $\chi_j = 0$ (i.e., $j$ is immune) implies $\zeta_{j \to i} = 0$. 

To simplify the notation, we adopt the following symbols to represent probabilities related to these indicator functions:
\begin{gather*}
\psi_i = P(\chi_i = 1), \\
\lambda_i = P(\zeta_i = 1 \mid \chi_i = 1), \\
\lambda_{j \to i} = P(\zeta_{j \to i} = 1 \mid \chi_j = 1), \\
g_{j \to i} = P(\chi_j = 1 \mid \chi_i = 1).
\end{gather*}
The message-passing equations for the giant residual component are now written as
\begin{equation}
\lambda_i = \prod_{j \in \mathcal{N}_i} (1 - g_{j \to i} + g_{j \to i} \lambda_{j \to i}),
\label{eq:giant_residual_receive_message}
\end{equation}
where $\lambda_{j \to i}$ satisfies a consistency equation
\begin{equation}
\lambda_{j \to i} = \prod_{\ell \in \mathcal{N}_j \setminus i} (1 - g_{\ell \to j} + g_{\ell \to j} \lambda_{\ell \to j}).
\label{eq:giant_residual_transmit_message}
\end{equation}
Since Eq.~\eqref{eq:giant_residual_transmit_message} can be written for every ordered pair of nodes $i$ and $j$ connected by an edge, it represents a system of $2E$ equations, where $E$ is the number of edges. Usually, they cannot be solved analytically but can easily be solved numerically by recursion. We can then compute the proportions of the immune individuals and the residual giant component:
\begin{gather}
\pi_\mathrm{R} = \frac{1}{N}\sum_i (1 - \psi_i), 
\label{eq:immune_frac}\\
\pi_\mathrm{S_G} = \frac{1}{N}\sum_i \psi_i (1 - \lambda_i).
\label{eq:giant_residual_frac}
\end{gather}

The expected size of a post-immunity epidemic can be computed in similar steps. Let $\omega_{i}$ denote the conditional probability that node $i$ is not infected in a post-immunity epidemic with transmission probability $T'$ given that $i$ is in the residual subgraph, and $\omega_{j \to i}$ the same probability for node $j$ in a network where node $i$ is absent. Here, transmission probability is the probability that the disease is transmitted from an infected node to a susceptible neighbor, and related to transmission rate $\beta'$ and recovery rate $\gamma$ as $T' = \beta' / (\beta' + \gamma)$. Node $i$ in the residual subgraph is not infected if every adjacent node $j$ is (i) connected by a non-transmissible edge, (ii) immune, or (iii) not infected in the absence of $i$. The final condition comes from the fact that no epidemic that spreads to a finite fraction of the residual graph can occur outside the giant residual component, because every smaller component occupies a vanishing fraction. 
The message-passing equations are formulated as
\begin{equation}
\begin{aligned}
\omega_i &= \prod_{j \in \mathcal{N}_i} [1 - T' + T' (1 - g_{j \to i} + g_{j \to i} \omega_{j \to i})] \\
&= \prod_{j \in \mathcal{N}_i} [1 - T' g_{j \to i} + T' g_{j \to i} \omega_{j \to i})]
\end{aligned}
\label{eq:second_epidemic_receive_message}
\end{equation}
and likewise, 
\begin{equation}
\omega_{j \to i} = \prod_{\ell \in \mathcal{N}_j \setminus i} [1 - T' g_{\ell \to j} + T' g_{\ell \to j} \omega_{\ell \to j}].
\label{eq:second_epidemic_transmit_message}
\end{equation}
Note that these equations have the same form as Eqs.~\eqref{eq:giant_residual_receive_message} and \eqref{eq:giant_residual_transmit_message}, except $g_{j \to i}$ is replaced by $T' g_{j \to i}$. After solving the consistency equations recursively, we obtain the fraction of the population infected by the post-immunity epidemic as
\begin{equation}
\pi_\mathrm{R_2} = \frac{1}{N}\sum_i \psi_i (1 - \omega_i).
\end{equation}

In the above pipeline, the only factors that depend on the immunity scenario are probabilities $\psi_i$ and $g_{j \to i}$. In the following, we focus on obtaining the expressions for these two probabilities for each scenario, which then can be plugged into the message-passing equations to compute the relevant quantities.

\paragraph*{Disease-induced immunity.} When the immunity of nodes is induced by natural infection with transmission probability $T$, the probability $\psi_i$ that node $i$ is not infected during an immunity-inducing epidemic and thus in the residual subgraph is given by
\begin{equation}
\psi_i = \prod_{j \in \mathcal{N}_i} (1 - T + T \psi_{j \to i}),
\label{eq:first_epidemic_receive_message}
\end{equation}
where $\psi_{j \to i}$ denotes the probability that node $j$ is not infected during the epidemic in the absence of $i$, which satisfies the following consistency equation:
\begin{equation}
\psi_{j \to i} = \prod_{\ell \in \mathcal{N}_j \setminus i} (1 - T + T \psi_{\ell \to j}).
\label{eq:first_epidemic_transmit_message}
\end{equation}
Probabilities $\psi_{j \to i}$ and $\psi_{i}$ can be computed by solving Eq.~\eqref{eq:first_epidemic_transmit_message} recursively. The joint probability that adjacent nodes $i$ and $j$ are both in the residual subgraph is given by
\begin{equation}
P(\chi_i = \chi_j = 1) = \psi_{i \to j} \psi_{j \to i}.
\end{equation}
Here, note that $P(\chi_i = \chi_j = 1) \neq \psi_{i} \psi_{j \to i}$ because probabilities $\psi_i$ and $\psi_{j \to i}$ are not independent.
This allows us to identify the conditional probability 
\begin{equation}
g_{j \to i} = \frac{\psi_{j \to i}}{1 - T + T \psi_{j \to i}}.\label{eq:cond_prob_residual_DII}
\end{equation}

\paragraph*{Random immunization.} If the immunity is instead induced by random immunization of coverage $\pi_\mathrm{R}$, the probabilities of interest are self-evident:
\begin{equation}
\psi_i = g_{j \to i} = 1 - \pi_\mathrm{R}.
\end{equation}

\subsection*{Expressions for configuration model networks}
The message-passing framework allows us to calculate probabilities associated with each node and edge for any locally tree-like network structure. However, we may be more interested in averaged quantities over nodes and edges across an ensemble of networks rather than individual nodes and edges in a specific network. To this end, we derive expressions for the configuration model network ensemble, where the degree distribution $p_k$ is the only determinant of the network structure. We take the same approach as the previous subsection: we first derive a generic formulation and then apply it to the two immunity scenarios.

Since the configuration model generates maximally random connection patterns for a given degree distribution, we can reasonably assume two things: (i) the probability that a node with degree $k$ is in the residual subgraph only depends on $k$ (which we represent by $\Psi_k$), and (ii) the probability that a random edge connects to the giant component or to an infected node is constant for any edge.
These two assumptions allow us to take the average of Eqs.~\eqref{eq:giant_residual_receive_message} and \eqref{eq:giant_residual_transmit_message}. Representing the probability that a neighbor of a residual node is also in the residual subgraph by $g$, we have
\begin{equation}
\lambda = \frac{\sum_k p_k \Psi_k (1 - g + g \tilde{\lambda})^k}{\sum_k p_k \Psi_k}
\label{eq:giant_residual_receive_message_config}
\end{equation}
and 
\begin{equation}
\tilde{\lambda} = \frac{\sum_k q_k \Psi_k (1 - g + g \tilde{\lambda})^k}{\sum_k q_k \Psi_k}.
\label{eq:giant_residual_transmit_message_config}
\end{equation}
Here, $\lambda$ denotes the probability that a random node in the residual subgraph is not in the giant residual component. Similarly, $\tilde{\lambda}$ is the probability that a random neighbor of the focal node, given that the neighbor is in the residual subgraph, would not be in the giant residual component if it were not adjacent to the focal node. By solving Eq.~\eqref{eq:giant_residual_transmit_message_config} recursively, we can compute relevant quantities through averaged versions of Eq.~\eqref{eq:immune_frac} and \eqref{eq:giant_residual_frac}:
\begin{gather}
\pi_\mathrm{R} = 1 - \psi, \\
\pi_\mathrm{S_G} = \psi (1 - \lambda), \label {eq:giant_residual_frac_config}
\end{gather}
where $\psi = \sum_k p_k \Psi_k$ is the probability that a random node is in the residual subgraph (i.e., not immune). The message passing equations for the post-immunity epidemic, involving $\omega$ and $\tilde{\omega}$, can be derived in a similar manner by replacing $g$ with $T'g$ in Eqs.~\eqref{eq:giant_residual_receive_message_config} and \eqref{eq:giant_residual_transmit_message_config}. It can then inform the post-immunity epidemic size:
\begin{equation}
\pi_\mathrm{R_2} = \psi (1 - \omega).
\end{equation}

To apply these generic expressions to a specific immunity scenario, we need to identify $\Psi_k$ and $g$. For later convenience, we introduce the probability generating function (PGF) for the degree distribution $p_k$ as $F_0(x) = \sum_k p_k x^k$. Similarly, the PGF for the excess degree distribution, defined as
\begin{equation}
q_k = \frac{(k + 1) p_{k + 1}}{\langle k \rangle}, 
\end{equation}
is denoted by $F_1(x) = \sum_k q_k x^k$.

\paragraph*{Disease-induced immunity.} Let $u$ denote the probability that an edge is not transmissible in an immunity-inducing epidemic with transmissibility $T$. We have 
\begin{equation}
u = 1 - T + T \tilde{\psi},
\end{equation}
where $\tilde{\psi}$ is the average probability that a neighbor of a focal node would not be infected by the immunity-inducing epidemic if it was not connected to the focal node. The probability that a node is not infected through any of the $k$ incident edges and thus remains susceptible is given by $\Psi_k = u^k$, which allows us to write the configuration model analog of  Eqs.~\eqref{eq:first_epidemic_receive_message} and \eqref{eq:first_epidemic_transmit_message} using PGFs:
\begin{equation}
\psi = F_0(u)
\label{eq:first_epidemic_receive_message_config}
\end{equation}
and 
\begin{equation}
\tilde{\psi} = F_1(u).
\label{eq:first_epidemic_transmit_message_config}
\end{equation}
Eq.~\eqref{eq:first_epidemic_transmit_message_config} is a self-consistent equation for $\tilde{\psi}$ and can numerically be solved to compute the averaged version of Eq.~\eqref{eq:cond_prob_residual_DII}:
\begin{equation}
g = \frac{\tilde{\psi}}{u} = \frac{F_1(u)}{u}.
\label{eq:cond_prob_residual_DII_config}
\end{equation}
By substituting Eq.~\eqref{eq:cond_prob_residual_DII_config} to Eqs.~\eqref{eq:giant_residual_receive_message_config} and \eqref{eq:giant_residual_transmit_message_config}, we arrive at 
\begin{gather}
\lambda = \frac{F_0(u - F_1(u)(1 - \tilde{\lambda}))}{F_0(u)}, \\
\tilde{\lambda} = \frac{F_1(u - F_1(u)(1 - \tilde{\lambda}))}{F_1(u)}.    
\end{gather}
This is consistent with formulas derived by Newman~\cite{newman2005threshold}. Likewise, the averaged message passing equations for the post-immunity epidemic are given by
\begin{gather}
\omega = \frac{F_0(u - T'F_1(u)(1 - \tilde{\omega}))}{F_0(u)}, \\
\tilde{\omega} = \frac{F_1(u - T'F_1(u)(1 - \tilde{\omega}))}{F_1(u)}.    
\end{gather}

This framework allows us to count the nodes in each state as well as the edges between nodes in each state. For example, a node with degree $k$ is immune after an epidemic with probability $1 - u^k$, from which we obtain the expected degree of an immune node as
\begin{equation}
\begin{aligned}
\langle k \rangle_\mathrm{R} &= \frac{\sum_k k p_k (1 - u^k)}{1 - F_0(u)} \\
&= \frac{F_0'(1) - u F_0'(u)}{1 - F_0(u)}.
\end{aligned}
\end{equation}
Another quantity of interest is the number of edges at the susceptible-immune interface. For this, we calculate the probability that a random edge connects a susceptible node and an immune node, given by
\begin{equation}
\begin{aligned}
\rho_\mathrm{SR} &= 2 F_1(u) (1 - F_1(u)) (1 - T) \\
&= 2 F_1(u) (u - F_1(u)).
\end{aligned}
\end{equation}
Here, $F_1(u)$ and $1 - F_1(u)$ represent the probability that a node would be susceptible or immune, respectively, after the immunity-inducing epidemic if not for the edge in question, $1 - T$ is the probability of no transmission on the edge, and the factor $2$ accounts for the arbitrariness in choosing the susceptible end node. 

\paragraph*{Random immunization} The probabilities for random immunization can also be derived as an averaged version of the message-passing equations. In this case, we simply have $\Psi_k = g = \psi = 1 - \pi_\mathrm{R}$, leading to 
\begin{gather}
\lambda = F_0(1 - (1 - \pi_\mathrm{R})(1 - \tilde{\lambda})), \\
\tilde{\lambda} = F_1(1 - (1 - \pi_\mathrm{R})(1 - \tilde{\lambda})), \\
\omega = F_0(1 - T'(1 - \pi_\mathrm{R})(1 - \tilde{\omega})), \\
\tilde{\omega} = F_1(1 - T'(1 - \pi_\mathrm{R})(1 - \tilde{\omega})).
\end{gather}
The number of edges at the interface can be easily calculated:
\begin{equation}
\rho_\mathrm{SR} = 2 \pi_\mathrm{R} (1 - \pi_\mathrm{R})
\end{equation}

\paragraph*{Proportional immunization} In the manuscript, we also consider proportional immunization, where the degree distribution of immune nodes is exactly the same as the disease-induced case, but otherwise they are chosen randomly in the network. In this case, the probability that a node with degree $k$ is in the residual subgraph is given by $\Psi_k = u^k$, and the probability that a neighbor of a node is in the residual subgraph (regardless of the node being in the residual subgraph or not) is given by
\begin{equation}
g = \frac{\sum_k k p_k u^k}{\langle k \rangle} = u F_1(u).
\end{equation}
Note the difference from Eq.~\eqref{eq:cond_prob_residual_DII}. We obtain the averaged message-passing equations as
\begin{gather}
\lambda = \frac{F_0(u - u^2 F_1(u)(1 - \tilde{\lambda}))}{F_0(u)}, \\
\tilde{\lambda} = \frac{F_1(u - u^2 F_1(u)(1 - \tilde{\lambda}))}{F_1(u)}.
\end{gather}
Since in this case the dynamical states of neighbor nodes are not correlated, the number of edges at the interface is given by
\begin{equation}
\rho_\mathrm{SR} = u F_1(u) (1 - u F_1(u)).
\end{equation}

\section{Proof for equally strong herd immunity after natural infection and random immunization in Erd\H{o}s-R\'{e}nyi graphs}
\label{section:proof}
Based on Eq.~\eqref{eq:giant_residual_frac_config}, we see that the necessary and sufficient condition for equal-sized disease-induced immunity and random immunization to induce equally strong structural herd immunity in a configuration model network is that $\lambda$ is equal in both scenarios. For $\lambda$ to be always equal, $\tilde{\lambda}$ must also be equal and the PGFs of the degree distribution must satisfy, for any $u$ and $x$ in $[0, 1]$, 
\begin{gather}
F_0(u - x F_1(u)) = F_0(u) F_0(1 - x F_0 (u)), 
\label{eq:equality_condition_f0} \\
F_1(u - x F_1(u)) = F_1(u) F_1(1 - x F_0 (u)).
\label{eq:equality_condition_f1}
\end{gather}
Our goal is to identify the degree distribution generated by $F_0$ that satisfies these conditions. For that, we expand the two sides of Eq.~\eqref{eq:equality_condition_f0} as
\begin{equation}
\begin{aligned}
\mathrm{lhs}
&= \sum_k p_k [u - x F_1(u)]^k \\
&= \sum_m (-x)^m \left[\frac{F_1(u)}{u}\right]^m \sum_{k=m}^\infty \binom{k}{m} p_k u^k,
\end{aligned}  
\end{equation}
and 
\begin{equation}
\begin{aligned}
\mathrm{rhs} 
&= F_0(u) \sum_k p_k [1 - x F_0(u)]^k \\
&= \sum_m (-x)^m \left[F_0 (u)\right]^{m+1} \sum_{k=m}^\infty \binom{k}{m} p_k.
\end{aligned}
\end{equation}
Seeing each side as a polynomial function of $-x$, it follows from the polynomial identity that the corresponding coefficients must be equal, that is
\begin{equation}
\left[\frac{F_1(u)}{u}\right]^m \sum_{k=m}^\infty \binom{k}{m} p_k u^k \\
= \left[F_0 (u)\right]^{m+1} \sum_{k=m}^\infty \binom{k}{m} p_k
\label{eq:polynomial_identity}
\end{equation}
for any $m$. For $m=0$, Eq.~\eqref{eq:polynomial_identity} is an identity; both sides equals $F_0(u)$. For $m=1$, the lhs of Eq.~\eqref{eq:polynomial_identity} is equal to $F_1(u) {F_0}'(u)$ while the rhs is equal to $[F_0(u)]^2 {F_0}'(1)$. Equating the two sides and using $F_1(u) = {F_0}'(u) / {F_0}'(1)$, we get
\begin{equation}
F_0(u) = F_1(u) = {F_0}'(u) / \langle k \rangle,
\end{equation}
where $\langle k \rangle = {F_0}'(1)$ is the mean degree. The only case where this relation holds for any $u$ is when 
\begin{equation}
F_0(u) = \exp[(u - 1) \langle k \rangle], 
\end{equation}
that is, the degree distribution is Poisson. In fact, one can easily show that this generating function satisfies both Eqs.~\eqref{eq:equality_condition_f0} and \eqref{eq:equality_condition_f1}. Therefore, the Poisson distribution is the only degree distribution of the configuration model that results in equally strong structural herd immunity in the two immunity scenarios. Following the same steps for $\omega$, we can show that herd immunity characterized by the post-immunity epidemic size is equally strong for the two scenarios only when the degree distribution is Poisson.

\section{Spatiality and localization}
\label{section:spatiality_localization}
\begin{figure*}[tb]
\centering
\includegraphics[width=0.75\linewidth]{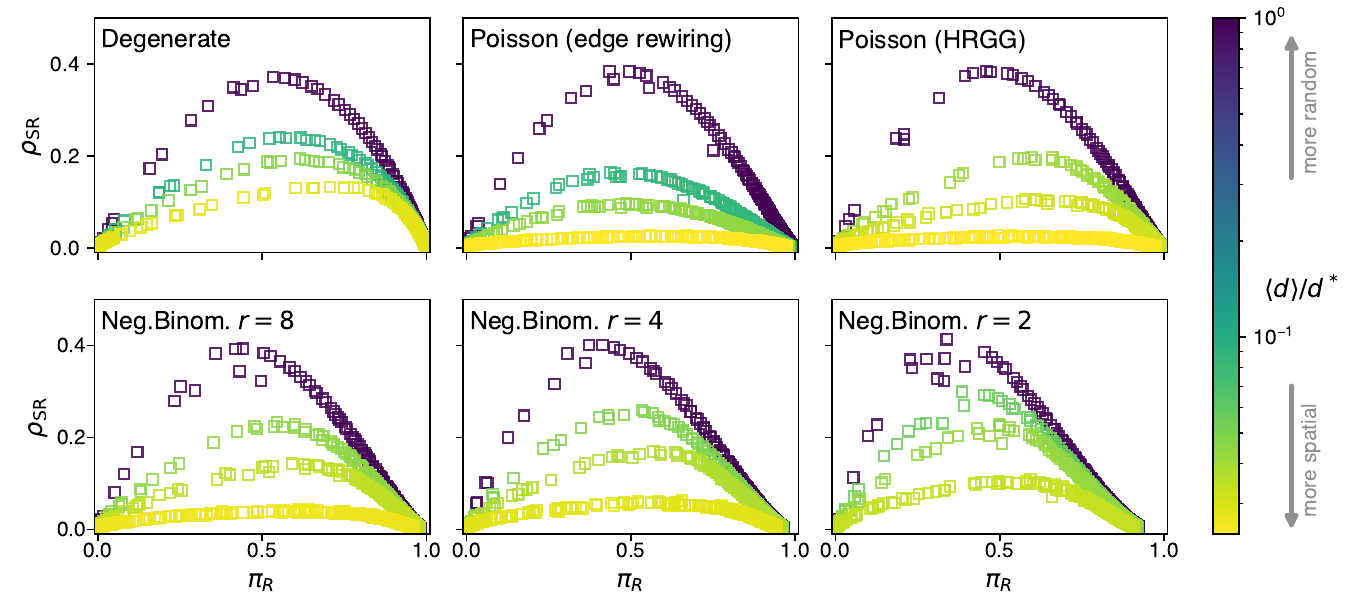}
\caption{\textbf{Interface size as a function of the immune fraction in networks with varying degrees of spatiality.} Spatiality is quantified by normalized mean edge length $\langle d \rangle / d^*$, which is represented by different colors. }
\label{fig:rho_sr}
\end{figure*}

In the main text, we argue that the spatiality of the contact network enhances the localization of immunity. Fig.~3B shows that the maximum interface size $\max(\rho_\mathrm{SR})$ as a function of immune fraction $\pi_\mathrm{R}$ decreases as the network becomes more spatial and the normalized mean edge length $\langle d \rangle / d^*$ decreases. To further support this claim, Fig.~\ref{fig:rho_sr} shows the full functional form of $\rho_\mathrm{SR}(\pi_\mathrm{R})$ for networks with different values of $\langle d \rangle / d^*$. For each degree distribution, we observe that a smaller mean edge length results in a smaller interface size for all values of $\pi_\mathrm{R}$, not just its maximum. This implies that the trend observed with the maximum interface size in Fig.~3B can be generalized to the interface size at any value of $\pi_\mathrm{R}$.

\section{Difference in structural herd immunity thresholds}
\label{section:diff_structural_hit}
\begin{figure}[tb]
\centering
\includegraphics[width=\linewidth]{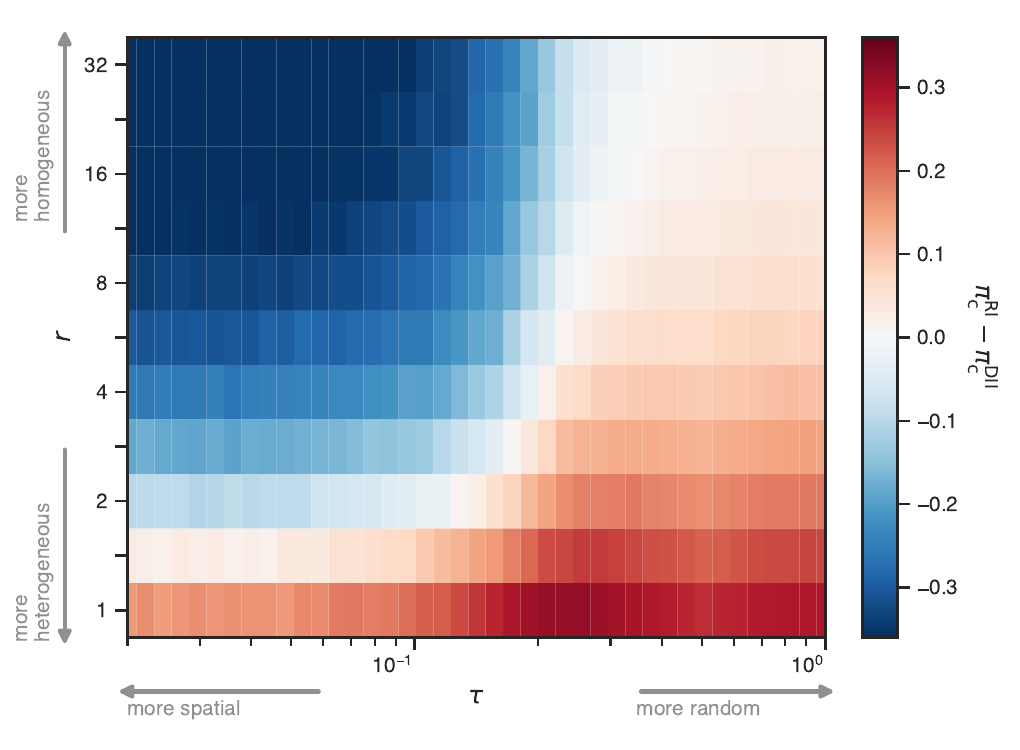}
\caption{\textbf{Difference between the herd immunity thresholds as a function of degree heterogeneity and spatiality of the contact network.} Contact networks are generated by the HRGG model with negative binomial degree distributions, parameterized by temperature $\tau$ (horizontal axis) and dispersion parameter $r$ (vertical axis). Colors represent the difference between the herd immunity thresholds of random immunization and disease-induced immunity, $\pi_\mathrm{c}^\mathrm{RI}-\pi_\mathrm{c}^\mathrm{DII}$. Blue regions represent $\pi_\mathrm{c}^\mathrm{DII} > \pi_\mathrm{c}^\mathrm{RI}$, while red regions represent $\pi_\mathrm{c}^\mathrm{DII} < \pi_\mathrm{c}^\mathrm{RI}$. }
\label{fig:diff_thr_nbinom}
\end{figure}

We present in Fig.~\ref{fig:diff_thr_nbinom} a systematic examination of the difference between the thresholds $\pi_\mathrm{c}$ after disease-induced immunity and random immunization for heterogeneous random geometric graphs with negative binomial distributions. As we vary the dispersion parameter $r$ (smaller $r$ means degrees are more heterogeneous) and the temperature parameter $\tau$ (smaller $\tau$ means contacts are more spatial), we find three regions: When node degrees are more heterogeneous and the network is more random (i.e., small $r$ and large $\tau$), disease-induced immunity has a lower threshold ($\pi_\mathrm{c}^\mathrm{DII} < \pi_\mathrm{c}^\mathrm{RI}$); when the degrees are more homogeneous and the network is more spatial (i.e., large $r$ and small $\tau$), the threshold for random immunization is lower ($\pi_\mathrm{c}^\mathrm{DII} > \pi_\mathrm{c}^\mathrm{RI}$); when the network is random and homogeneous (large $r$ and $\tau$), there is no significant difference between the two thresholds ($\pi_\mathrm{c}^\mathrm{DII} \simeq \pi_\mathrm{c}^\mathrm{RI}$). This further corroborates our findings that degree heterogeneity and spatiality are competing structural factors of the contact network, the tradeoff between which determines which of the two immunity scenarios induces stronger structural herd immunity.

\section{Post-immunity epidemic sizes}
\label{section:post_immunity_episize}
In the main text, we discuss the strength of herd immunity in network models by quantifying it using the size of the giant residual component. Because the giant residual component acts as the substrate on which a post-immunity epidemic can unfold, it represents the population still at risk of a disease of unknown transmissibility after immunity is induced for a fraction of individuals.

If the transmissibility of the disease is known, we can quantify herd immunity by the expected size $\pi_{\mathrm{R}_2}$ of the post-immunity epidemic, as it is typically done in the literature. The post-immunity epidemic size is obtained numerically by running epidemic simulations on the residual subgraph using the same method described in \ref{section:epn}. For configuration model networks, we can also compute it analytically as described in \ref{section:message_passing}. 

\begin{figure*}[tb]
\centering
\includegraphics[width=\linewidth]{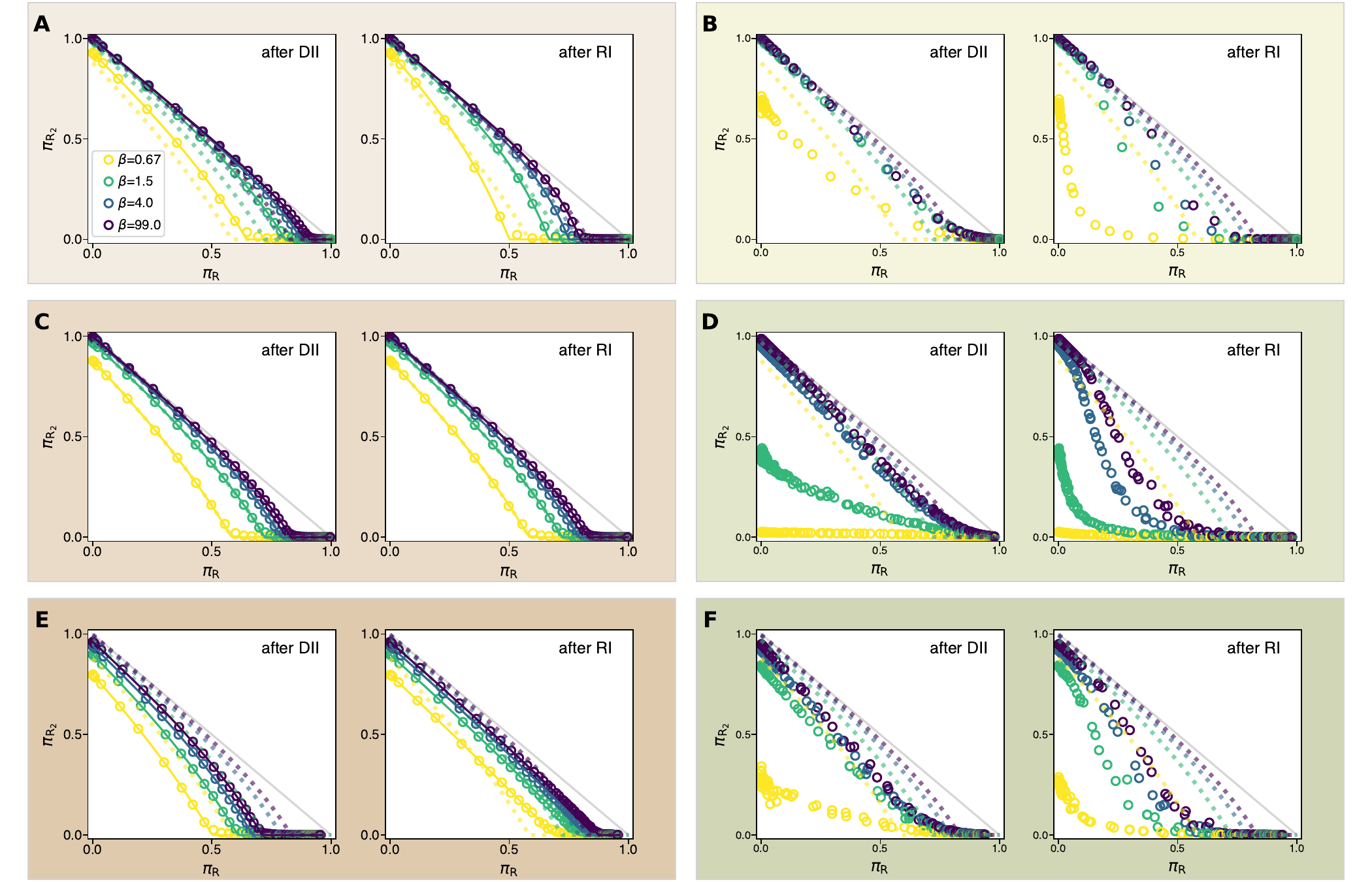}
\caption{\textbf{Post-immunity epidemic size $\pi_{\mathrm{R}_2}$ as a function of immune fraction $\pi_\mathrm{R}$.} Transmission rate $\beta$ is represented by different colors. Solid lines (analytical) and symbols (numerical) denotes the results of the network model, while the results based on the homogeneously mixed mean-field model are shown in dotted lines as a reference. The network structures are (A) random regular graphs (B) regular lattices, (C) Erd\H{o}s-R\'{e}nyi graphs, (D) random geometric graphs, (E) configuration model networks with a negative binomial degree distribution, and (F) heterogeneous random geometric graphs with a negative binomial degree distribution and temperature parameter $\tau = 0.05$. All networks are of size $N = 10^4$ and have average degree $\langle k \rangle = 6$. The negative binomial degree distribution in (E) and (F) are parameterized by dispersion parameter $r= 3$. }
\label{fig:second_episizes_vs_immune}
\end{figure*}

Figure~\ref{fig:second_episizes_vs_immune} shows post-immunity epidemic size $\pi_{\mathrm{R}_2}$ as a function of immune fraction $\pi_\mathrm{R}$. The analytical calculations of epidemic size for configuration model networks show good agreement with the simulation results for different values of transmissibility. 
As a simple illustration of the model dependence of herd immunity estimates, we compare these results to a mean-field epidemic model that assumes homogeneous mixing. In the homogeneous mean-field model, the epidemic size in a population where a fraction $\pi_\mathrm{R}$ is immune is given by $\pi_{\mathrm{R}_2} = (1 - \pi_\mathrm{R}) z$, where $z$ is the solution of the following implicit assumption~\cite{kermack1927contribution}:
\begin{equation}
z = 1 - e^{- R_0 (1 - \pi_\mathrm{R}) z}, 
\end{equation}
where $R_0$ is the basic reproduction number of the post-immunity epidemic. For given values of transmission rate $\beta$ and recovery rate $\gamma$ (which is set to one without loss of generality) for the network model, the corresponding value of $R_0$ for the mean-field model is chosen to be $R_0 = \beta \langle k \rangle / (\beta + \gamma)$.

\begin{figure}[tb]
\centering
\includegraphics[width=\linewidth]{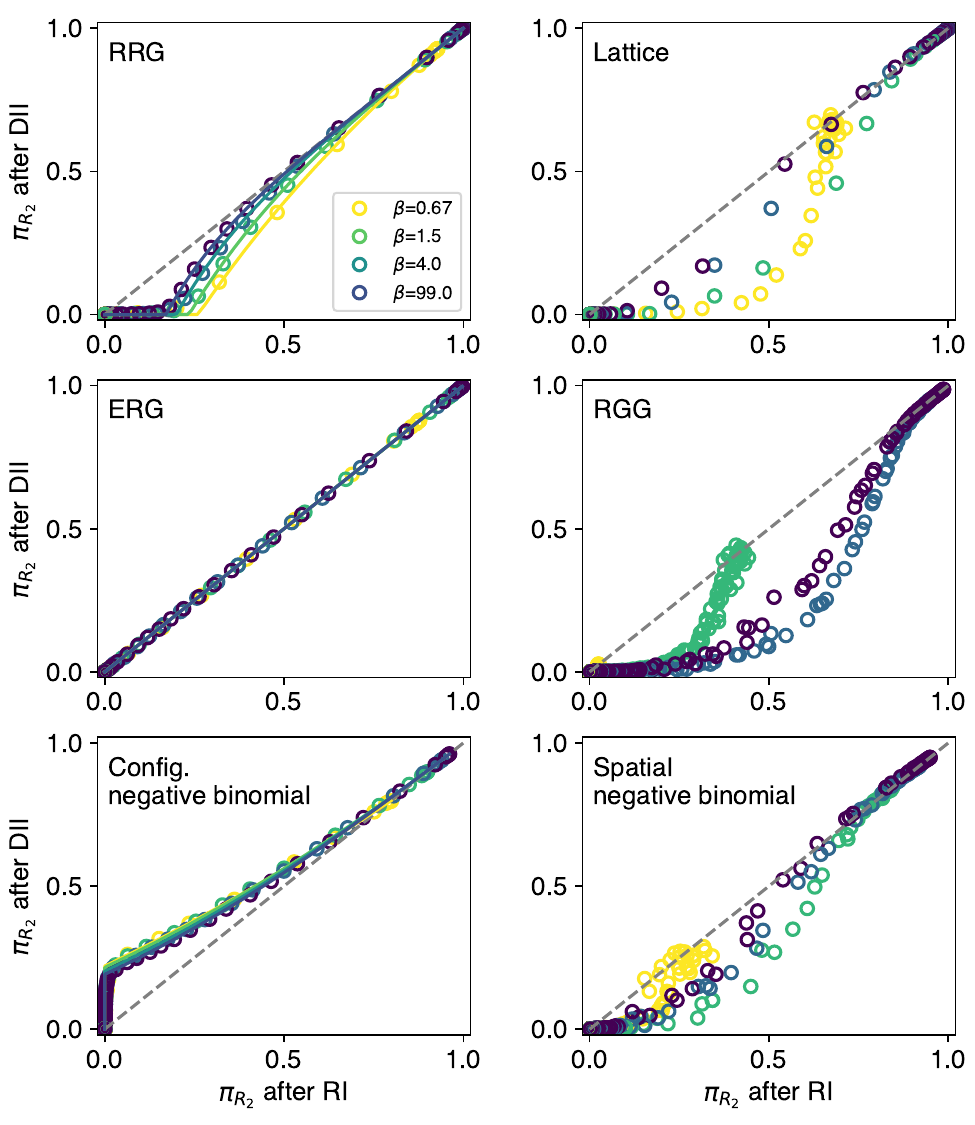}
\caption{\textbf{Comparison of post-immunity epidemic sizes $\pi_{\mathrm{R}_2}$ after disease-induced immunity and after random immunization in different networks.} Each data point corresponds to a pair of the fraction $\pi_\mathrm{R}$ of immune individuals and transmission rate $\beta$ of the post-immunity epidemic. Solid lines denotes analytical calculations. Each contact network has $N = 10^4$ nodes with average degrees $\langle k\rangle = 6$ for all models. The negative binomial degree distribution is parameterized by dispersion parameter $r = 3$. Spatial negative binomial networks are generated with $\tau = 0.05$. }
\label{fig:second_episize_dii_vs_ri}
\end{figure}

Figure~\ref{fig:second_episize_dii_vs_ri} shows the comparison between $\pi_{\mathrm{R}_2}$ after random immunization and after disease-induced immunity for different network structures. We find that they are remarkably consistent with the conclusions from the results for structural herd immunity. In random regular graphs, random immunization results in smaller post-immunity epidemics, and the difference between the two scenarios is more prominent for less infectious pathogens (smaller $\beta$). In Erd\H{o}s-R\'{e}nyi graphs, the two scenarios lead to exactly the same post-immunity epidemic sizes as long as immune proportion $\pi_\mathrm{R}$ and the transmission rate $\beta$ are equal. When the degrees of the contact network are more heterogeneously distributed with a negative binomial distribution, the disease-induced immunity proves to be more effective in creating herd immunity. In all configuration model networks, the analytical method provides an accurate prediction of the numerical results. For spatial networks, random immunization consistently leads to a smaller value of $\pi_{\mathrm{R}_2}$, inducing stronger herd immunity. This is the case even in degree-heterogeneous networks, as illustrated by the heterogeneous random geometric graph with a negative binomial degree distribution. 

\section{Robustness of the results}
\label{section:size_degree_dependence}
In this section, we provide a sensitivity analysis of our results presented in the main text by checking its dependence on the population size and average degree.

\subsection*{Size dependence of the synthetic network model results}
\begin{figure*}[tb]
\centering
\includegraphics[width=0.75\linewidth]{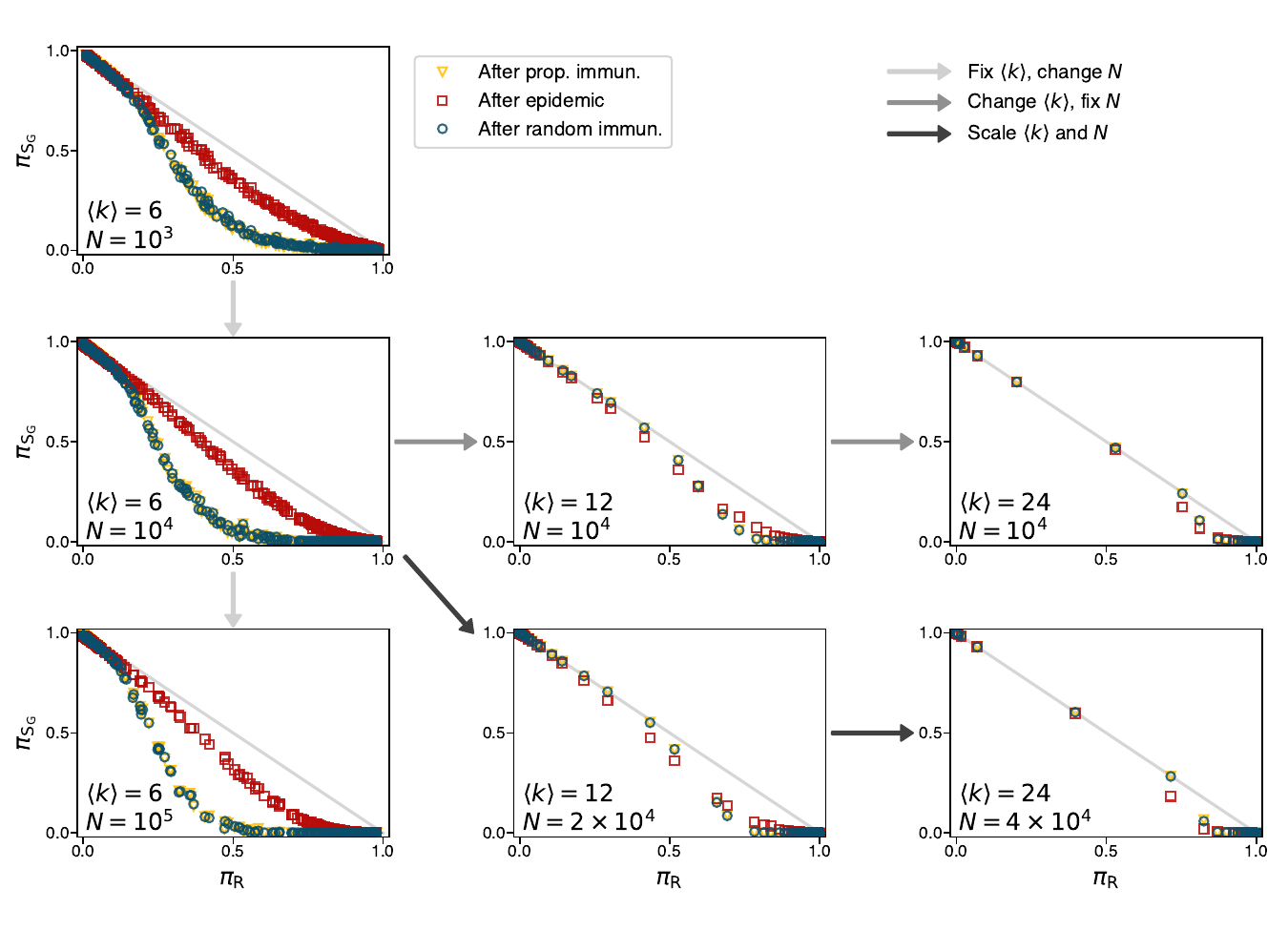}
\caption{\textbf{Size and degree dependence of structural herd immunity in random geometric graphs.} Light gray arrows represent changing the number of nodes $N$ with a fixed average degree $\langle k\rangle = 6$. Medium gray arrows highlight changing the average degree $\langle k\rangle$ for networks of fixed size $N=10^4$. Dark gray arrows shows the case where the number of nodes is scaled linearly with the average degree. Each panel shows the size $\pi_\mathrm{S_G}$ of the giant residual component for $\pi_\mathrm{R}$ the size of the immunity-inducing epidemic. The colors of the symbols indicate natural infection (red), random immunization (blue), and proportional immunization (yellow).}
\label{fig:rgg_change_degree_size}
\end{figure*}

In the main text, the analytical calculations have been performed by assuming networks of infinite size; nevertheless, the results are in good agreement with our simulations on finite networks of $N = 10^4$. As we only have analytical solutions for the configuration model, we have explored the effects of network size on the residual giant component size in random geometric graphs (RGGs) as shown in Fig.~\ref{fig:rgg_change_degree_size}. We see that when RGGs grow in size, the transition at the onset of a giant residual component becomes sharper in general, implying finite-size effects. However, the finite-size effects are more pronounced for random immunization than for disease-induced immunity. Consequently, the differences between the structural herd immunity between the two immunity scenarios become even larger for sparse networks. 

\begin{figure}[tb]
\centering
\includegraphics[width=\linewidth]{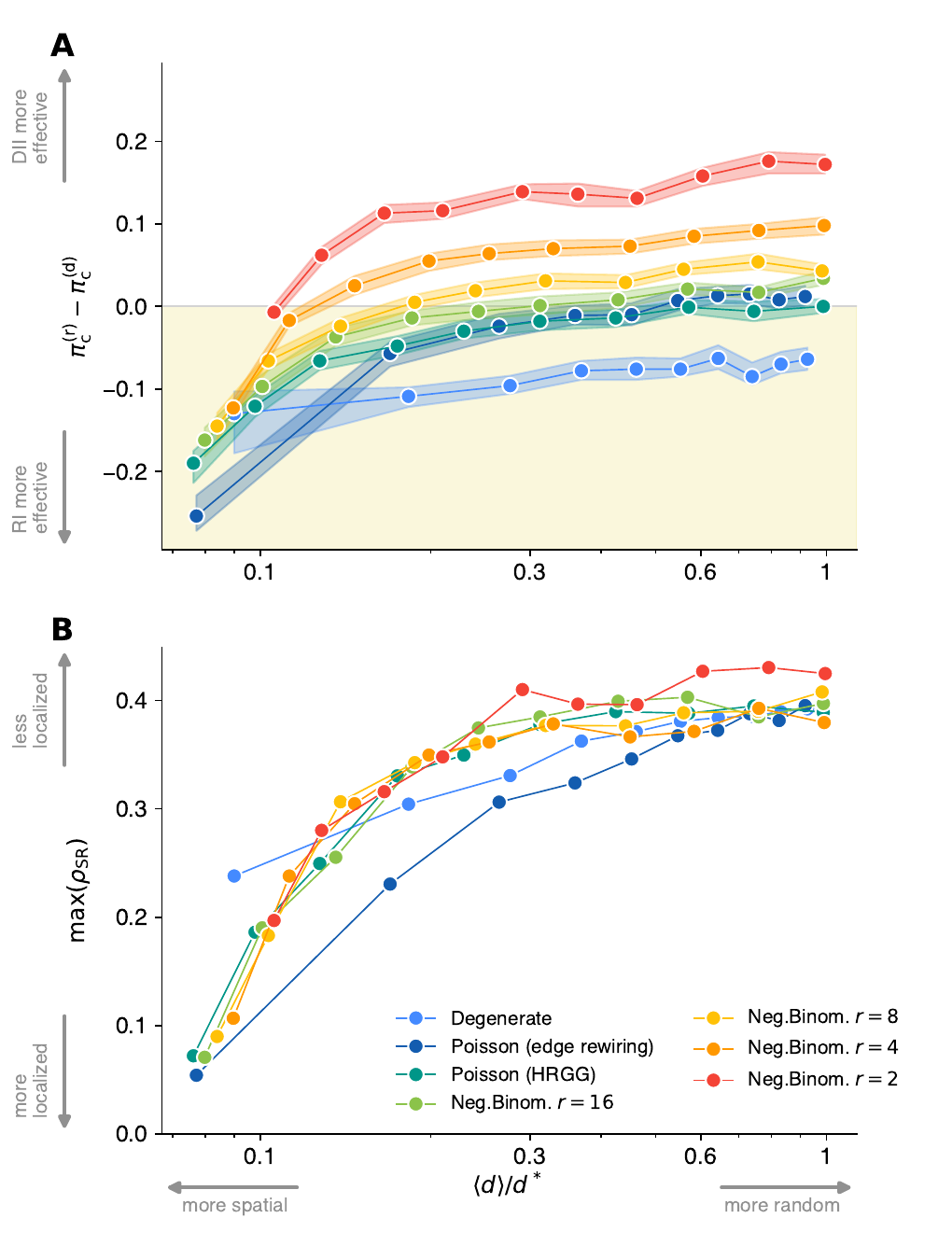}
\caption{\textbf{Disease-induced herd immunity as a function of spatiality in networks of size $N=10^3$.} Spatiality is measured by the normalized mean edge length $\langle d \rangle / d^*$. \textbf{(A)} Difference in the structural herd immunity thresholds $\pi_\mathrm{c}^\mathrm{RI} - \pi_\mathrm{c}^\mathrm{DII}$. Different degree distributions are indicated by line colors. The shaded area along each line represents the $95\%$ confidence interval. \textbf{(B)} The maximum (peak value) of $\rho_\mathrm{SR}$ as a function of $\pi_\mathrm{R}$ for disease-induced immunity.}
\label{fig:difference_thr}
\end{figure}

We also check the robustness of our results on the difference between structural herd immunity thresholds (Fig.~3 in the main text) against population size. Figure~\ref{fig:difference_thr} shows that the results are qualitatively unchanged by changing the system size by an order of magnitude.

\subsection*{Dependence on average degree}
\begin{figure*}[tb]
\centering
\includegraphics[width=0.75\linewidth]{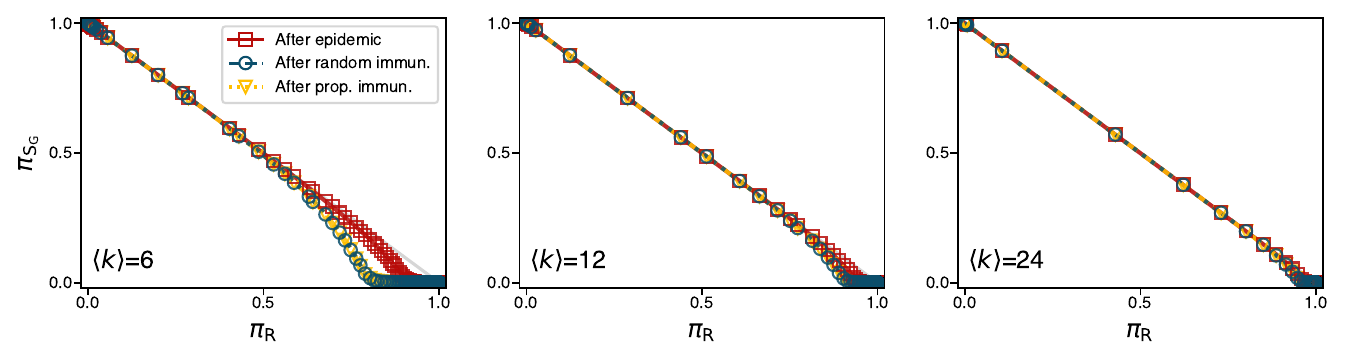}
\caption{\textbf{Comparison of herd immunity effects in random regular graphs.} Each contact network has $N = 10^4$ nodes with average degrees $\langle k\rangle = 6, 12, 24$. Each panel shows the size $\pi_\mathrm{S_G}$ of the giant residual component for $\pi_\mathrm{R}$ the size of the immunity-inducing epidemic. The colors of the symbols indicate natural infection (red), random immunization (blue), and proportional immunization (yellow). The symbols represent numerical results, and the lines indicate theoretical predictions.}
\label{fig:rrg_change_degree}
\end{figure*}

In the main text, all synthetic network models have the same average degree $\langle k\rangle=6$. Here we repeat the same analysis of structural herd immunity with denser networks where the average degrees are $\langle k\rangle=12$ and $\langle k\rangle=24$. Generally, increasing the average degree of the contact network leads to a larger giant residual component, i.e., it diminishes the effect of structural herd immunity. 
This is seen for random regular graphs (Fig.~\ref{fig:rrg_change_degree}) and for random geometric graphs (Fig.~\ref{fig:rgg_change_degree_size}), the two types of networks where the effect of immunity localization is most prominent. For denser networks, we observe that all the curves that represent different immunity scenarios are pushed close to the diagonal, where no indirect protection is provided. In this regime, how immunity is induced becomes irrelevant, since random, proportional, and disease-induced immunity all provide only a negligible amount of structural herd immunity, in particular for random regular graphs. In other words, the network model becomes similar to a mean-field model where every individual interacts with all other individuals. 

We also see that the curves for disease-induced immunity are only weakly affected, while the curves for random and proportional immunization are significantly shifted. For random immunization, where immunity is uniformly distributed, a larger average degree means an increasing number of paths connecting different regions of the network, making the network more robust to node removal (immunization). In contrast, because disease-induced immunity is highly localized, there is a clear separation between immune and susceptible regions. Susceptible regions are well connected even for $\langle k \rangle = 6$; additional edges in a denser network are largely redundant for the connectivity of the residual subgraph and contribute little to its improvement.

As noted above, the agreement between analytical and numerical results for configuration model networks implies that the effect of average degree on herd immunity outcomes is independent of population size. However, the lack of an analytical method prompts us to investigate whether the same is true for spatial networks. In Fig.~\ref{fig:rgg_change_degree_size} we show the case where the network size and the average degree are scaled linearly, i.e., the network density is held constant. Interestingly, the results are almost identical to the case where only the average degree is altered while the network size is fixed. This suggests that structural herd immunity in an RRG is largely determined by local connectivity rather than global connectivity.

We remark that the degrees of this static network model should not be interpreted literally as the number of contacts individuals have, but rather as an effective constraint acting on the pathways of the disease. In reality, contact frequencies are very heterogeneous, with, e.g., many short, recurrent contacts with family members and random, one-off encounters with strangers in public places; for model parameters to have a realistic interpretation, these heterogeneities should be incorporated. 

\subsection*{Finite size effects in the network model of the Finnish population}

\begin{figure}[tb]
\centering
\includegraphics[width=\linewidth]{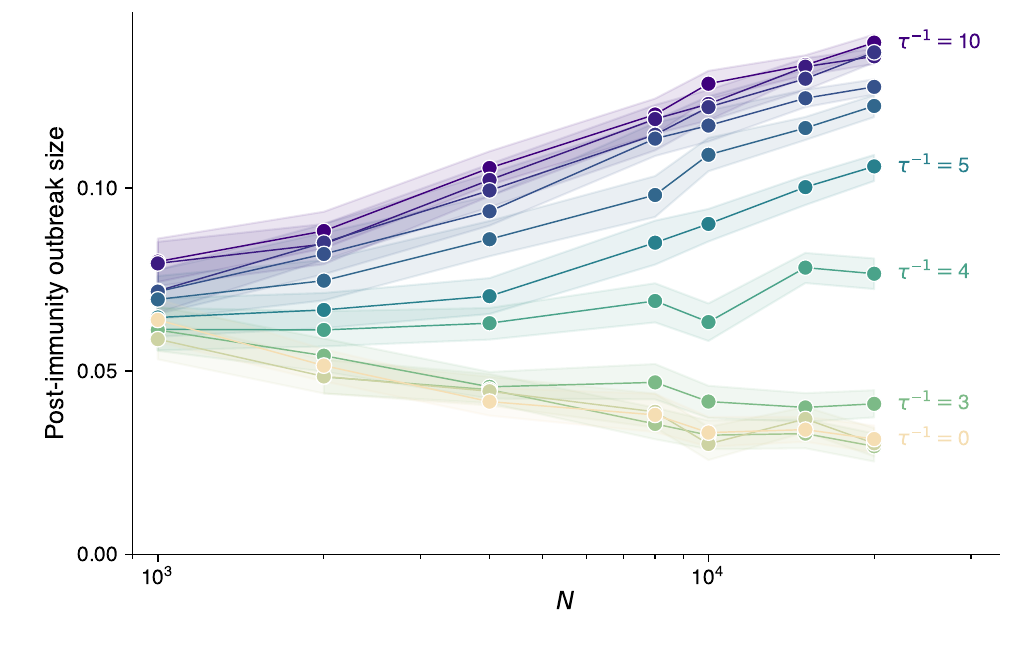}
\caption{\textbf{Effect of the population size on the size of the outbreak in a population that reached the mean-field herd immunity threshold.} The shaded areas represent the 95\% confidence interval. }
\label{fig:Finland_finite_size}
\end{figure}

To test the robustness of our claim that the mean-field model estimates the herd immunity threshold lower than the network model, we check how the results of epidemic simulations based on the population of Finland depend on the network size $N$. Figure~\ref{fig:Finland_finite_size} shows outbreak sizes in spatial SBM networks after immunity is induced by an epidemic terminated at the mean-field herd immunity threshold. For networks with inverse temperature $\tau^{-1} = 0$ (no spatiality, pure SBM), the post-immunity outbreak becomes smaller with increasing population. This could imply that the nonzero outbreak size after reaching the mean-field threshold is due to finite size effects, and the difference between the two model estimates will disappear in the limit of infinite population. However, the decrease is relatively slow; in a realistically large population, we can still expect that the post-immunity outbreak to be significantly large. For larger values of $\tau^{-1}$, the decrease in outbreak size as a function of $N$ becomes more gradual, and, remarkably, the slopes are positive for $\tau^{-1} \geq 5$. This implies that in a spatial network, the outbreak size increases as the population grows larger. 

\section{Herd immunity threshold for the network model of Finnish population}
\label{section:network_thr}
We compare the disease-induced herd immunity threshold of the mean-field model and the network (spatial SBM) model of the age-stratified population of Finland. We first let an immunity-inducing epidemic of $R_0 = 3$ evolve in a fully susceptible spatial SBM network of $N = 20000$ nodes, starting from a random initially infected node, until it spreads to a $\pi_\mathrm{R}$ fraction of the population. After terminating the infection process and allowing all infected individuals to recover, we compute three quantities: (i) the expected size $\pi_\mathrm{R_2}$ of a large post-immunity epidemic induced by reintroducing the pathogen with the same value of $R_0$; (ii) the variance-to-mean ratio of $\pi_\mathrm{R_2}$, usually referred to as \emph{susceptibility} in statistical physics ~\cite{colomer-de-simon2014double, radicchi2015predicting, hebert-dufresne2019smeared} and denoted by $\chi$; (iii) effective reproduction number $R_\mathrm{eff} = R_0 \Lambda\big(\operatorname{diag}(\mathbf{s}) M\big) / \Lambda(M)$, where notations are as defined in the main text. Each quantity is used for a different definition of the herd immunity threshold in terms of the value of $\pi_\mathrm{R}$: (i) the post-immunity epidemic spreads to 1\% of the population, i.e., $\pi_\mathrm{R_2} = 0.01$; (ii) the peak of susceptibility $\chi$; (iii) the effective reproduction number equals one, i.e., $R_\mathrm{eff} = 1$. The first two definitions represent the threshold of the network model while the third definition amounts to the threshold of the mean-field model. 

\begin{figure}[tb]
\centering
\includegraphics[width=\linewidth]{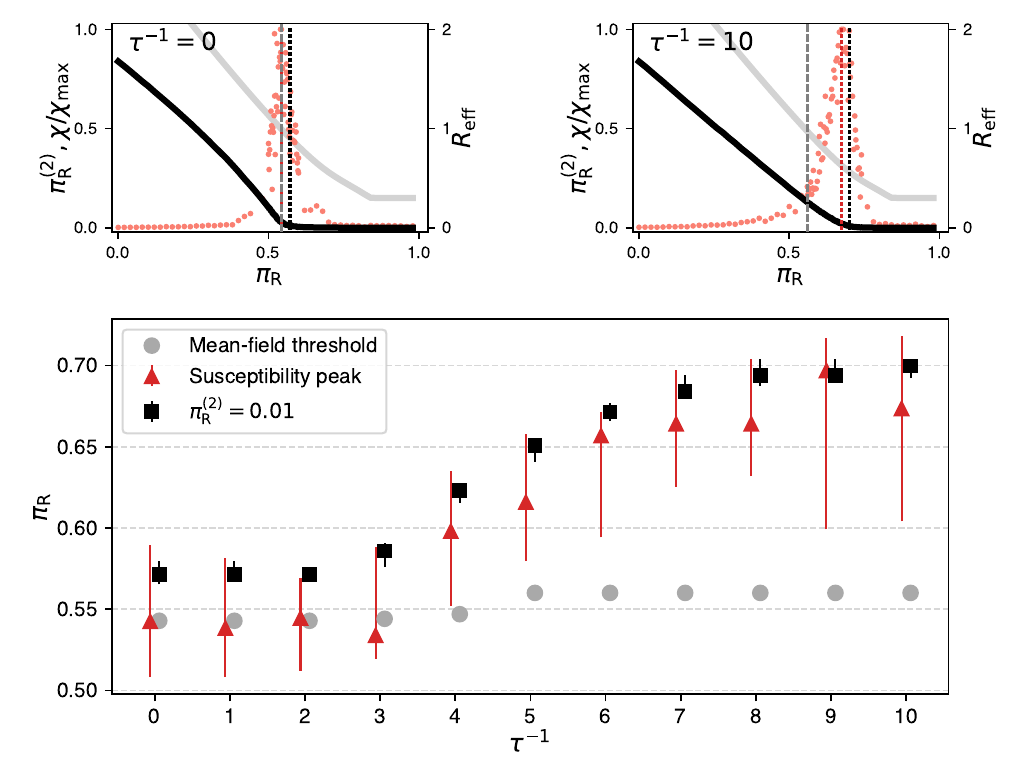}
\caption{\textbf{Comparison of the disease-induced herd immunity thresholds of the mean-field and network models of the Finnish population.}
Top: The post-immunity epidemic size $\pi_\mathrm{R_2}$ (black solid line), susceptibility normalized by its maximum value $\chi / \chi_{\max}$ (light red points), and effective reproduction number $R_\mathrm{eff}$ (gray solid line) as a function of immunity-inducing epidemic size $\pi_\mathrm{R}$. 
The vertical lines indicate the herd immunity thresholds using the three definitions: $\pi_\mathrm{R_2} = 0.01$ (black dotted line), maximum of $\chi$ (red dotted line), and $R_\mathrm{eff} = 1$ (dashed gray line). 
The network has $N = 20000$ nodes and the inverse temperature of the spatial SBM is set to $\tau^{-1} = 0$ (left panel) and $\tau^{-1} = 10$ (right panel).
Bottom: Herd immunity thresholds computed by the three definitions: $\pi_\mathrm{R_2} = 0.01$ (black squares; bars represent the 95\% confidence interval), maximum of $\chi$ (red triangles; bars represent the half-amplitude width of $\chi$), and $R_\mathrm{eff} = 1$ (gray circles; 95\% confidence intervals are smaller than the symbol size).}
\label{fig:Finland_thr_comparison}
\end{figure}

Figure~\ref{fig:Finland_thr_comparison} shows the behavior of each quantity as a function of $\pi_\mathrm{R}$ and the values of the thresholds computed using the three definitions as a function of the inverse temperature parameter $\tau^{-1}$. Note that $\tau^{-1} = 0$ means that the connections are completely random with respect to the spatial position of the nodes, while the larger the values of $\tau^{-1}$, the more likely the nodes are to make contact with spatially nearby nodes. The three threshold values are similar to each other when $\tau^{-1} = 0$, i.e., in a pure SBM network without spatiality, although the threshold defined using $\pi_\mathrm{R_2} = 0.01$ is significantly higher than the mean-field threshold. As the network becomes more spatial, the two thresholds of the network model become higher than the mean-field threshold and the gap between them becomes wider, particularly for $\tau^{-1} \geq 5$. This implies the mean-field model provides a lower estimate of the herd immunity threshold if contacts between individuals are spatially constrained. 

